\documentclass[a4paper,12pt]{article}
\pdfoutput=1
\usepackage{epsfig}
\usepackage{amssymb}
\usepackage{amsfonts}
\usepackage{amsmath}
\usepackage{amsthm}
\usepackage{euscript}
\usepackage{verbatim}
\usepackage{latexsym}
\usepackage{graphicx}
\usepackage{caption}
\usepackage{float}
\usepackage{subcaption}
\usepackage{enumitem}

\usepackage{listings}

\newif\ifdtup

\catcode`\@=11

\@addtoreset{equation}{section}

\def\@normalsize{\@setsize\normalsize{15pt}\xiipt\@xiipt
\abovedisplayskip 14pt plus3pt minus3pt%
\belowdisplayskip \abovedisplayskip
\abovedisplayshortskip \z@ plus3pt%
\belowdisplayshortskip 7pt plus3.5pt minus0pt}

\def\small{\@setsize\small{13.6pt}\xipt\@xipt
\abovedisplayskip 13pt plus3pt minus3pt%
\belowdisplayskip \abovedisplayskip
\abovedisplayshortskip \z@ plus3pt%
\belowdisplayshortskip 7pt plus3.5pt minus0pt
\def\@listi{\parsep 4.5pt plus 2pt minus 1pt
     \itemsep \parsep
     \topsep 9pt plus 3pt minus 3pt}}

\relax

\catcode`@=12

\catcode`\@=11

\def\section{\@startsection{section}{1}{\z@}{3.5ex plus 1ex minus
   .2ex}{2.3ex plus .2ex}{\large\bf}}

\def\SymBoxes#1#2#3#4{\newdimen\un@t \un@t#3%
\raisebox{#1}{\rule{#2\un@t}{#4}\hskip-#2\un@t% lower horizontal
\@tempdimb\un@t \advance\@tempdimb by-#4\@tempcntb#2\relax%
\@whilenum{\@tempcntb>0}\do{%                         % #2 vertical lines
\rule{#4}{\un@t}\hskip\@tempdimb \advance\@tempcntb by\m@ne}%
\hskip-#2\un@t \rule[\un@t]{#2\un@t}{#4}%
\rule[\un@t]{#4}{#4}\hskip-#4%             % upper horizontal line
\rule{#4}{\un@t}}\hskip-#4}                % rightest vertical line
\begin{document}
%\begin{letter}{~}

%%%%%%Define some new commands and  macros
\newcommand{\beq}{\begin{equation}}
\newcommand{\eeq}{\end{equation}}
\newcommand{\bea}{\begin{eqnarray}}
\newcommand{\eea}{\end{eqnarray}}
\newcommand{\beas}{\begin{eqnarray*}}
\newcommand{\eeas}{\end{eqnarray*}}
\newcommand{\defi}{\stackrel{\rm def}{=}}
\newcommand{\non}{\nonumber}
\newcommand{\bquo}{\begin{quote}}
\newcommand{\enqu}{\end{quote}}
%%%%%%%%%%%%%%%%
\renewcommand{\(}{\begin{equation}}
\renewcommand{\)}{\end{equation}}
%%%%%%%%%%%%%%%%%%%%%%%%%%%%%%%%%% definitions
\def \eqn#1#2{\begin{equation}#2\label{#1}\end{equation}}

\def\e{\epsilon}
\def\IZ{{\mathbb Z}}
\def\IR{{\mathbb R}}
\def\IC{{\mathbb C}}
\def\IQ{{\mathbb Q}}
\def\de{\partial}
\def\Tr{ \hbox{\rm Tr}}
\def\H{ \hbox{\rm H}}
\def\HE{ \hbox{$\rm H^{even}$}}
\def\HO{ \hbox{$\rm H^{odd}$}}
\def\K{ \hbox{\rm K}}
\def\Im{ \hbox{\rm Im}}
\def\Ker{ \hbox{\rm Ker}}
\def\const{\hbox {\rm const.}}
\def\o{\over}
\def\im{\hbox{\rm Im}}
\def\re{\hbox{\rm Re}}
\def\bra{\langle}\def\ket{\rangle}
\def\Arg{\hbox {\rm Arg}}
\def\Re{\hbox {\rm Re}}
\def\Im{\hbox {\rm Im}}
\def\exo{\hbox {\rm exp}}
\def\diag{\hbox{\rm diag}}
\def\longvert{{\rule[-2mm]{0.1mm}{7mm}}\,}
\def\a{\alpha}
\def\dag{{}^{\dagger}}
\def\tq{{\widetilde q}}
\def\p{{}^{\prime}}
\def\W{W}
\def\N{{\cal N}}
\def\hsp{,\hspace{.7cm}}

\def\br{\nonumber}
\def\IZ{{\mathbb Z}}
\def\IR{{\mathbb R}}
\def\IC{{\mathbb C}}
\def\IQ{{\mathbb Q}}
\def\IP{{\mathbb P}}
\def \eqn#1#2{\begin{equation}#2\label{#1}\end{equation}}

\newcommand{\C}{\ensuremath{\mathbb C}}
\newcommand{\Z}{\ensuremath{\mathbb Z}}
\newcommand{\R}{\ensuremath{\mathbb R}}
\newcommand{\rp}{\ensuremath{\mathbb {RP}}}
\newcommand{\cp}{\ensuremath{\mathbb {CP}}}
\newcommand{\vac}{\ensuremath{|0\rangle}}
\newcommand{\vact}{\ensuremath{|00\rangle}                    }
\newcommand{\oc}{\ensuremath{\overline{c}}}
\newcommand{\psizero}{\psi_{0}}
\newcommand{\phizero}{\phi_{0}}
\newcommand{\hzero}{h_{0}}
\newcommand{\psiin}{\psi_{\rh}}
\newcommand{\phiin}{\phi_{\rh}}
\newcommand{\hin}{h_{\rh}}
\newcommand{\rh}{r_{h}}
\newcommand{\rb}{r_{b}}
\newcommand{\psibnd}{\psi_{0}^{b}}
\newcommand{\psibndp}{\psi_{1}^{b}}
\newcommand{\phibnd}{\phi_{0}^{b}}
\newcommand{\phibndp}{\phi_{1}^{b}}
\newcommand{\gbnd}{g_{0}^{b}}
\newcommand{\hbnd}{h_{0}^{b}}
\newcommand{\zh}{z_{h}}
\newcommand{\zb}{z_{b}}
\newcommand{\man}{\mathcal{M}}
\newcommand{\hbr}{\bar{h}}
\newcommand{\tbr}{\bar{t}}

\begin{titlepage}
\begin{flushright}
CHEP XXXXX
%ULB-TH/09-10\\
%hep-th/yymmnnn\\
\end{flushright}
\bigskip
\def\thefootnote{\fnsymbol{footnote}}

\begin{center}
{\Large
{\bf Page Curve and the Information Paradox in Flat Space\\
}
}
\end{center}

\bigskip
\begin{center}
Chethan KRISHNAN$^a$\footnote{\texttt{chethan.krishnan@gmail.com}}, \ Vaishnavi PATIL$^a$\footnote{\texttt{vaishnaviptl@gmail.com}}, \ Jude PEREIRA \footnote{\texttt{judepereira25@gmail.com }} 
\vspace{0.1in}

\end{center}

\renewcommand{\thefootnote}{\arabic{footnote}}

\begin{center}
%\vspace{0.2cm}

$^a$ {Center for High Energy Physics,\\
Indian Institute of Science, Bangalore 560012, India}\\

\end{center}

\noindent
\begin{center} {\bf Abstract} \end{center}
Asymptotic Causal Diamonds (ACDs) are a natural flat space analogue of AdS causal wedges, and it has been argued previously that they may be useful for understanding bulk locality in flat space holography. In this paper, we use ACD-inspired ideas to argue that there exist natural candidates for Quantum Extremal Surfaces (QES) and entanglement wedges in flat space, anchored to the conformal boundary. When there is a holographic screen at finite radius, we can also associate entanglement wedges and entropies to screen sub-regions, with the system  naturally coupled to a sink. The screen and the boundary provide two complementary ways of formulating the information paradox. We explain how they are related and show that in both formulations, the flat space entanglement wedge undergoes a phase transition  at the Page time in the background of an evaporating Schwarzschild black hole.  Our results closely parallel recent observations in AdS, and reproduce the Page curve. That there is a variation of the argument that can be phrased directly in flat space without reliance on AdS, is a strong indication that entanglement wedge phase transitions may be key to the information paradox in flat space as well. Along the way, we give evidence that the entanglement entropy of an ACD is a well-defined, and likely instructive, quantity. We further note that the picture of the sink we present here may have an understanding in terms of sub-matrix deconfinement in a large-$N$ setting.

\vspace{1.6 cm}
\vfill

\end{titlepage}

\tableofcontents

\setcounter{footnote}{0}

%%%%%%%%%%%%%%%%%%%%%%%%%%%%%%%%%%%%%%%%%%%%%%%%%%%%%%%%%%%%%%%%%%%%%%%%%%%%%%%%%%%%%%%%%%%%%%
%%%%%%%%%%%%%%%%%%%%%%%%%%%%%%%%%%%%%%%%%%%%%%%%%%%%%%%%%%%%%%%%%%%%%%%%%%%%%%%%%%%%%%%%%%%%%%
\section{Introduction}

It is often said that the AdS/CFT correspondence \cite{Maldacena, Witten} resolves the information paradox \cite{Hawking, Mathur, AMPS} in anti-de Sitter space. This is a plausible claim, because it is hard to imagine a version of the correspondence that holds at infinite $N$, but does not hold at finite $N$ \cite{Polchinski}. If one views information loss as an infinite-$N$ problem, then since infinite-$N$ is merely a useful limit of the unitary finite-$N$ ${\cal N}=4$ SYM theory, it would certainly be surprising if information were lost at finite but large $N$. Indeed, we know that unitary CFT correlators can look thermal when $N$ is taken to infinity, see eg., \cite{Anous}. %\footnote{We find these arguments reasonable, but not everyone does: see \cite{Hampton} for a counter-claim to that in \cite{Anous}.}.

But even though it is plausible that AdS/CFT resolves the information paradox, the situation is unsatisfactory. There do exist potentially holographic theories that are non-unitary at finite-$N$, but are seemingly unitary at any perturbative order in $1/N$ \cite{Vafa}.  More generally, it is the finite-$N$ effects of AdS/CFT that are the least understood, and maybe it is these effects that are needed for resolving the information paradox.  A more pragmatic concern is that even if AdS/CFT were true, we do not have a {\em bulk} understanding of the resolution of the information paradox. In other words, what is wrong with Hawking's original calculation? This point was first phrased sharply by Mathur using entanglement arguments in \cite{Mathur}, and it was later popularized in a closely related setting in the firewall paradox \cite{AMPS}. 

The firewall argument phrases the problem as a paradox that arises for old black holes after their {\em Page time}. Page time is the time at which the entanglement entropy of the Hawking radiation that has left the black hole equals its present Bekenstein-Hawking entropy. A smooth horizon for such a black hole would require that the Hawking radiation that is being emitted now, be maximally entangled with the modes behind the horizon\footnote{This can be viewed as a manifestation of the Principle of Equivalence. If the horizon is non-singular, one should be able to go to a free-fall coordinate system there (essentially Kruskal). In that frame, physics should look like conventional effective field theory (EFT) and the horizon should look like the vacuum state of the EFT. A basic feature of the vacuum of a local EFT is that it is entangled in terms of the modes in the local basis (ie., Hartle-Hawking quanta).}. But since we expect these interior modes to also be (maximally) entangled with the early Hawking radiation because the older Hawking radiation was also emitted by a smooth horizon, we have an immediate conflict with the monogamy of entanglement.

A simple (but highly nonlocal) way to fix the paradox would be to  declare that the interior modes are somehow secretly the same degrees of freedom as the early Hawking radiation. This possibility was recognized since the early days of the firewall paradox, see eg. \cite{AMPS, Nomura, Suvrat1, Verlinde, ER=EPR, QEC} for a representative list of discussions. But despite the simplicity of this idea, a mechanism for realizing such a proposal after the Page time, was not known. In particular, the {\em dynamical} reason behind such an identification was obscure, so it remained a prescription to protect the equivalence principle at the horizon.

Recently, a new idea has appeared in the work of \cite{Penington, Almheiri} (see also follow-ups in eg., \cite{coredump}). They argue that the key new ingredient is the notion of an {\em entanglement wedge phase transition}. The idea is to let an AdS black hole Hawking radiate onto a large sink Hilbert space $H_R$, while keeping track of the bulk entanglement wedges of both the CFT Hilbert space $H_{CFT}$ and $H_R$. The entanglement wedge of a sub-region of the CFT is usually defined as the bulk region (more precisely, the bulk domain of dependence) where bulk local operators can be reconstructed from the CFT sub-region. In the present scenario, this is generalized to allow the possibility that some sub-regions of the bulk might actually be supported in the reservoir $H_R$, since the radiation is allowed to leak out of the CFT Hilbert space. Before the Page time, the entanglement wedge of the boundary contains the entire bulk, including the black hole interior. But after the Page time, it is possible to argue (and we will, in a closely related flat space context) that the entanglement wedge of the CFT does not contain the black hole interior, instead the interior belongs to the wedge of the radiation $H_R$. This is an explicit  realization of the idea that the black hole interior can now be identified as belonging to the early radiation Hilbert space, providing a new dynamical mechanism for tackling the information paradox. 

One of the main points of these results is that  they aim to give a {\em bulk} understanding of the origin of the Page curve. Interestingly, it is only the general expectations about entanglement and holography that go into these arguments. The details of the CFT do not seem to play a role, except some of the consequences of the fact that it is a local theory. In fact, we will see that even conformal invariance is not essential! This is interesting, because it raises the possibility that these arguments might generalize beyond AdS. Another relevant observation is that the arguments of \cite{Penington, Almheiri} rely on generalized entropy, which is expected to be a sensible quantity to all orders in bulk semi-classical perturbation theory \cite{QFC, EngelWall}. The bulk metric in this approach is quantum corrected and therefore respects the Generalized Second Law (GSL) and not the Null Energy Condition (NEC), but there is nothing particular about these facts that is tied to AdS. It is therefore not implausible that there is a variation of the generalized entropy idea that can be useful in the bulk of flat space.

These speculations naturally lead us to the question: is it possible to adapt enough of the relevant AdS structures to flat space, so that we can apply the arguments of \cite{Penington, Almheiri} to flat space black holes? If true, this would imply that (variations of) entanglement wedge phase transitions are the key to the Page curve and information paradox in flat space as well. Clearly, this is a question of interest because black holes in flat space are generally believed to be of significance to the ``real world"\footnote{It is perhaps worth emphasizing here that the real world is actually cosmological and not asymptotically flat. But since we expect quantum gravity in flat space  to be a well-defined theory (at the very least in 10 dimensions where perturbative string theory seems clearly well-defined), and since we know that in the semi-classical limit it contains black holes, the information paradox {\em needs} a resolution in flat space. Black holes in cosmologies do not offer such a sharp paradox, because they are inextricably tied to the difficulties with time-dependent backgrounds and cosmological horizons in quantum gravity.}. In the following sections, we will see that even though locality in the usual sense almost certainly does not hold in the hologram of flat space, enough of the necessary features remain, so that a suitable chain of arguments parallel to those in \cite{Penington, Almheiri}, hold. %It is perhaps worth emphasizing here at this stage, that the real world is actually cosmological and not asymptotically flat, even though this does not seem to bother many purists. Our results can be viewed as evidence that the distinction between asymotically AdS and asymptotically flat, are less drastic than the distinction between asyptotically flat and cosmological.  

We start by presenting our ACD/holographic screen infrastructure first, and give a quick summary. The reader who is primarily interested in our claims about the information paradox might want to skip directly to the conclusion section.

\subsection{The Big Picture} 

%\footnote{Note that a generic subregion can be obtained as a possibly infinite union of spherical subregions. Arbitrary unions are well-defined in point set topology, it is only intersections that are required to be finite.} 

\begin{figure}[h]\centering
\hspace{-5mm}
\includegraphics[angle=0,width=110mm]{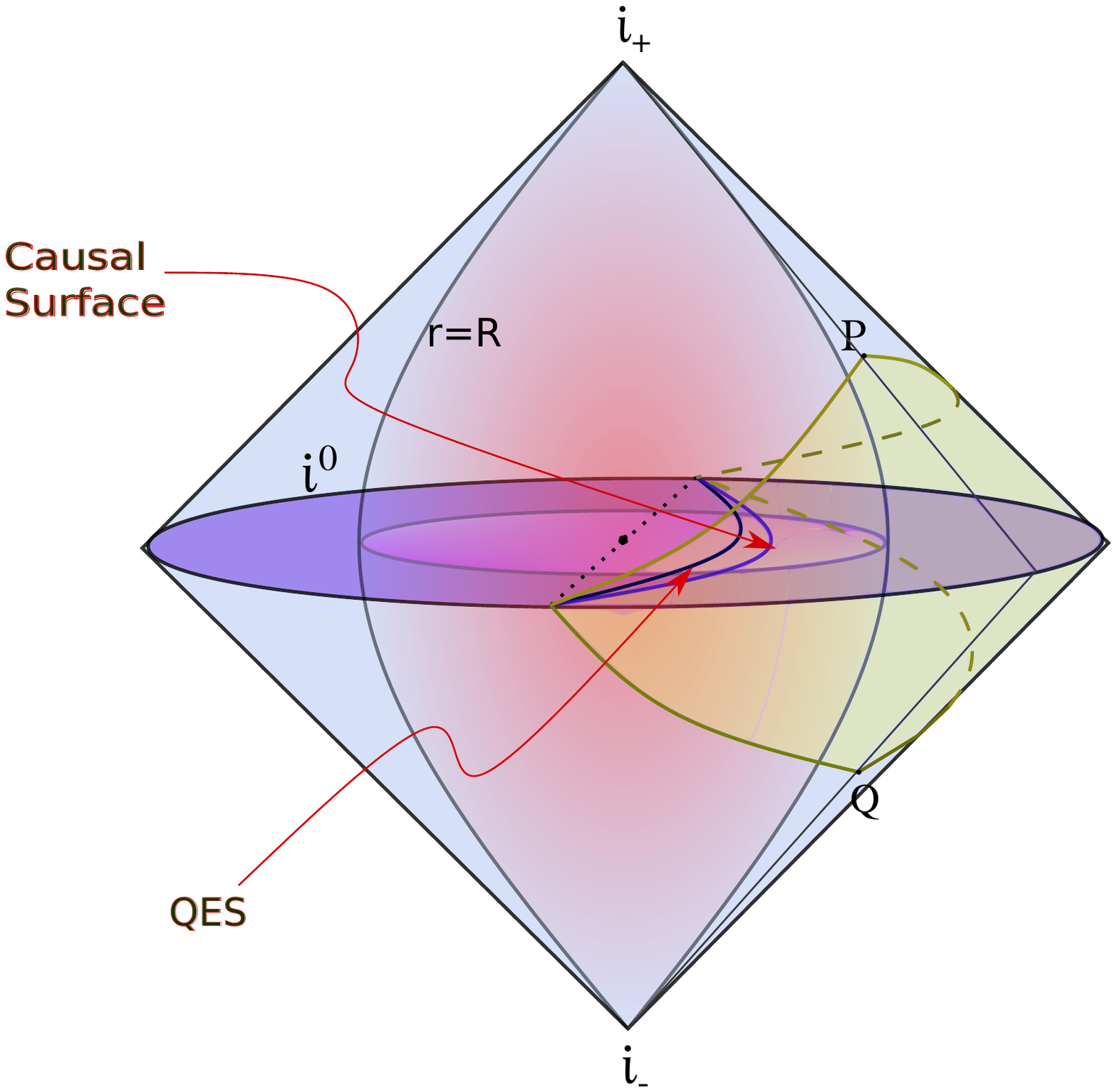} \\
\vspace*{-4cm}
\caption{The ACD defined by vertices $P, Q$, its causal surface and its Quantum Extremal surface (QES) are shown. This data defines the entanglement/reconstruction wedge completely. We have also shown the holographic screen. The shadow of the ACD is its intersection with the conformal boundary.}
\label{entwedge}
\end{figure}

A key ingredient in our work will be the notion of an Asymptotic Causal Diamond (ACD), introduced in \cite{CK}, and the entanglement wedge associated to it. ACDs are bulk causal diamonds whose vertices are attached to points on the null boundary of flat space. These vertices define them. As is discussed in detail in \cite{CK, future}, the data described by these objects on the conformal boundary of flat space is parallel to the data of a spherical sub-region on the boundary of AdS. A spherical sub-region on the AdS boundary defines a boundary causal diamond, and vice versa. A key fact is that (quantum) extremal surfaces in AdS can be defined by the boundaries of sub-regions to which they are anchored to. This implies that these extremal surfaces can {\em also} be defined by the vertices of the boundary causal diamonds that define these sub-regions. In flat space, we do not have a notion of boundary causal diamonds, but the observation of \cite{CK} was that ACDs can be used to define equivalent data on the boundary. See figure \ref{entwedge} for some quick intuition.
 In particular, by analogy with AdS, we can define the domain of dependence of the region between the QES and the shadow of the ACD to be the entanglement wedge\footnote{In the bulk, this object is the precise analogue of the AdS entanglement wedge. The holographic dual of flat space quantum gravity is not explicitly known. But we will briefly allude to evidence that its entanglement/locality structure is quite different from that of the CFT in AdS/CFT, see \cite{future} for details.}  associated to the ACD. These objects will be one of the key ingredients in our discussions in this paper. When the vertices of the ACD climb up/down to the future/past timelike infinity, the entanglement wedge of Minkowski space becomes the entire spacetime.  Sometimes we will use the phrase ``entanglement wedge" without any qualifiers to refer to this   type of a ``global" object.
 
Along with the conformal boundary, we will also find it useful to have the notion of a holographic screen (of finite but large radius) in flat space, introduced previously in \cite{CK}. See also a closely related, but different aspect of the screen in \cite{Budha}. We will present evidence that to describe the evolution of the black hole states that we are after, we can treat the Hilbert space of the full quantum gravity to be an approximate tensor factorization between the interior and the exterior of the screen. The exterior factor is a sink; it has effectively infinite phase space volume. For large values of the cut-off size $R$ with
\bea
R \gtrsim \frac{M}{ M_{\rm Pl}^2}  \label{large}
\eea
where $M$ is the initial mass of the black hole (that formed by collapse) and $M_{\rm Pl}$ is the Planck scale, this approximation of tensor factorization is a good one. A stronger assumption $R \gg \frac{M}{ M_{\rm Pl}^2}$ may be necessary if one wants to distinguish between Hawking radiation near the cut-off and near the zone region, but this is a technicality for us in most of our discussions. Note that the semi-classical limit corresponds to 
\bea
R M_{\rm Pl} \gg 1, \ \ \frac{M}{ M_{\rm Pl}} \gg 1
 \eea
which is conceptually distinct. We will typically assume that these relations also hold -- after all, we want to be working in a context where a well-defined semi-classical information paradox exists, before we can discuss its solution. Hawking radiation fits into this picture: the factorization formalizes the prejudice that when radiation is far from the black hole, non-gravitational effective field theory (which we will take to include free gravitons) should be able to describe it.  More generally, there exists an asymptotically flat chart in a finite neighborhood of infinity, where  bulk effective field theory is sufficient to describe the states under consideration, and the deviation of the metric from flatness is small\footnote{Eqn \eqref{large} can also be viewed as a type of scale separation in the dynamics one expects in flat space quantum gravity. We expect that when describing the evolution/evaporation of a black hole that satisfies \eqref{large}, we can ignore the effects of states with $ \frac{M}{ M_{\rm Pl}^2} \gtrsim R $ (which clearly violate the tensor factorization). }. Note that we expect the asymptotic region to be low curvature during the entirety of Hawking evaporation. Finally, let us emphasize that our screen is expected to be relevant only when describing the dynamics of specific classes of states, so it should not be confused with other uses of the word ``holographic screen" in the literature (eg., \cite{Sanches Weinberg}). See \cite{CKcore, Ajay} for some other approaches to screens in the context of gravity/holography.

%One might worry that a condition like \eqref{large} will require further conditions to ensure that the within-the-screen region ``contains" all the relevant gravitating objects. Are boosted black holes included? Wouldn't they at some point cross the screen? Perhaps one should ensure that we are working in the frame of the center of mass of the bulk system? What if we have a pair of black holes with zero net momentum? Wouldn't they cross the screen? Remarkably all of these problems go away with the simple demand that the spacetime is asymptotically flat, and that the screen lies in the asymptotically flat chart. We will discuss  this further later, but let us emphasize here that Hawking radiation is consistent with this.  More generally, the tensor factorization picture can be viewed as the statement that there exists an asymptotically flat chart where  bulk (non-holographic) effective field theory is well-defined for all bulk states under consideration.  In other words, asymptotic flatness is taylor-made to localize gravitating systems. 

A key point about the holographic screen in flat space that distinguishes it from a cut-off in AdS, is that the cut-off going to infinity limit is a tame limit in AdS. This is because in AdS the boundary is only a ``finite distance" away, and therefore placing a reflecting boundary condition on the cut-off can approximate the reflecting boundary condition at the conformal boundary. Indeed, if we set transparent boundary conditions at an AdS cut-off, because of the finite propagation time to the boundary, it will reflect back into the cut-off soon enough anyway. This is not the case at all in flat space, because the conformal boundary is ``infinitely far away". This makes it natural to view the outside-the-screen tensor factor in flat space, as a sink for the interior. When one is dealing only with states satisfying \eqref{large}, the region outside the cut-off is simply an infinite, non-gravitational Minkowski space into which (say) Hawking radiation from the interior can free-stream. Note that the backreaction in the outside-the-screen region is arbitrarily small, when \eqref{large} holds. 

A distinct heuristic argument and motivation for our flat space screen/cut-off, emerges from intuition about AdS/CFT and large-$N$ gauge theories. This is in terms of deconfinement in a submatrix of the $N \times N$ matrix. Since the motivation here is a bit different, we will elaborate on it in the Conclusion section.

It turns out that the interior tensor factor has many holographic features familiar from AdS. We will be able to associate  extremal/maximin surfaces to sub-regions on the screen as well\footnote{Note that entanglement wedges on the screen should be distinguished from the entanglement wedges we described earlier, on the boundary. The distinction between these two types of quantities will be a running theme in this paper. The context should hopefully be enough to clarify which one we are referring to. Occasionally we will call the screen-based quantities, {\em relative} quantities to emphasize the distinction.}. The areas of classical extremal surfaces associated to these screen sub-regions turn out to satisfy strong sub-additivity. These facts are suggestive that suitably defined sub-region entanglement entropy on the screen may be a key element in understanding flat space holography. But how precisely should one define such an entropy? Is it simply the sub-region entropy as it is in AdS?

To get a hint, we first observe that sub-regions and their entropies are only approximately defined quantities at finite cut-off\footnote{But they are in some sense easier to picture than the exact entanglement wedges we discussed earlier, because they anchor to the timelike holographic screen much like in AdS.}, and therefore should be viewed as regulated versions of a more fundamental quantity that is defined with respect to the conformal boundary.  Note that this is true for sub-regions on a cut-off in asymptotically AdS spaces as well. But there are a couple of key differences here, when compared to AdS. One is that as we discussed above, the region outside the cut-off does not qualify as a sink tensor factor in AdS. Therefore, removing the regulator does not qualitatively change the physics. A manifestation of this is that in AdS, we can define sub-regions not just at the timelike cut-off, but also at the conformal boundary which is timelike and has well-defined time evolution. The situation is clearly quite different in flat space, where (unlike the screen) the asymptotic boundary has no well-defined causal structure at all. In fact the structure of $i^0$ does not allow a meaningful notion of sub-regions. These are hints that when we remove the regulator in flat space with a screen, we are in fact including the sink as well, and the system is qualitatively changing.  We claim that the natural object that one should associate entanglement entropies to at the asymptotic boundary of flat space, are ACDs (or more precisely, shadows of ACDs) and not sub-regions of $i^0$. Note that this is consistent with our previous statement that ACDs are the analogues of AdS boundary sub-regions. We will also see that ACDs provide us with a natural way to pick on-screen sub-regions associated to them. This involves some technicalities and approximations (just as in AdS) because the screen is at finite cut-off, but the interpretation that the on-screen quantity is a regulated version of the more fundamental quantity defined at the boundary, remains intact. 

In AdS, the natural sub-region entropy one defines is that of the reduced density matrix of the sub-region with its complement. In bulk effective field theory, this is given by the prescription that one has to compute the so-called generalized entropy of the quantum extremal surface anchored to the sub-region. On the flat space holographic screen, the natural generalization is to define sub-region entropy on the screen, where the reduced density matrices are computed after tracing both over the complementary sub-region as well as the sink Hilbert space outside the screen\footnote{Note that such quantities are interesting to define in AdS as well, but one will have to explicitly couple a sink Hilbert space to the AdS/CFT system before they make sense.}. A very natural bulk EFT prescription for evaluating this can be given: it takes the form \eqref{screenF}, the details will be elaborated there. This can be viewed as a generalization of the generalized entropy idea when the screen is attached to a sink. Compelling evidence can be given that it satisfies strong sub-additivity. In the classical limit, it also gives us an understanding why extremal surfaces on the screen showed up even though we are in flat space, and why they satisfy strong sub-additivity. 

But it goes further. According to our earlier philosophy, this entropy should be viewed as a regulated version of the entanglement entropy of a causal diamond at the conformal boundary of flat space. This means that we can try to find a renormalization prescription to remove the regulator. The problem of regulating and renormalizing areas of extremal surfaces has not been developed too much in AdS (but see \cite{Sorce}). This is largely because as we mentioned earlier, the AdS boundary can be fairly intuitively related to the AdS cut-off. But it seems clear that in flat space, developing an analogous formalism will be a very useful piece in our understanding of flat space holography. We plan to come back to this problem in future work \cite{future}, but since our primary objective in this paper is the information paradox, we will settle for a crude background subtraction prescription with respect to Minkowski space to define finite quantities. We do not believe this is the final word on renormalizing these quantities, but it has two virtues that make it interesting: 
\begin{itemize}
\item A part of the subtraction we do has a very intuitive interpretation as including the modes from the sink outside the screen. Note that the entropy should decrease when we include the degrees of freedom from the sink because it purifies the interior, which is indeed what subtraction does. Note also that geometrically, including the sink is precisely what we need to get to the conformal boundary from the screen.
\item The potential subtleties of the renormalization prescription really only affect the extremal surfaces anchored to proper sub-regions on the screen (or proper ACDs). When the sub-region in question is the entirety of the screen spatial slice, these issues can be easily seen to be irrelevant. To make the argument about the phase transition at the Page time, we only need this ``global" entanglement wedge, so the details of renormalization will not matter for that.  
\end{itemize} 

One of the punchlines of this paper will be that we can phrase and resolve the information paradox (and obtain the Page curve) for flat space black holes from these constructions. We will demonstrate this via two formulations. The first formulation uses the holographic screen to split the Hilbert space into two (approximate) tensor factors like we discussed above. The first factor is the interior and contains the black hole, the second is the exterior and can be viewed as the sink. We will show that the approximate entanglement wedge (the one defined from the screen) undergoes a phase transition at the Page time. In other words, unlike in AdS where the system had to be coupled to a heat sink, flat space can be viewed as coming with its own sink. We will further elaborate on this later, but this captures the intuition that Hawking radiation can ``leave the system".

It is also possible to treat the full quantum gravity Hilbert space in flat space as a single tensor factor, and extract Hawking radiation by coupling this system to an external sink, via a holographic source. We call this, Formulation 2. This is similar to the picture in \cite{Penington, Almheiri} where the Hawking radiation left the AdS/CFT system via a coupling at the boundary. Though perhaps less familiar to the reader, a parallel construction can be done in asymptotically flat spaces where the source lives on a codimension-1 holographic screen \cite{Budha}\footnote{The situation where the coupling to the sink is at null infinity can be viewed as a limiting version of Formulation 2  -- this is a set up that may be a bit more familiar to the reader who has not seen \cite{Budha}. The key point in either case is that without a coupling to an outside tensor factor, we cannot see the radiation escape in this picture, because the system is defined by the full Penrose diagram. Let us also emphasize that the role of the screen in our two formulations is quite different. These matters will be discussed in great detail, later. }. This leads again to the correct Page curve, but now we are extracting Hawking radiation out of the quantum gravity Hilbert space. We elaborate on this picture as well in a later section. In some ways Formulation 2 is cleaner because it deals directly with the exact entanglement wedge defined with respect to the conformal boundary, but because the sink is external to the system, interest in it is perhaps more formal. 

In later sections, we will elaborate on various aspects of both these approaches and their inter-relationship. The main goal of our paper is to establish the necessary flat space infrastructure in terms of ACDs and related ideas. Once they are in place, the calculation of the phase transition goes through automatically as in \cite{Penington, Almheiri}. 

% we only need the exact entanglement/reconstruction wedge we discussed at the beginning of this section, so a reader who prefers to be blissfully unaware of the screen may wish to skip to section 5. But much of our intuition comes from discussions of the screen, so we suspect even the purely business-like reader may find something of interest in what goes before section 5. We also suspect, in light of the discussions in \cite{CK, future, Budha}, that the screen is of some intrinsic significance for flat space holography, even though the correlators on it are non-local. 

\subsection{The Main Technicality}

Together with a holographic screen at a finite but large radial cut-off, it was shown in \cite{CK} that ACDs can reproduce (in Minkowski space) many of the causal/entanglement aspects of holography, including quantum error correction \cite{QEC}. A key point is that in flat space, many questions become clearer when we do {\em not} conflate the two logically distinct objects: the asymptotic boundary and the holographic screen. In AdS, picking the screen to coincide with the asymptotic boundary leads to perfect decoupling of gravity, and therefore to simplifications. In flat space on the other hand, we expect the holographic dual to be non-local anyway, so it is less clear that using the asymptotic boundary as a screen is advantageous {\em a-priori}\footnote{Note however from our previous discussion that there is a useful generalization of the ``boundary sub-region" idea, the shadow of the ACD, that does make sense directly in the conformal boundary of flat space. But the dual theory is almost certainly non-local.}. In fact, since we will be demanding only that gravity is weak and not zero on the screen, there is some freedom in the choice of the holographic screen (just like there is some freedom in choosing a finite cut-off in AdS). This means that we expect this description of holography to be ``screen-covariant" in some suitable sense. Indeed we will see that the ideas and properties (eg., strong subadditivity) that are described in \cite{CK} and this paper, do not depend on many of the details of the screen, but merely on the fact that there {\em is} a screen. It should also be noted that just as quantum extremal surfaces and entanglement wedges are supposed to make sense in semi-classical perturbation theory to all orders in the bulk, we expect these screens to also make sense in the same approximation. The implicit assumption here is that the bulk metric, though quantum corrected, is still a well-defined quantity. In passing, let us also note that the semi-classical holographic correspondence between the bulk and the screen can be developed quite a bit \cite{Budha}, with striking parallels to the conventional AdS/CFT correspondence \cite{Witten, HKLL}. The key difference of course in many of these discussions being that the on-screen correlation functions one finds are not those of a conventional local theory. The results of this paper can be viewed as a hint that this non-locality is of a relatively special type.

The arguments of \cite{CK} were limited to the ground state,  aka empty Minkowski space. In this paper, we want to extend it to more general asymptotically flat settings. As mentioned, we find that there exist natural adaptations of extremal surfaces \cite{HRT} and entanglement wedges \cite{EngelWall}, that can be associated to both the holographic screen and the conformal boundary. The former are {\em approximate} in the sense that defining them in terms of screen sub-regions is meaningful only upto subleading corrections in (powers of) the radius of the cut-off. We wish to relate the two ideas, further strengthening the connection to related AdS objects. In AdS, the extremal objects are defined via sub-regions on the asymptotic boundary at some $t=0$ slice, and the anchoring surfaces for bulk extremal surfaces are the boundaries of sub-regions. As we will see,  when we are working with a holographic screen at finite cut-off, there is a bit of subtlety in picking an anchoring surface. When the cut-off is large however, all possible choices coincide up to subleading corrections. 

% because of the presence of bulk diffeomoprhisms that die down only at the boundary

Let us elaborate this with the following comments. While it is straightforward enough to pick an asymptotically Minkowski time coordinate $t$ and then use it to cut out a spatial slice on the holographic screen, the boundaries of sub-regions on the cut-off on the $t=0$ slice come with some unique challenges. One such feature we will emphasize is tied to the definition of a Quantum Extremal Surface (QES). To make the Page curve argument, we need to work with QES and not just classical extremal surfaces. Classical extremal surfaces can be anchored to boundaries of the sub-regions on large enough cut-offs, and they are well-defined by extremizing the area. But QES are defined as bulk codimension-2 surfaces that extremize the generalized entropy, which includes contributions not just from the (regulated) geometric area of the surface but also from the bulk entanglement entropy of quantum fields. To define the latter the surface needs to split a bulk Cauchy surface into two disconnected pieces.  This leads to two observations: 
\begin{itemize}
\item For defining sub-region entropies on the screen, we can use the sub-region itself to close the extremal surface anchored to the boundary of the sub-region. This splits the bulk into two disconnected pieces and we can calculate the bulk entanglement entropy. We will see that this approach automatically leads to a natural EFT prescription \eqref{screenF} for computing the entanglement entropy of the sub-region when it is coupled to a sink (ie., the outside-the-screen part of the spacetime).
\item For connecting the sub-region extremal surfaces to the more Platonic entanglement wedges, we need to have a prescription for relating the latter to on-screen quantities. We will now argue that for an asymptotically flat space with a screen, the true entanglement wedge of an ACD can be related to an approximate screen-dependent entanglement wedge, which can be thought of as a regulator for the former.   
\end{itemize}

%This will be our operational approxination to the reconstruction wedge, which is naturally assoicated to ACDs and {\em not} to subregions at $i^0$ (as we will explain). The Page curve will be obtained via a phase transition in the reconstruction wedge. Entanglement/reconstruction wedges are bounded by QES, so we {\em must} address the problem outlined in the previous paragraph to even get off the ground.

The key input we will use is to first go back to AdS and to note that sub-regions on the AdS boundary can be defined in an alternate way using {\em boundary} causal diamonds, and then use that as the jumping off point for a flat space generalization. Note that boundary causal diamonds\footnote{To be fully precise, we should consider symmetric boundary causal diamonds with respect to some $t=0$ slice of the boundary. These have vertices that are at $t=T$ and $t=-T$ for arbitrary $T$, and spatial coordinates identical.} \cite{Czech,Myers} in AdS live in Minkowski space and so they can be used to define spherical sub-regions, and through their (potentially infinite) unions, arbitrary sub-regions. Moreover, the moment we move the boundary to a finite cut-off, the problem we mentioned in the flat space case, arises {\em also} in AdS. It becomes unclear how boundaries of sub-regions on this cut-off can be used as anchoring surfaces for QES, because they also require a prescription for being extended to the conformal boundary. Note that this point is typically under-emphasized in AdS, because there exists perfectly well-defined sub-regions on the conformal boundary, so the subtleties at the cut-off can usually be glossed over\footnote{If one wants to define sub-regions at the cut-off, the approach we outline in this paper is a very natural one, both in AdS and flat space. This can be viewed as a ``covariant'' approach to the problem, because we rely on the waists of ACDs to do this. When discussing QES, we found this to be natural. See \cite{Sorce} for alternate approaches to defining sub-regions and entropies on cut-off AdS that are more ``canonical''. }.

A natural and simple workaround is to first note that a spherical sub-region on the asymptotic AdS boundary can also be defined via a {\em bulk} AdS causal wedge attached to the boundary causal diamond. The intersection of the boundary with the waist of the bulk causal wedge gives the boundary of the boundary sub-region. This suggests that instead of considering boundaries of sub-regions on a cut-off, we consider codimension-3 surfaces on the cut-off that are cut out by the waist of the bulk causal wedge. To define quantum extremal surfaces anchored to such surfaces, there exists a very natural prescription: extend the screen QES to the conformal boundary beyond the cut-off {\em along} the waist of the causal wedge. The idea here is that since the cut-off is large, the error one incurs by doing this, as opposed to anchoring directly to the conformal boundary will vanish in the limit of infinite cut-off\footnote{Note that the error here is from multiple sources, all related to the fact that we are working at finite cut-off. The first is that one needs to choose a prescription for what kind of variations one is allowing, when one is varying the surface to extremize the generalized entropy. For the screen QES, we will limit our attention to only variations in the bulk within the cut-off. This is one source of error. Note that because extremal surfaces coincide with causal surfaces in empty AdS, this error becomes smaller and smaller as we make the cut-off larger and larger. A second error appears when we work with a non-trivial geometry (instead of empty AdS) in the bulk. This also leads to errors that are sub-leading in the cut-off in asymptotically AdS spaces. A related issue is that the waist of the bulk casual wedge need not lie entirely on a bulk $t=0$ slice even if the waist of the boundary causal diamond does -- this is because the $t=0$ slice has some leftover diff freedom in the bulk.  For asymptotically AdS coordinates, this error dies down at large cut-off.}. %Note also that since the cut-off is large, one also expects to be able to approximate (boundaries of) arbitrary subregions on the cut-off, using unions of causal wedges intersecting with the cut-off. This is the cut-off version of the statement that arbitrary subregions on the conformal boundary can be viewed as unions of spherical subregions.

This entire chain of reasoning can be transliterated {\em mutatis mutandis} to asymptotically flat space, when we have a finite cut-off. The only difference is that there is no analogue of a boundary causal diamond in the conformal boundary of flat space, because the boundary of flat space does not have an intrinsic causal structure. But this does not prevent us from working with the flat space analogue of a bulk causal wedge, the ACD\footnote{For an asymptotically Minkowski time coordinate $t$, we should introduce the notion of a symmetric ACD whose vertices will be symmetric with respect to $t=0$. This is a fairly obvious generalization of the symmetric ACD in Minkowski space \cite{CK}, and we will elaborate on this idea further in the next section.}. The fact that this object is eminently well-defined was the message of \cite{CK}, and it can be used to define QES anchoring surfaces on the cut-off in precise analogy with our AdS discussion. This then provides us with a way to specify our approximate QES.  The key point which we are relying on in all of these discussions is that ACDs are well-defined objects in asymptotically flat space. %especially outside a large enough cut-off.

% Our approximate entanglement wedges will be defined via the bulk domains of dependence associated to these QES, and when there are unions, they will be the unions of these entanglement wedges.

One subtlety that one might worry about in this context is that unlike in AdS, the asymptotic limits of bulk extremal surfaces in flat space must tend to straight lines (or hyperplanes in $d+1$-dimensions with $d>2$), and therefore there is no meaningful sense in which one can associate a boundary sub-region to the spatial boundary at $i^0$ \cite{Ashtekar}. But this is okay, because the key object that needs to be well-defined is the entanglement wedge, and we will argue that this object is as well-defined as the AdS entanglement wedge of a spherical sub-region is. See also \cite{future} for further discussions\footnote{Note that there are differences in detail -- for example, in flat space the entanglement wedge of the unions of two ACDs can be the entire spacetime even when they are not complimentary. This is related to the non-locality of flat space holography.}. The key observation that gets us off the ground is the one made in \cite{CK}, that pairs of points on the future and past null boundaries of flat space contain the same information as a spherical sub-region contains at the boundary of AdS.

Much of the content of the next two sections will be about elaborating on the discussions above: about approximate sub-regions on a holographic screen. We will argue that various ideas that are familiar from AdS, eg., maximin surfaces, extremal surfaces, strong sub-additivity, and the like can be given suitable flat space counterparts on the screen when the spacetime is asymptotically flat and the screen is suitably large. Section 4 contains both a discussion of entanglement wedges for ACDs as well as the details of the various aspects of the information paradox in this setting. In the final section, we will present a general discussion of information paradox in light of the recent developments, and how one should think of our work in the broader context.

\section{Asymptotic Causal Diamonds and Holographic Screens}

We start by setting up the context of the paper by recalling the definition of an Asymptotic Causal Diamond (ACD) from \cite{CK}, and then formally defining a radial cut-off/holographic screen in flat space. 

\noindent
{\bf Definition 2.1:} An {\em Asymptotic Causal Diamond}, $\mathfrak{C}(p,q)$, is defined as the intersection of the past light cone of a point $p$ at future null infinity $\mathfrak{I}^+$ and the future light cone of a point $q$ at past null infinity $\mathfrak{I}^-$. That is
\bea
\mathfrak{C}(p,q)={\cal I}^-(p) \cap {\cal I}^+(q), \ {\rm where} \ p \in \mathfrak{I}^+, \ q \in  \mathfrak{I}^-.
\eea
We define the {\em shadow} of an ACD to be its intersection with the entire conformal boundary.

In \cite{CK} this definition was applied to the vacuum Minkowski space to suggest that ACDs are useful for  understanding bulk local questions from a holographic point of view. In this paper, we will apply the same idea to more general asymptotically flat spacetimes, in particular one containing an evaporating Schwarzschild black hole. Unless otherwise specified, by flat space we will always mean asymptotically flat space and not merely Minkowski space in the present paper. In empty Minkowski space, ACDs can be understood as a generalization of (spherical) Rindler wedges, but in more general asymptotically flat spaces they do not have such an interpretation. It should also be kept in mind that ACDs are defined via pairs of points at null infinity and not via (say) trajectories of accelerated observers in the bulk or via boost generators and such, so conceptually the two are quite different. Again, these differences are clearest when one is not in the vacuum Minkowski space.

A key observation of \cite{CK} was that a radial cut-off played a significant role in many of the arguments. Despite its usefulness, the precise details of the cut-off did not seem to matter too much for the claims of \cite{CK}, and  the arguments would go through for large classes of cut-offs. It was suggested that the cut-off should be understood as an operational tool for defining a localized gravitating system in asymptotically flat space\footnote{In other words, one can talk about localizing a gravitating system when the (curvature) length scales involved ($\ell$) are all small compared to the cut-off radius $R$, ie., $\ell/R \ll 1$.}. In this paper we will call this cut-off, a {\em holographic screen}. The fact that large classes of cut-offs can be used to describe the bulk physics in isomorphic ways will be viewed as a manifestation of {\em screen covariance} of holography for sufficiently asymptotic screens. Our holographic screen has clear analogies to the cut-off in AdS. The thing that is distinct about AdS is that there exists a (for many purposes, non-degenerate) limit where the holographic screen becomes the conformal boundary. A special thing about this limit in AdS is that gravity decouples. Just like the finite cut-off situation in AdS, we do not expect gravity to decouple on the holographic screen in flat space\footnote{Let us emphasize therefore that the reader should not blindly attribute to our flat space holographic screen, features that might be intuitive in the AdS/CFT context.}. Note however that to all orders in perturbation theory in the Newton constant, the bulk metric is still a well-defined object (even though it is not a solution of tree level Einstein equations).  Quantum extremal surfaces are also defined \cite{EngelWall} to all orders in perturbation theory. %, and this is why the holographic screen will be a useful tool for us in defining related ideas. 

%Therefore the screen is a perfectly well-defined bulk surface to all orders in perturbation theory.

 %where the goal was to put the arguments of \cite{Penington, Almheiri} on a solid footing. We will go back and forth in using the Dyson sphere terminology in what follows when referring to our cut-off. 

\begin{figure}
        \begin{subfigure}[h]{0.45\textwidth}
                \centering
                \includegraphics[width=.85\linewidth]{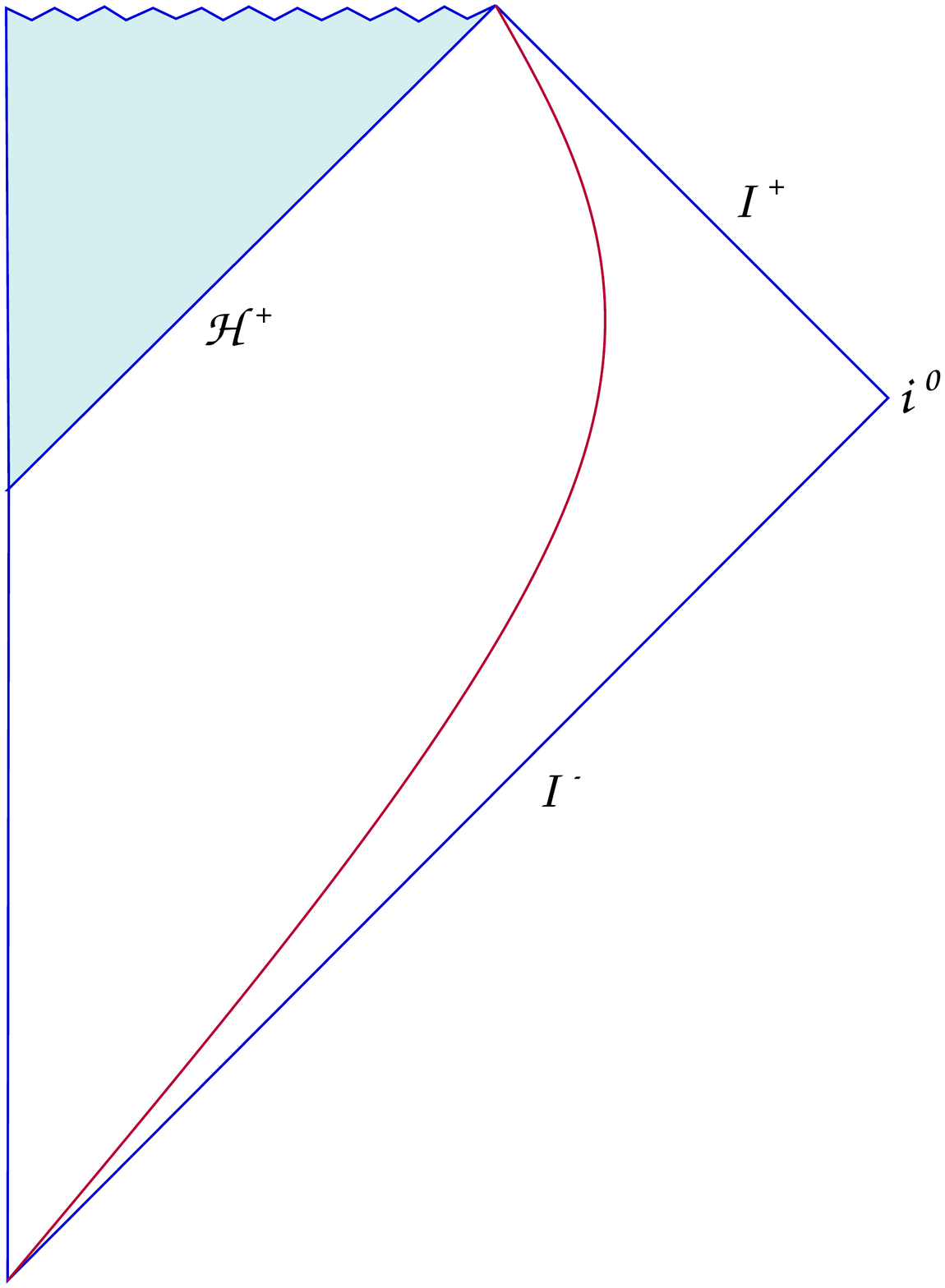}
              
                \caption{A surface visible from  boundary.}
                \label{XX}
        \end{subfigure}%
        \begin{subfigure}[h]{0.45\textwidth}
                \centering
                \includegraphics[width=.85\linewidth]{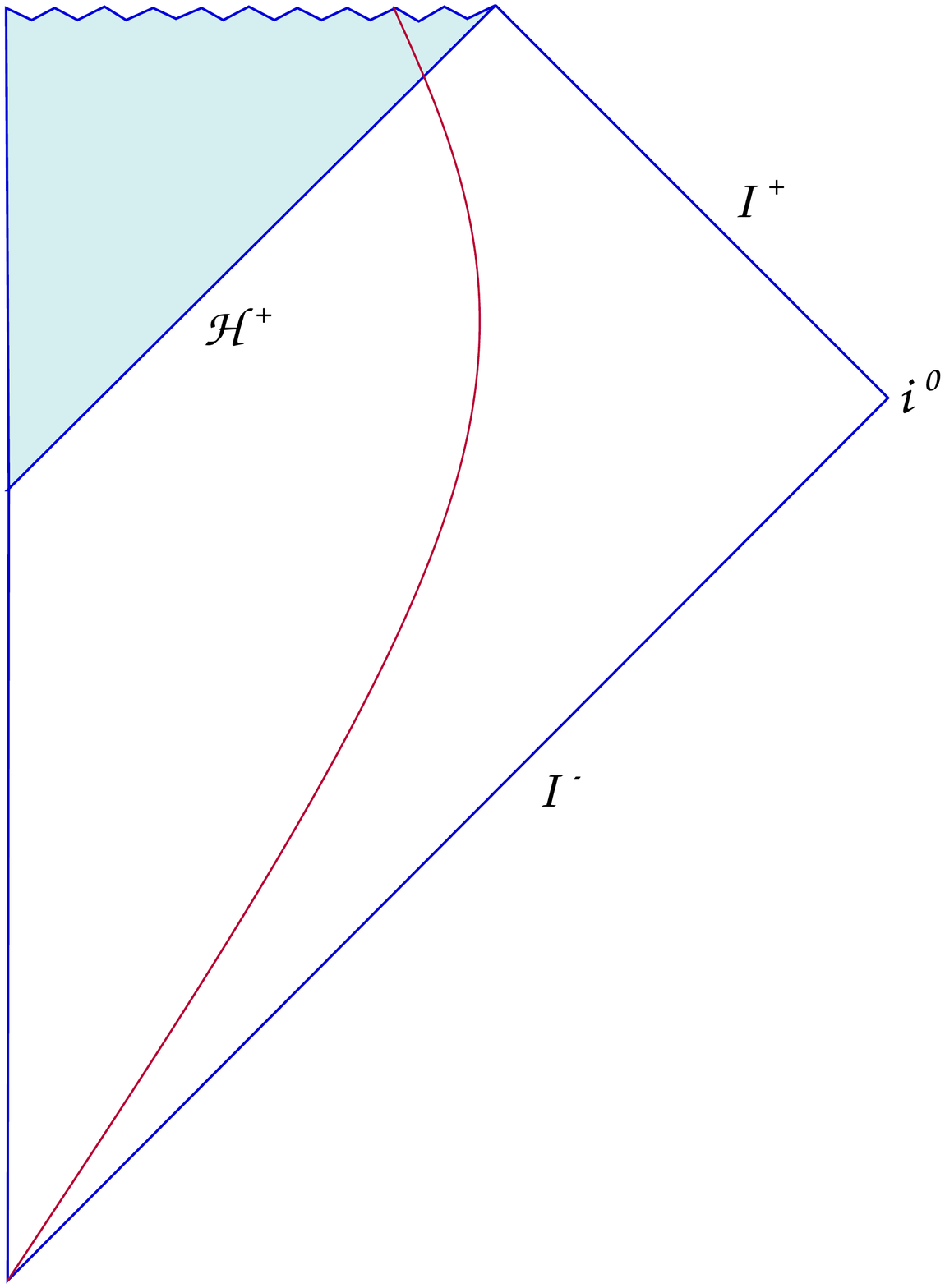}
                
                \caption{A surface that violates 2.2 (c).}
                \label{YY}
        \end{subfigure}%
        \caption{Comparing acceptable and unacceptable holographic screens.}\label{Dind}
\end{figure}

Following this philosophy, we will consider cut-off surfaces which are to be viewed as enclosing the localized gravitating system, i.e, the evaporating Schwarzschild black hole (except for the Hawking radiation that radiates out to the conformal boundary). A corollary of this requirement is that the cut-off/screen should  be sufficiently ``asymptotic". It should {\em always} be visible from the null boundaries and should never go behind trapped surfaces:  the outgoing null geodesics from them should not have caustics. In practice, this means that at any particular epoch during Hawking radiation extraction (we are using the terminology of \cite{Penington} here), the geometry we consider can be treated to have the Penrose diagram\footnote{This picture is approximate and is useful only during a given epoch, because once the black hole fully evaporates, there is no singularity left. Unitary black hole evaporation means that all black hole Penrose diagrams are identical to that of Minkowski space. Similar caveats apply to the Penrose diagrams of \cite{Penington} in an AdS context.} of figure \ref{XX} and not figure  \ref{YY}.
This should be compared to figure 4 of \cite{Penington}. In our discussions, we just want to have a box that encloses our black hole, but it may also be interesting to consider a more general system. One which contains a gravitationally bound system that can radiate to infinity. We discuss some interesting aspects of such systems in an Appendix.

The holographic screen will be a crucial part of our discussions. For causal structure purposes, we can define it as follows. 

\noindent 
{\bf Definition 2.2:} We define $\mathfrak{S}$ to be a {\em holographic screen} or {\em radial cut-off} in an asymptotically flat\footnote{For our immediate purposes, a  spacetime asymptotically flat when it is Weakly Asymptotically Simple and Empty (WASE) and is future asymptotically predictable, according to the definitions of  \cite{HawkEllis}.} spacetime $\cal{M}$ iff 
\begin{itemize}
\item[{\bf a.}] $\mathfrak{S}$ is a co-dimension 1 connected surface that is topologically $\IR\times S^2$ that separates $\cal{M}$ into disconnected interior and exterior regions, ${\rm Int}(\cal{M})$  and ${\rm Ext}(\cal{M})$, 
\item[{\bf b.}] $\mathfrak{S}$ is an everywhere timelike surface, and the normal that defines the extrinsic curvature of $\mathfrak{S}$ is taken to point towards ${\rm Ext}(\cal{M})$,
\item[{\bf c.}] Future(past) directed causal curves from any point on $\mathfrak{S}$ must reach $\mathfrak{I}^+$ ($\mathfrak{I}^-$),
%\item[{\bf d.}] $\mathfrak{S}$ has no boundary, $\partial \mathfrak{S}=\emptyset$\footnote{Some of these items may not be entirely independent. Eg., it is conceivable that item (a) and the fact that $\mathfrak{S}$ is everywhere timelike, implies item (d), at least under some further technical assumptions about $\cal{M}$. Our goal is not to make the most general statement, but to make the most concrete one for our purposes.}.
\end{itemize}

To prove most our basic theorems and to make our claims, the above definition will be enough  but less formally we will think of this radial cut-off as living far away: the radius of the cut-off should satisfy \eqref{large}. It should be kept in mind however that there still is an ``infinite amount of space" between $\mathfrak{S}$ and the conformal  boundary of $\cal{M}$.

Note that the definition is essentially kinematical but not entirely so. This is because of the presence of item (c). It basically captures our intuition that the shell is sufficiently far away from localized gravitational phenomena. With these we can already prove some simple facts that will be useful for making some of our arguments. The following statement is a verification of our intuitive sense that an asymptotic cut-off surface should not fall inside a black hole. 

{\bf Theorem 2.1:} Outward directed future null geodesic congruences that emanate from spatial slices of $\mathfrak{S}$ cannot converge. 

{\bf Proof:} Otherwise the spatial slice in question would be  a closed trapped surface, which in strongly asymptotically predictable spacetimes $\cal{M}$ are behind event horizons (proposition 9.2.1 of \cite{HawkEllis}). This would violate item (c) in the definition of $\mathfrak{S}$. $\hfill \Box$
\\

\noindent
We will define the cylinder $\cal{M}_\mathfrak{S} \equiv {\rm Int} (\cal{M}) \cup \mathfrak{S}$, which makes $\mathfrak{S} = \partial \cal{M}_\mathfrak{S}$. A basic observation regarding the nature of these cut-offs is the following:

{\bf Theorem 2.2:} Consider an Asymptotic Causal Diamond $\mathfrak{C}(p,q)$ and the radial cut-off $\mathfrak{S}$. If $D^{induced} \equiv \mathfrak{C}(p,q) \cap \mathfrak{S}$, then ${\cal I}^-(D^{induced}) \cap {\cal I}^+(D^{induced}) \cap {\cal M}_\mathfrak{S} = 
{\mathfrak{C}}(p,q)\cap  {\cal M}_\mathfrak{S} \equiv {\mathfrak{C}}(p,q)_{\mathfrak{S}}$. %We will call ${\mathfrak{C}}(p,q)_{\mathfrak{S}}$ the {\em Inner Causual Diamond} inside the cut-off $\mathfrak{S}$.

\begin{figure}[h]\centering
\hspace{-5mm}
\includegraphics[angle=0,width=80mm]{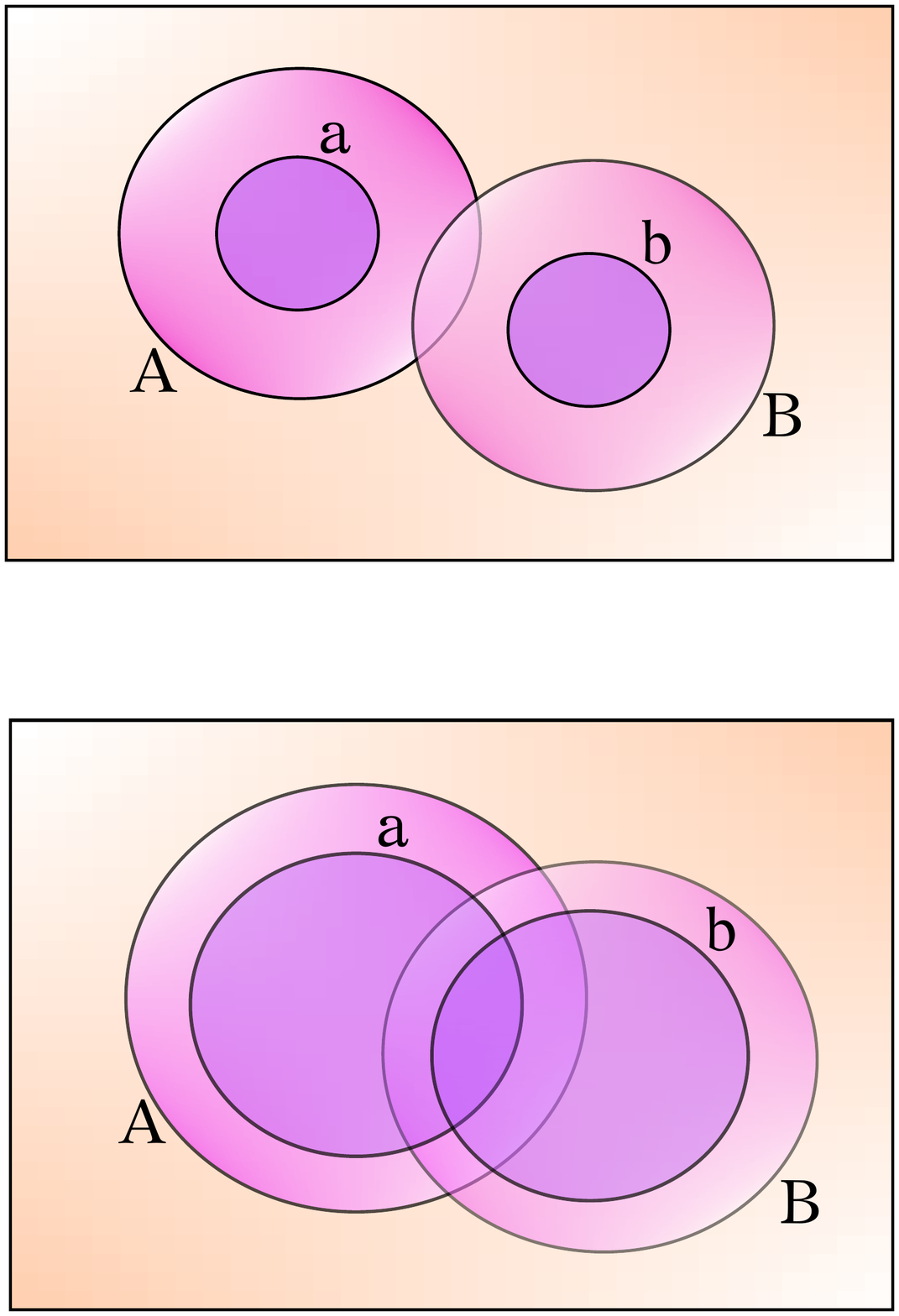} \\
%\vspace*{-7cm}
\caption{Points in non-trivial intersections of subsets lie in intersections of the parent sets as well. A version of this fact, when the sets are certain ACDs, is used in the proof of theorem 2.2. We present this picture to emphasize that the idea is trivial.}
\label{VENN}
\end{figure}

{\bf Proof:} If ${\mathfrak{C}}(p,q)_{\mathfrak{S}}$ is empty, the result follows trivially. So we will assume otherwise, and prove the claim in two steps, by showing first that LHS $\subset$ RHS and then that RHS $\subset$ LHS. First we prove that ${\cal I}^-(D^{induced}) \cap {\cal I}^+(D^{induced}) \cap {\cal M}_\mathfrak{S} \subset
{\mathfrak{C}}(p,q)\cap  {\cal M}_\mathfrak{S}$. Proof: If $z\in$ LHS, it means  that $\exists \ x, y \in D^{induced}$ such that $z \in {\cal I}^-(x) \cap {\cal I}^+(y) \cap {\cal M}_\mathfrak{S} \neq \emptyset$. Now, ${\cal I}^-(x) \subset {\cal I}^-(p)$ because $x$ is in the past of $p$. Similarly,  ${\cal I}^+(y) \subset {\cal I}^+(q)$. Since ${\cal I}^-(x) \cap {\cal I}^+(y) \cap {\cal M}_\mathfrak{S}$ is necessarily non-empty, this means that $z \in {\cal I}^-(p) \cap {\cal I}^+(q) \cap {\cal M}_\mathfrak{S}$, see figure \ref{VENN}. 
This proves LHS $\subset$ RHS. Next we prove that ${\cal I}^-(D^{induced}) \cap {\cal I}^+(D^{induced}) \cap {\cal M}_\mathfrak{S} \supset {\mathfrak{C}}(p,q)\cap  {\cal M}_\mathfrak{S}$. Proof: Assume $z \in$ RHS. This means that either $z \in {\rm Int}({\cal M})$ or $ z \in \mathfrak{S}$. If the latter, since $z \in {\mathfrak{C}}(p,q)$ as well, we have shown what needs to be shown. If the former, since $z \in  {\mathfrak{C}}(p,q)$ there exists a future-directed causal curve from $z$ to $p$ that must intersect $\mathfrak{S}$ at some $x$ by Jordan's curve theorem. This $x \in D^{induced}$ because the curve is causal. Now $z \in {\cal I}^-(x) \subset {\cal I}^-(D^{induced})$. A similar argument in the past direction shows  $z \in  {\cal I}^+(D^{induced})$. Together with $z \in{\cal M}_\mathfrak{S}$,   this proves RHS $\subset$ LHS, proving the theorem. $\hfill \Box$
\\

%that $z \in {\cal I}^-(D^{induced})$ and $z \in {\cal I}^+(D^{induced})$ and $z \in {\cal M}_\mathfrak{S}$ simultaneously. This implies

\begin{figure}[h]\centering
\hspace{-5mm}
\includegraphics[angle=0,width=90mm]{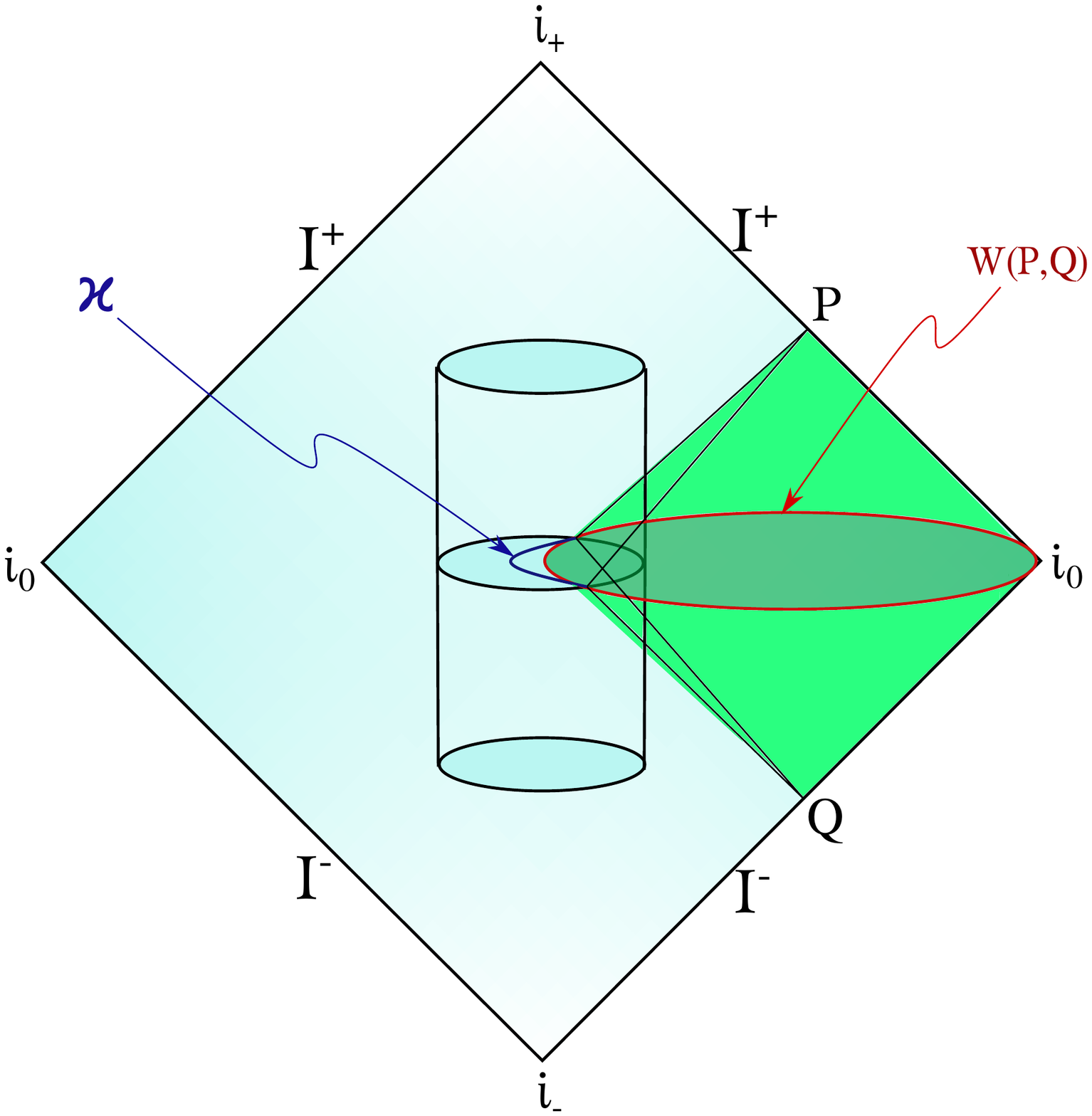} \\
\vspace*{-2cm}
\caption{A cartoon of various points and surfaces associated to an ACD and the holographic screen. We emphasize that this really is a cartoon: it mixes some features of conformal coordinates and some features of spacetime coordinates. Note in particular, that the circle is actually of infinite radius and therefore the waist should be a straight line inside the cut-off. The ``cylinder" should really be a cigar ending in future and past timelike infinity (assuming that the background contains no horizons etc.).}
\label{cartoon}
\end{figure}

\noindent 
Given the previous definitions and observations, the above statement is again almost kinematical, see cartoon \ref{cartoon}. 
The utility of the construction is that it provides us with intuition on how to connect the asymptotic picture with the screen based picture. %We will see, eg., that the intersection of the ``waist" of the ACD  with $\mathfrak{S}$ is the natural place to which we can anchor our extremal surfaces to. 

\subsection{Anchoring Surfaces and Sub-regions}

Let us compare the relevant structures in AdS and flat space. In AdS, one usually starts with a spatial region $A$ on the asymptotic boundary and then considers $\partial A$ as the natural place where the extremal surface anchors to. In flat space on the other hand, the conformal boundary has an entirely different structure, and the AdS picture does not immediately have an analogue. In fact, it is easy to see that spacelike geodesics (and similar extremal surface generalizations in higher dimensions) stretch out between diametrically opposite points on the celestial sphere in Minkowski space, which naively would seem to preclude the possibility of a useful notion of boundary sub-region at spatial infinity\footnote{This is a hint of the non-locality of the hologram of flat space. See \cite{CK} for some closely related discussions.}. A second observation is that in flat space we have two interesting surfaces: the conformal boundary and the holographic screen. In AdS, the two can be taken to be the same, and when defining boundary sub-regions in AdS/CFT, this is precisely what we do. It is in fact the coincidence of these two logically distinct entities that makes AdS holography simpler in many ways. Since we wish to keep the holographic screen to be a distinct entity for much of our discussions, here we will need to rethink some of this. We have aimed our discussions below to have some redundancy with parts of the discussions in the Introduction -- but this time over, we will elaborate some details which were glossed over in the first iteration.

A key ingredient in defining a screen sub-region is that we need a $t=0$ slice on the screen. In AdS, this is a simple matter because the screen coincides with the asymptotic boundary, and the choice of $t$ is related to the choice of a Poincare frame in the boundary Minkowski space. Once one chooses such a time coordinate, it is trivial to define sub-regions on the $t=0$ slice at the screen/boundary. In flat space however, since our screen is in the bulk and does not coincide with the conformal boundary,  a choice of asymptotically Poincare time coordinate will only fix it in the bulk upto diffeomorphisms that vanish at infinity\footnote{The allowed class of diffeomorphisms and fall-offs fixes the asymptotic symmetry group. The details will not be important for us, the existence of the ambiguity is all that will be significant.}. 

As mentioned in the introduction, a second problem arises when we try to define Quantum Extremal Surfaces (QES). A quantum extremality condition is most naturally defined for surfaces that separate a bulk Cauchy slice into two disconnected pieces, because it involves a bulk entanglement entropy contribution.  This means that when we  wish to relate extremal surfaces anchored on screen sub-regions with entanglement wedges anchored at the boundary, we need to give a prescription for how this is to be done. 

%We will discuss this point in more detail when we discuss flat space quantum extremal surfaces, at the moment we merely want to note that there {\em is} a problem.

The way we will approach this problem is by first formulating the idea of a sub-region in AdS in a way that lends itself more naturally to being adapted to flat space.  A key piece in our intuition will come from finding a suitable definition for a sub-region in AdS at a {\em finite cut-off}. At finite cut-off, both problems that we mentioned above are present in AdS as well. Therefore, a suitable construction in cut-off AdS that has natural resolutions to both these problems, will act as a inspiration for our definitions in asymptotically flat space. 

We will do this in four steps. First, we note that a spherical sub-region on a $t=0$ slice of the Minkowski space at the asymptotic boundary of AdS can be defined via the waists of symmetric boundary causal diamonds\footnote{As noted before, these are causal diamonds whose vertices have identical spatial coordinates, and time coordinates $t=T$ and $t=-T$ for some $T$. Note that the waists of causal diamonds in Minkowski space are spheres.}. Second, we note that an arbitrary sub-region can be defined in terms of a (possibly infinite) union of such solid spheres, and therefore in terms of symmetric causal diamonds. Third, we note that a causal diamond on the boundary Minkowski space is the boundary restriction of a bulk AdS causal wedge. This means that we can view a general sub-region at the asymptotic boundary using the boundary restriction of a (possibly infinite) number of such symmetric AdS causal wedges. Fourth, now if we want to define a sub-region at finite cut-off, we simply have to consider the restriction of these causal wedges to the cut-off instead of to the asymptotic boundary\footnote{Note that the causal wedge are still {\em anchored} at the boundary. }.

In asymptotically flat space, the conformal boundary does not have a causal structure unlike the Minkowski boundary of AdS. So we cannot do the analogues of the first two steps. But the remarkable fact noticed in \cite{CK} was that there exists a bulk object that has all the relevant properties of a causal wedge, namely the ACD. This means that we can use steps 3 and 4 above to define sub-regions on cut-off surfaces in flat space precisely in the same way that we did in AdS. This is the path that we will take,  after we formalize some basic (and obvious) terminology.

{\bf Definition 2.3:} We define the waist of an Asymptotic Causal Diamond $\mathfrak{C}(p,q)$ to be $W(p,q) \equiv \partial {\cal I}^-(p) \cap \partial {\cal I}^+(q)$. Where the waist cuts the cut-off surface defines the {\em Special Anchoring Surface (SAS)}, which we define via $S \equiv W(p,q) \cap \mathfrak{S}$. 

Note that the waist is codimension-2 and the SAS is a codimension-3 (topological) sphere. This has the curious consequence that when ${\cal M}$ is 2+1 dimensional, the SAS is actually a pair of points. In higher dimensions SAS are always connected. 

%See eg., appendices of \cite{Prabhu} for a recent pragmatic review of the Ashtekar-Hansor definition \cite{Ashtekar} of asymptotic flatness at spatial infintiy. 

\begin{figure}[h]\centering
\hspace{-5mm}
\includegraphics[angle=0,width=80mm]{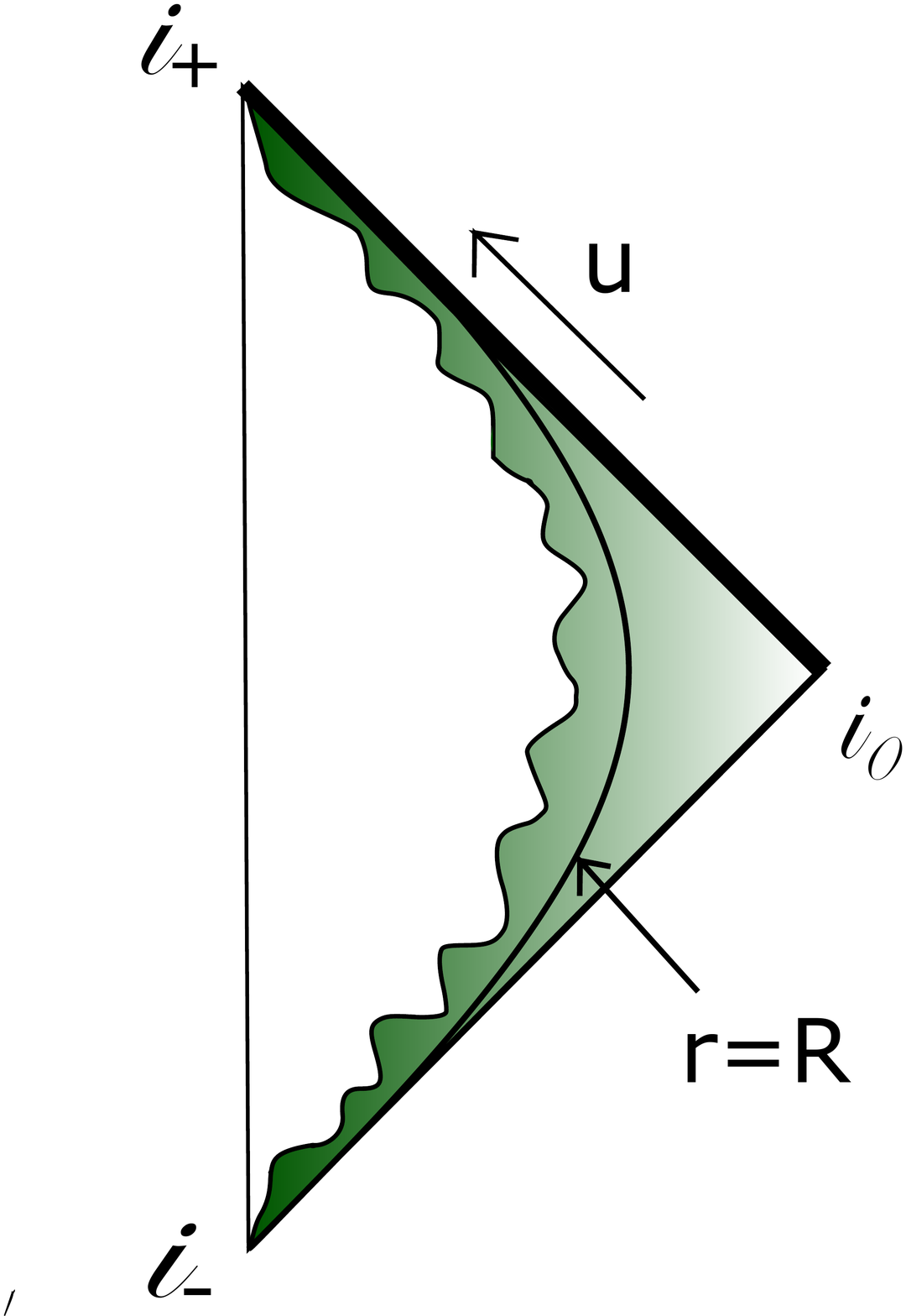} \\
%\vspace*{-7cm}
\caption{The olive shaded region shows a chart where the asymptotically flat $u, r$ coordinates are well-defined on this geometry.}
\label{asympchart}
\end{figure}

Now in order to talk about sub-regions on the screen, we should first introduce an asymptotically flat coordinate system near the boundary\footnote{In somewhat more detail: we will take the existence of an asymptotically flat chart to imply that there exists some coordinate $r$ such that the metric takes the asymptotically flat form when $r \in (\Lambda, \infty)$ for some positive (and  presumably large) $\Lambda$. We will assume that the holographic screen lies entirely within this chart. The standard radial coordinate in Schwarzschild spacetime provides an example of such a coordinate.}. Such a choice results in a time coordinate choice in the asymptotic Poincare frame\footnote{In AdS, the analogous choice will be an asymptotically Minkowski time coordinate on the boundary.}. In the bulk, the precise set of fall-offs and allowed diffeomorphisms for an asymptotically flat spacetime can be found in standard references. For concreteness, let us take a BMS condition (see eg. \cite{Laddha}) to be our definition of asymptotic flatness in $d+1$ dimensions:
\begin{eqnarray}
d s^{2}=-d u^{2}-2 d u\ d r+r^{2} \gamma_{A B}\ d \Omega^{A}\ d \Omega^{B} + \hspace{2in}\nonumber \\ + \ r C_{A B}\ d \Omega^{A}\ d \Omega^{B}+\frac{2 m_{\mathrm{B}}}{r} \ d u^{2}+\gamma^{D A} D_{D}\ C_{A B}\ d u \ d \Omega^{B}+\ldots \label{BMS} \end{eqnarray}
where $u,r, \Omega$ are the coordinates of the chart, $\Omega$'s stand for angle coordinates and $\gamma_{AB}$ is the unit $(d-1)$-sphere metric. The unknown functions are independent of $r$ ($r \rightarrow \infty$ is the asymptotic region). They satisfy some further conditions, which will not affect our discussion. The first three terms define Minkowski space in these coordinates, and the rest are the sub-leading fall-offs. Note that (see figure \ref{asympchart}) 
one can indeed choose the cut-off to be large enough to be within the $(u,r)$ chart: a simple choice would be $r=R$ with large enough $R$. An example of an asymptotically Minkowski (non-unique!) time coordinate is $t \equiv u+r$, and sub-regions can be defined on the cut-off, using such a $t=0$ slice. As we noted previously, one point to be careful about is that the asymptotically flat coordinates that we have defined above in \eqref{BMS} are only unique upto diffeomorphisms that retain the form of the fall-offs. A further point is that we need a way of relating extremal surfaces anchored to screen sub-regions, and those anchored at the conformal boundary (see section \ref{rec} for a more detailed discussion). To do this, we will use SAS  as we defined above: these are a class of codimension-3 surfaces on the cut-off, built from ACDs, and they can act as anchoring surfaces on the screen. Because the metric is asymptotically flat and the cut-off is large, SAS can be used to approximate sub-regions. Let us see how these two steps work concretely.

We consider ACDs anchored at the conformal boundary, that are symmetric with respect to the asymptotic $t=0$ coordinate. The waists of these ACDs will cut out codimension-3 surfaces on the cut-off (a set of SAS, according to our definition above). Note that just because these ACDs have been picked to be symmetric with respect to the $t=0$ time on the boundary, does {\em not} mean that their waists lie on the $t=0$ slice in the bulk: the metric is only {\em asymptotically} flat. But we do expect that we can do a diffeomorphism that keeps us within the \eqref{BMS} class, so that any individual SAS can be brought to the $t=0$ slice of a new set of coordinates\footnote{\label{allACD}It seems possible that the SASs of {\em all} symmetric ACDs can be made to lie on the same bulk $t=0$ slice, by an appropriate choice of asymptotically flat bulk time. We have not been able to prove this, so we will proceed in the rest of this section and the next, without assuming it. But if this claim happens to be true, some of our arguments will simplify -- once we work with this specific choice of time. The claim is true for stationary spacetimes (with no further assumptions). It is also straightforward that a common time slice can be found for spherically symmetric spacetimes even when they are non-stationary (though we have not proved that the choice can be made asymptotically flat in the technical sense.).}. This in turn means that the error in the metric if we simply replace a point $(t \neq 0, r=R, \Omega)$ on the SAS in the original coordinates by $(t=0, r=R, \Omega)$ is sub-leading in $R$. Note that this replacement associates with every SAS a sub-region on the original $t=0$ slice. The sub-regions defined this way we will call {\em approximate sub-regions}, and they depend on the SAS and therefore the ACD. The process outlined above, of obtaining the approximate sub-region from the SAS of an ACD, we will call {\em projection}. Since the error in the metric due to this approximation is sub-leading, it also means that the relative error one is making in geometric quantities like area are also suppressed in powers of the cut-off size $R$. In other words, if we can show various properties (eg., strong sub-additivity) for extremal surfaces anchored to screen sub-regions, they can be viewed as being approximately true for the SAS anchored extremal surfaces as well\footnote{In particular, it means that even though these SAS-anchored extremal surfaces do not necessarily lie on the chosen $t=0$ slice on the cut-off, it is meaningful to view them  as (approximately) defining $t=0$ sub-regions with associated entropies to them. }. In the next sub-sections, we will show that indeed, for sub-regions on the screen, many of the features familiar from AdS do hold.

The advantage of SAS as opposed to the boundaries of generic $t=0$ sub-regions is that (the would-be-QES) surfaces anchored on the former can be extended along the waist of the ACD to $i^0$. We can view it as an approximation/regulator to the ``true'' QES that defines the ``true'' entanglement/reconstruction wedge we presented in figure (\ref{entwedge}) and will discuss in section 4. Note that the true entanglement wedge is simple, elegant and anchored to the conformal boundary. But when we regulate/approximate that Platonic object using the holographic screen, as often in physics, we have to deal with some technicalities. This is what we encountered above.

 %holographic screen The precise definition of asymptotic flatness has some ambiguities in the literature, let us for where the metric takes the form \begin{equation} d s^{2}=-\left(1+\mathcal{O}\left(\rho^{-(d-3)}\right)\right) d t^{2}+\left(1+\mathcal{O}\left(\rho^{-(d-3)}\right)\right) d r^{2}+r^2\left(1+\mathcal{O}\left(\rho^{-(d-3)}\right)\right) d \Omega^2_{d-2}\end{equation} with $\rho$ defined via $\rho^{2}=r^{2}-t^{2}$. In the last term involving angles there is a slight abuse of notation: the subleading piece can in principle depend on the angle directions as well, we refer the reader to \cite{MannMarolf} instead of dwelling on notation that will not affect our discussion. The coordinates $t, r$ and the angles define an asymptotically Minkowski coordinate system is the only fact we will need: there is some flexibility in the precise choice of fall-off, we adopt one in \cite{MannMarolf} only for concreteness. The existence of the asymptotically Minkowski coordinates allows us to consider holographic screens defined by (for example) $r=R$.

Note that when defining screen sub-regions, we have to choose some asymptotically flat coordinates. This should be compared to what we do at the asymptotic boundary of AdS when we pick a Minkowski time coordinate to define sub-regions on $t=0$. The extra bit here is that in the bulk, there is an extra diffeomorphism freedom which is not fully fixed. This is entirely physical, and an indication that gravity is weak, but {\em not} entirely non-dynamical at the screen. Note that in defining the exact entanglement/reconstruction wedge, this issue does not arise because it can be defined entirely via data on the conformal boundary.

%reason why this theorem is useful, is because ${\mathfrak{C}}(p,q)_{\mathfrak{S}}$
\section{Classical Extremal Surfaces and Maximin Surfaces on the Screen}

With these ingredients, now we are ready to define various types of classical extremal surfaces. We will sometimes call them {\em relative} extremal surfaces to emphasize that they depend on the holographic screen -- they are anchored to codimension-3 surfaces on the screen. Ultimately our interest is in understanding their areas in terms of classical limits of entanglement entropies of screen sub-regions coupled to a sink. But this we will get to only in the next section, in the broader context of discussions about (quantum) extremal surfaces and entanglement wedges anchored to the conformal boundary. In this section, which is self-contained, we will simply study the classical properties of screen-anchored extremal/maximin surfaces. 

\subsection{Relative Extremal Surfaces}

%se can be chosen either by a subregion lying on the cutoff surface or by the choice of the ACD, but we will not emphasize it because an analogous statement is true in AdS as well: for spherical entangling regions, instead of thinking of extremal surfaces as being defined by boundary sub-regions, one can equivalently think of them as being defined by the vertices of the boundary domains of dependence. The new ingredient in flat space is that to define subregions, we need to have a cut-off. Now we have two approaches to extremal surfaces and subregions:

As discussed in the last section, we will be interested in two distinct kinds of anchoring surfaces on the screen: the SAS, and the boundary of a $t={\rm fixed}$ sub-region. Lets elaborate on this and the distinction between the various specimens in the zoo of sub-regions we will come across.
\begin{enumerate}
    \item To define a notion of sub-region on a screen in a globally hyperbolic spacetime $\mathcal{M}$, we need to intersect the screen with a Cauchy slice $\Sigma$. In the asymptotically flat coordinates, these will be the $t={\rm fixed}$ slices. On  the intersection of this slice and the screen, we can define a {\em naive sub-region} $\mathcal{A}$ with a boundary $S\equiv\partial \mathcal{A}$. Naive sub-regions satisfy properties such as strong subadditivity (SSA) for areas of extremal surfaces anchored to them, as we will later prove. When we use the word sub-region without qualifiers, we will mean naive sub-regions unless explicitly stated otherwise.
    
 %Such a definition of a subregion is "nice" in the sense that it will satisfy the properties such as strong subadditivity, nesting of entanglement wedges, etc which we will prove later. Like in the case of AdS, the boundary of this subregion can then be used to define relative extremal surfaces inside the cutoff.
    \item Consider the Special Anchoring Surface $S$, as defined in the previous section, for a single symmetric ACD. Since the screen is large and the spacetime is asymptotically flat, we expect to find an asymptotically flat coordinate system such that the SAS lies on its $t= {\rm fixed}$ Cauchy slice. We define a {\em SAS sub-region} $\mathcal{A}$ as the subset of the intersection of this Cauchy slice with the holographic screen, bounded by $S$. This is loosely analogous to the case of spherical entangling surfaces constructed using vertices of a causal diamond on the boundary of AdS. Note however that here we are working with large but finite cut-off.
%Even in that case one can define subregions as the subset of the $t=0$ slice of the boundary of AdS, bounded by the spherical entangling surface.  %Another way to define the relative extremal surfaces is using the prescription of Asymptotic Causal Diamonds (ACD) which we have previously defined. In this case, we anchor the extremal surface on the cutoff at the Special Anchoring Surface $S$.
\end{enumerate}{}
In the rest of this section\footnote{It is quite unwieldy to talk about ``the relative classical entanglement wedge anchored to a SAS sub-region" and such, so outside of this section, we will often rely on context to make it clear what kind of sub-regions we are talking about. We apologize to the reader for our lack of imagination in coming up with simpler names.}, we shall use $\mathcal{A}$ to denote both naive and SAS sub-regions when the definition/property/theorem being discussed is independent of the approach being adopted to define sub-regions. Similarly $S$ will be used to refer to both the SAS and the boundary of the naive sub-region. When it does depend on the approach, we shall explicitly state the type of sub-region/anchoring surface that we are referring to. Let us also note that the projection that we defined towards the end of the last section can be viewed as a way of associating a naive sub-region to a SAS sub-region. In other words, approximate sub-regions are naive sub-regions that approximate SAS sub-regions. 

{\bf Definition 3.1:} Given a codimension-3 anchoring surface $S$ on the holographic screen $\mathfrak{S}$, the {\em Relative Extremal Surface} $m(S)$ is the codimension-2 surface with extremal area anchored at $S$. If the spacetime has non-trivial homology, we demand that $m(S)$ is homologous to sub-region $\mathcal{A}$ on the holographic screen.

Classically, the area of a surface that is anchored on the screen is perfectly well-defined and finite. The non-trivial question is whether there are meaningful extremal surfaces among them that are stationary under variations of the surface. Note that in empty Minkowski space with a radial cut-off, we certainly know that such surfaces exist, because they are simply the flat hyperplanes that arise when the waist of an ACD intersects the cutoff spacetime. Another piece of circumstantial evidence is that in AdS, it is expected that extremal surfaces anchored to a holographic screen exist \cite{Sorce}. We will discuss the existence question further after defining the relative HRT surface, to which we now turn.
%But we will see later that when we include quantum corrections to the entropy, in some formulations it is useful to have surfaces that pass through SAS, but extend all the way to the asymptotic spatial boundary. The present classical prescription will then emerge as the classical limit of such an object, after a suitable IR-regulator (which we will elaborate upon) is used to define the entropy of such a surface.

The {\em Relative HRT Surface} $\chi_c$ is the relative extremal surface with minimum area. If there are multiple extremal surfaces having the same minimum area, then we can choose any one of them. In what follows, we will develop a version of the maximin construction \cite{Wall}, prove that maximin surfaces exist, and also show that these relative maximin surfaces and the relative HRT surfaces are the same. This will be further evidence for the existence of these surfaces.

\begin{figure}[H]\centering
	\hspace{-5mm}
	\includegraphics[angle=0,width=75mm]{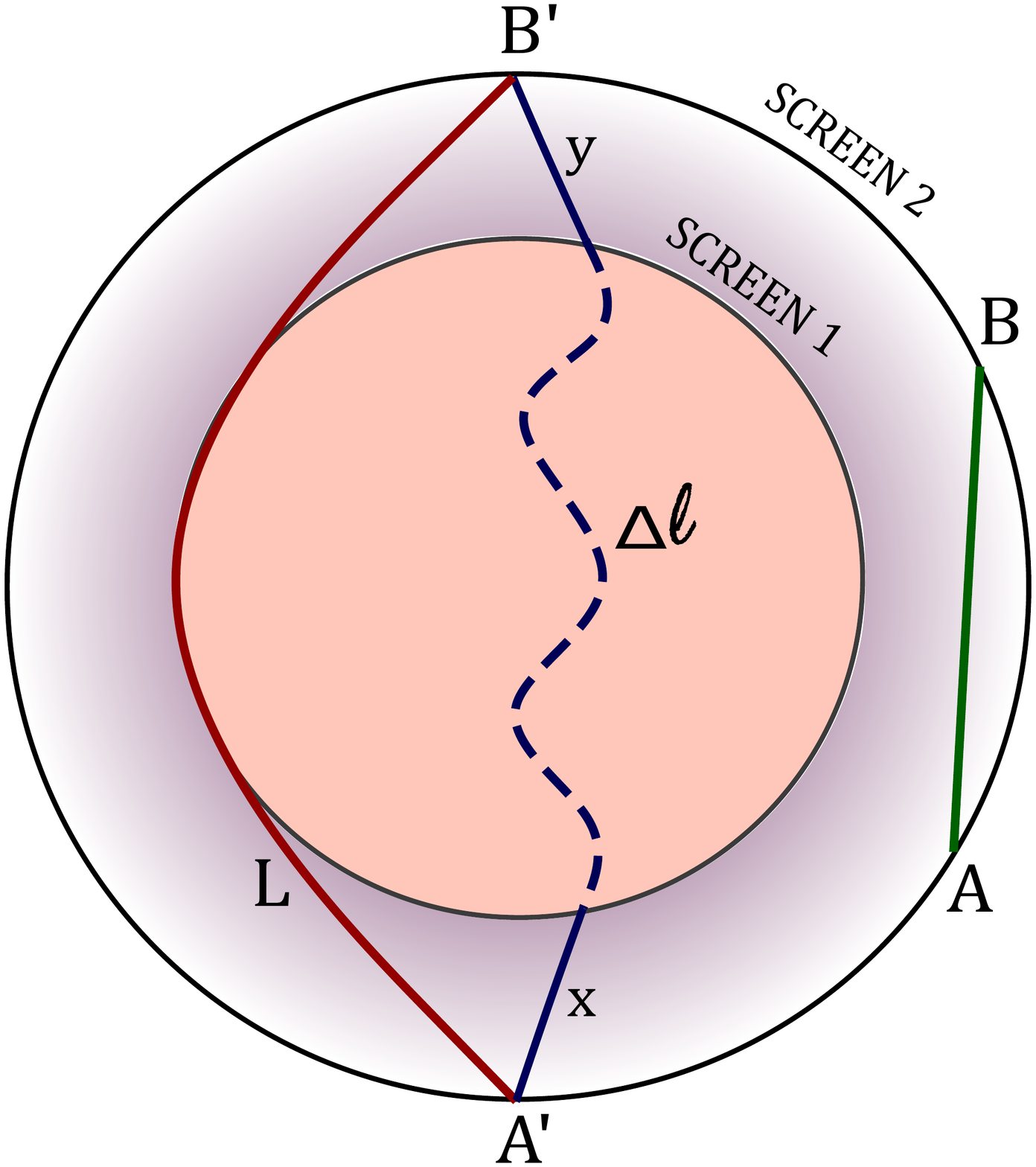} \\
	\caption{Containment argument.}
	\label{SCREENS}
\end{figure}

In order to prove the equivalence between the relative maximin and relative HRT surfaces, we will need to argue that they are {\em contained} within the screen. The key observation here is that for large enough screens, the metric outside is (approximately) Minkowski. Therefore just like in the case of a screen in Minkowski space \cite{CK}, we can argue that the minimal extremal surfaces must lie inside the screen even though dramatic things (eg., black holes,...) can lie deeper inside the screen. Even though this is obvious, let us follow the argument through to completion to make some related comments as well. We start by introducing two screens -- both large and in the asymptotically flat chart, see figure \ref{SCREENS}. 
Since our goal here is just to capture the main idea, we will work with 2+1 dimensions. The first observation is that for curves anchored on the screen, there are no extremal curves outside the screen, because their lengths can always be decreased by deforming appropriately into the screen. This is a property of flat space, and proves  containment. Note that sub-leading $1/R$ corrections to the metric cannot change this. It is clear that for boundary regions which are ``small" in an obvious sense, the relative HRT surfaces with respect to screen 2 are approximately those of Minkowski space, ie., they are close to straight lines. Let us also make a few comments about minimal surfaces, because they will also be useful when discussing strong sub-additivity. Let us first note that if we demand that we only look for minimal curves that stay outside screen 1, then we can find a conditional minimal curve of length $L$: this is marked in red in the figure. Now for a curve that passes inside screen 1, the contribution to its length from the interior $\Delta l$ has to be non-negative. As long as the total length $x+y+ \Delta l < L$, the true minimal curve will be one such curve, and not the red curve in the figure. But if not, the minimal curve will be the red curve. %In either case, we see that there is a minimal extremal curve that is contained within screen 2. The only point of this trivial argument is that when looking for a minimal curve connecting $A'$ and $B'$, there exists a curve (the red curve) that is of less length than {\em any} curve that goes outside screen 2. Therefore what happens inside screen 1 does not improve the viability of curves outside screen 2. 

% The concern is regarding ``big" subregions which cover an ${\cal O}(1)$ fraction of the perimeter of screen 2: can it be the case that for a minimal surface going into the interior of screen 1, containment can be violated for screen 2? To argue that it cannot, 

Note that our arguments in this regard are all based on the fact that we are working in an asymptotically flat spacetime and the screen is living in approximately Minkowski space. A key point about the screen of the kind we introduced is that in Minkowski space, it acts as type of barrier for extremal surfaces. Note that more typically, extremal surface barriers are implemented \cite{ESB, Sanches Weinberg} via  surfaces with conditions on the expansion $\theta$. This is because often the spacetimes one considers are much more general. We on the other hand are explicitly taking advantage of the flatness of spacetime together with eg., condition (c) in definition 2.2 to avoid null rays re-focusing outside the screen. This enables us to bypass the kind of problems alluded to in footnote 4 of \cite{Wall}. For example, we are not shooting light rays back from the cut-off surface, but from null infinity when we work with ACDs. Note also that the asymptotically flat coordinate system gives us spatial slices and since we are only looking for extremal surfaces anchored to sub-regions on a $t= {\rm fixed}$ slice, the kind of problem outlined in (III.1) of \cite{Sanches Weinberg} is also avoided. 

\subsubsection{Relative HRT Surface Lies Outside the Causal Surface}

%Unlike in the case of AdS, where we start with a boundary subregion and then define the HRT surface anchored to it, in the second method of defining extremal surfaces in flat space, we first defined the relative HRT surface using the Special Anchoring Surface $S$ as it is more natural. Now we can define the allowed subregions on the cutoff-surface $\mathfrak{S}$ using the homogeneity condition.

%{\bf Definition 3.2:} An {\em Allowed Subregion} $\mathcal{A}$ is a codimension-2 surface anchored at the Special Anchoring Surface (SAS) i.e. $\partial \mathcal{A} = W(p,q) \cap \mathfrak{S}$ and $\mathcal{A} \in \mathfrak{S}$ such that the hypersurface $\sigma$ with boundary $\partial \sigma = m(S) \cup \mathcal{A}$ is everywhere spacelike (codimension-1).

In AdS, the causal wedge is the bulk domain of dependence for a boundary sub-region $\mathcal{A}$ with an associated boundary domain of dependence $D_A$ i.e., causal wedge of $\mathcal{A}$ is $\cal{I}^-(D_A) \cap \cal{I}^+(D_A)$. The causal surface is defined as the edge of this causal wedge. We will treat the waist $W(p,q)$ of the ACD as the flat space analogue of the causal surface. The definition of a causal wedge and causal surface is sensible only in the second approach using ACDs and SAS's. Hence in the following theorem, $S$ refers to the SAS and $\mathcal{A}$ refers to SAS sub-regions. We want to show that the relative HRT surface lies further in the bulk than $W(p,q)$. We construct a proof similar to that of Theorem 6 of Wall \cite{Wall}. Note that both of these surfaces meet on the holographic screen at $S$. 

{\bf Theorem 3.1:} A relative HRT surface $\chi_c$ anchored to the Special Anchoring Surface $S$ lies spacelike separated from $W(p,q)$ and on the side away from SAS sub-region $\mathcal{A}$.

\begin{figure}[H]\centering
	\hspace{-5mm}
	\includegraphics[angle=0,width=75mm]{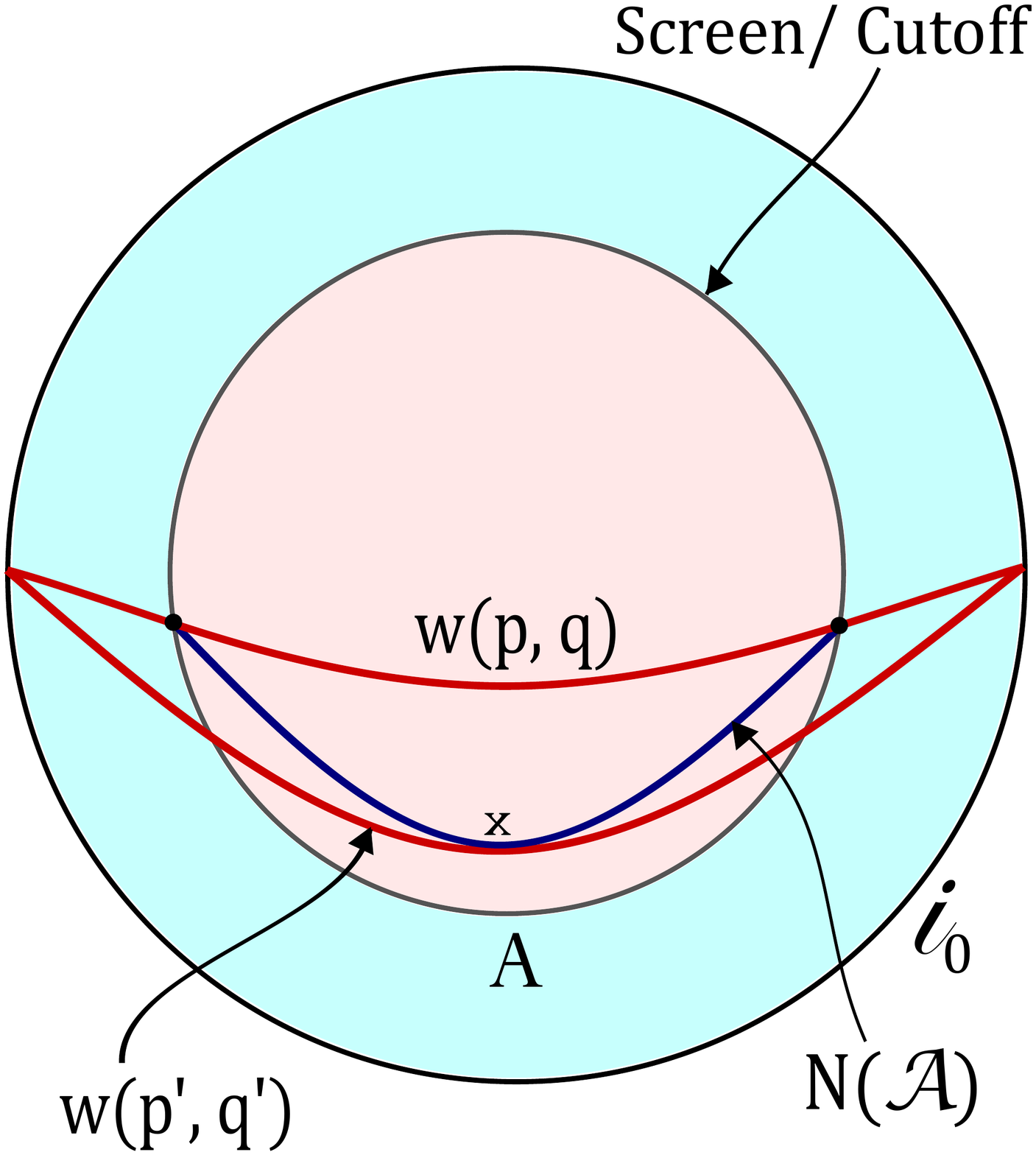} \\
	\caption{A schematic of the HRT surface being inside the causal surface. The blue line is the $N(\mathcal{A})$ surface and $W(p,q), W(p^\prime,q^\prime)$ are the causal surfaces. Note that the proof is by contradiction, so this is an impossible configuration.}
	\label{thm3.1}
\end{figure}

{\bf Proof:} We prove this by contradiction. Assume that $\chi_c$ lies inside $W(p,q)$ (i.e. on the side of $\mathcal{A}$). We also assume the Null Convergence Condition ($R_{ab}k^ak^b \geq 0$ for all null vectors $k$) and the generic condition \cite{Wall}.
\begin{itemize} 
\item{(i)} Since $\chi_c$ is an extremal surface we can shoot null congruence $N(\mathcal{A})$ (codimension 1) from it towards $\mathcal{A}$ which has expansion $\theta < 0$. This can be seen by noting that the extremal surface $\chi_c$ has $\theta = 0$ and the Raychaudhuri equation along with NCC and generic condition will make $\theta < 0$ everywhere else on the null congruence. $\partial I^-(p)$ is a causal horizon, so according to the Second Law its area must increase as we move in the future away from the causal surface and similarly area of $\partial I^+(q)$ will increase as we move in the past away from the causal surface. If $\theta < 0$ then by the Raychaudhuri equation, the rays would have to focus which is not possible since they are shot back from the null infinity. Hence on the causal surface $\theta >0$. 
\item{(ii)} Now we can move the points $p,q$ on the null boundary $\mathfrak{I^{\pm}}$ continuously such that the new causal surface $W(p^\prime, q^\prime)$ is nowhere in the exterior of $N(\mathcal{A})$ and touches it at a point $x$ (see Figure \ref{thm3.1}). According to Theorem 4 of \cite{Wall}, if two null congruences $(N_1,N_2)$ touch at a point and $N_2$ lies nowhere in the past of $N_1$ then in the neighbourhood of the  point of contact either they coincide or $N_2$ expands faster than $N_1$ $(\theta[N_2] > \theta[N_1])$. Therefore we get $\theta[N(\mathcal{A})] > \theta[W(p^\prime,q^\prime)]$. 
\end{itemize}
(i) and (ii) gives a contradiction hence we conclude that $\chi_c$ lies in the exterior of $W(p,q)$.  

\subsection{Relative Maximin Surfaces}

Now we introduce analogues of maximin constructions  \cite{Wall} on large holographic screens of the type that we introduced in the last section. These maximin constructions allow us to prove strong subadditivity.

\subsubsection{Definition}

Due to global hyperbolicity\footnote{Global hyperbolicity  makes the spacetimes under consideration somewhat less general than what they could be (eg., charged black holes are no longer allowed). But it does retain the crucial points that we wish to capture. Note that similar sacrifices are typical in AdS as well \cite{Wall}.} of the spacetime $\cal{M}$, we know that there exists a (complete) Cauchy surface $\Sigma$ which extends all the way up to spacelike infinity $i^{0}$.  Consider a codimension-3 anchoring surface $S$ on $\mathfrak{S}$. In the asymptotically flat setting, we will take the $S$ to be anchored to the screen on some $t={\rm fixed}$ slice. We will restrict our attention to Cauchy slices that coincide with $t={\rm fixed}$, at and outside the cut-off. In other words, $S$ is the boundary of a sub-region $\mathcal{A}$ of a $t={\rm fixed}$ slice on the screen. We will call these Cauchy slices $\mathcal{C}(S)$, to emphasize the anchoring surface $S$, even  though $\mathcal{C}(S)$ passes through not just $S$ and $\mathcal{A}$ but through the entire screen slice at $t={\rm fixed}$. %Note that such slices exist for both naive subregions as well as for SAS subregions, even though their time coordinates need not in general be the same.

%Note that this anchoring surface could be thought of as either the Special Anchoring Surface $S \equiv W(p,q) \cap \mathfrak{S}$ of Definition 2.3 or the boundary $S\equiv \partial\mathcal{A}$ of a subregion $\mathcal{A}$ on the holographic screen. 

%{\bf Definition 3.4:} A codimension-2 surface $\gamma$ anchored to a codimension-3 surface $\partial\gamma$ on the Dyson sphere $\mathfrak{S}$ (that is, $\partial\gamma\subset\mathfrak{S}$) is said to be {\em internally anchored} iff there exists a neighbourhood $\cal{O}$ of $\partial\gamma$ with $\cal{O}\subset\gamma$ such that $\cal{O}\subset\cal{M}_\mathfrak{S}$. Similarly once can define {\em externally anchored} iff $\cal{O}\subset{\rm Ext}(\cal{M})$.

{\bf Definition 3.2:} On any $\Sigma\subset\mathcal{C}(S)$, we define ${\rm min}(S,\Sigma)$ as the codimension-2 surface anchored to $S$ with minimum area\footnote{The area functional is explicitly finite because of the presence of the cut-off, thus allowing one to find the surface having the minimum area by comparison.} and homologous\footnote{In this case, by homologous we mean that there exists an achronal hypersurface $\sigma$ such that $\partial\sigma={\rm min}(S,\Sigma)\cup\cal{A}$. Note that $\sigma$ need not be a subset of $\Sigma$.} to the sub-region $\cal{A}$. Note that if there are multiple such surfaces which satisfy these conditions, then ${\rm min}(S,\Sigma)$ can refer to any one of them. The {\em relative maximin surface} $M(S)$ is then defined as the ${\rm min}(S,\Sigma)$ with maximum value of area for all $\Sigma\subset\mathcal{C}(S)$. Once again, if there are multiple such surfaces then we pick the $M(S)$ that is stable in the sense of Definition 8(b) of \cite{Wall}.

In the following, we will prove various mini-theorems regarding these maximin surfaces. Fortunately, most of the hard work has been done in \cite{Wall} in the context of AdS and we will be able to adapt them to our case. %The existence of the ACD and the fact that the screen is visible from null infinity, are crucial for us to define SAS subregions. %This is what lets us bypass the comments made in footnote 4 of \cite{Wall}. 

\subsubsection{Existence of Relative Maximin Surfaces}

Extremal/maximin surfaces in Lorentzian geometries are often difficult to explicitly construct when there is non-trivial matter/curvature, so this is one of those cases where a proof of existence is of some value. So we will present one for relative maximin surfaces. Our emphasis will only be on the places where our argument needs some modifications compared to \cite{Wall}. 

We start by proving the existence of a minimal area surface on a given Cauchy slice. To this end, we define $\mathcal{C}_\mathfrak{S}(S)$ as the set of partial Cauchy surfaces contained in the cutoff spacetime $\cal{M}_\mathfrak{S}$, that is, $\mathcal{C}_\mathfrak{S}(S)=\{\Sigma_\mathfrak{S}=\Sigma\cap\mathcal{M}_\mathfrak{S}\ |\ \Sigma\in \mathcal{C}(S)\}$. By definition, the partial Cauchy surface $\Sigma_\mathfrak{S}$ is compact and has the topology of a ball. This means that we can view it as a compact metric space. The metric can (but need not) be taken as the projection of the spacetime metric on $\Sigma_\mathfrak{S}$, note that this is Euclidean. Things are more straightforward here than \cite{Wall}, because we do not need a conformal compactification to make $\Sigma_\mathfrak{S}$ compact. By arguments exactly analogous to those made by \cite{Wall}, now it follows that the space $\mathcal{S}(\Sigma_\mathfrak{S})$ of all codimension-2 surfaces on $\Sigma_\mathfrak{S}$ that are homologous to the sub-region $\cal{A}$ is compact. From our arguments previously about containment, we will believe that ${\rm min}(S,\Sigma)\subset \Sigma_\mathfrak{S}$ and hence ${\rm min}(S,\Sigma)\subset \mathcal{S}(\Sigma_\mathfrak{S})$. Using again  the \cite{Wall} argument that the area of the curves are lower semi-continuous, we find by extreme value theorem that ${\rm min}(S,\Sigma)$ exists.

Now that the min step has been done, we will prove that there exists a Cauchy slice that maximizes the min value. Like \cite{Wall}, we will not do this in full generality. We will prove it for horizonless spacetimes, and for spacetimes with horizons where the singularities beyond the horizons are of Kasner type. This is fairly general, but probably not the most general situation where extremal surfaces can exist. 

Since the asymptotically flat spacetime $\cal{M}$ is globally hyperbolic, there exists a global time function\footnote{Note that we are not demanding that this time is an asymptotically flat time coordinate. It merely has to foliate the spacetime.} that can be understood as a map $t:\cal{M}\rightarrow \mathbb{R}$ such that (i) surfaces of constant $t$ are Cauchy surfaces with the same topology and (ii) the topology of $\cal{M}$ is $\mathbb{R}\times X$ where $X$ can be thought of as a constant time slice. Using the spatial coordinates on this constant time slice $X$, a general Cauchy surface $\Sigma$ can be represented using a continuous function $t^{\Sigma}:X\rightarrow\mathbb{R}$ such that $t^{\Sigma}(x)$ indicates the time of the Cauchy surface $\Sigma$ at each spatial position $x\in X$. Next, consider the restricted domain $t^{\Sigma}:Y\equiv X\cap\cal{M}_\mathfrak{S}\rightarrow\mathbb{R}$. Such time functions with restricted domain correspond to partial Cauchy surfaces bounded by the cut-off $\mathfrak{S}$. Clearly, $Y$ is compact.  Consider the subset of the space of continuous functions $\{t^{\Sigma}\}$ over the compact domain $Y$, which have the same value on the anchoring sub-region: we can think of this value as the value of $t$ on the $t= {\rm fixed}$ slice that defines our sub-region.  Using the Ascoli-Arzela theorem, one can then prove that this subset is compact with respect to uniform topology iff it is (a) equibounded, (b) equicontinuous and (c) closed. Before we proceed to the proof, note that this subset is isomorphic to the set of partial Cauchy surfaces $\mathcal{C}_\mathfrak{S}(S)$ passing through $S$ as defined before.  Note also that so far we have not made any assumptions about the presence or absence of horizons in the spacetime.   

%Since $\cal{M}$ is compact (after conformal compactification), $X$ will be compact and thus
%, that is $\{t^{\Sigma}\ |\ t^{\Sigma}(x)=t^0 \ \forall\ x\in R \}$, where $R$ stands for the intersection of the $t={\rm fixed$ slice 
%We can choose $t^0(S)=0$ without loss of generality.

We only need to show that the subset is equibounded. The arguments for the (b) and (c) follow through as in Theorem 10 of \cite{Wall} with the only difference that they must be applied exclusively to points living inside the cutoff spacetime $\cal{M}_\mathfrak{S}$ since we are just considering the restricted domain $Y$ of the function $t^{\Sigma}$. First we consider the case when there are no past or future horizons inside the cutoff spacetime $\cal{M}_\mathfrak{S}$. This means that timelike signals sent towards ${\rm Int}(\cal{M})$ from a point on the screen will reach any point in the interior after a finite amount of time. We want our Cauchy slices $\Sigma_\mathfrak{S}\subset\mathcal{C}_\mathfrak{S}(S)$ to be achronal. Since $\Sigma_\mathfrak{S}$ includes $\mathcal{A}$, this means that we do not want timelike signals\footnote{Achronal surfaces may contain light like pieces.} sent to/from $\mathcal{A}$  from/to points in $\Sigma_\mathfrak{S}$ to reach their destination. This puts a bound on the time function (note that we have fixed the value of the time function at $\mathcal{A}$) and it takes the form $t^{\Sigma}_{{\rm min}}(y)\leq t^{\Sigma}(y)\leq t^{\Sigma}_{{\rm max}}(y)\ \forall\ y\in Y$.  This is the statement of equiboundedness, and this proves the theorem for the horizon-less case. When there are horizons with Kasner singularities in their future, the proof is simply to argue that when maximizing the min area, the maximum surface never touches the singularity. This part of the proof goes through without any change in our case.

%$\Sigma_\mathfrak{S}$ in non-zero finite time iff $\Sigma_\mathfrak{S}$ were achronal. This gives us a ``test" for achronality which we term as {\em relative achronality}.
%\footnote{Note that we could have sent light signals from the null infinity of $\cal{M}$ instead. However it takes infinite time for any light signal shot from null infinity to reach any point in the interior. In this case, even if any $\Sigma\subset\mathcal{C}(S)$ were not achronal with respect to timelike geodesics shot from the point on $\Sigma$ at null infinity, they would appear as achronal since the light rays never reach and thus timelike geodesics never will. This is why we are bound to define achronality with respect to the Dyson sphere $\mathfrak{S}$.}%
%Since $\Sigma_\mathfrak{S}\subset\mathcal{C}_\mathfrak{S}(S)$ are relative achronal, we know that the value of time function for each $\Sigma_\mathfrak{S}$ will be bounded as $t^{\Sigma}_{{\rm min}}(y)\leq t^{\Sigma}(y)\leq t^{\Sigma}_{{\rm max}}(y)\ \forall\ y\in Y$. 

%Since $\Sigma_\mathfrak{S}\subset\mathcal{C}_\mathfrak{S}(S)$ includes $\mathcal{A}$, light (and timelike) signals sent to/from $\mathcal{A}$ from/to a points $\Sigma_\mathfrak{S}$ should not reach that point. 

Together this proves the existence of relative maximin surfaces in a large class of interesting spacetimes. Let us also note in passing how the relative maximin construction we outlined works out for Minkowski space with a screen \cite{CK}. This is easiest to visualize for the 2+1 dimensional case. Take the screen sub-region to be the one defined by the $t=0$ slice in some choice of Minkowski coordinate outside (and at) the screen. Inside the cut-off we let the Cauchy slice vary\footnote{Note that we are treating Minkowski just as any other geometry here inside the cut-off, and the Cauchy slices inside can take any shape as long as they are Cauchy slices. They do not have to respect the isometries that happen to be there in the geometry.}. It is easy enough to convince oneself that the max value of the min area happens on the Cauchy slice defined by the continuation of the $t=0$ slice into the interior of the screen, and that it happens on the straight line segment connecting the anchoring points.

%Using the Ascoli-Arzela theorem, we have thus been able to prove that $\mathcal{C}_\mathfrak{S}(S)$ is compact. Following the arguments in Theorem 10 of Wall, one can then prove that the relative maximin surface $M(S)$ anchored to $S\subset \mathfrak{S}$ exists.

\subsubsection{Properties}

\begin{itemize}
\item {\bf Equivalence with Relative HRT Surfaces:} The argument for the equivalence between relative maximin surfaces and relative  HRT surfaces is a trivial adaptation of that in \cite{Wall}. Though straightforward, this observation is what makes the relative maximin construction useful. 

%nd hence the relative HRT surface $\chi_c(S)$ by the previous bullet point

    \item When $S$ is SAS, the relative maximin surface $M(S)$ has lesser area\footnote{In \cite{Wall}, it was possible for the difference in areas of the causal and maximin surfaces to be infinite due to leading order divergences of the area functional near the AdS boundary. However in our case, there is no such problem since the areas of each of the surfaces are defined to be finite, thanks to the bulk cutoff, {\em aka} holographic screen.} than the waist of the asymptotic causal diamond $W(p,q)$. Here by the area of the waist of the ACD, we mean the area of the component of the waist lying inside the cutoff spacetime, that is, $W^0(p,q)\equiv W(p,q)\cap\mathcal{M}_\mathfrak{S}$. 
     
    \begin{proof} Consider $\Sigma_\mathfrak{S}\subset\mathcal{C}_\mathfrak{S}(S)$ such that the relative maximin surface $M(S)\subset\Sigma_\mathfrak{S}$. By definition, $M(S)$ is minimum on $\Sigma_\mathfrak{S}$. Without loss of generality, we choose  $\Sigma_\mathfrak{S}$ to lie to the future of $W^0(p,q)$, that is $\mathcal{I}^{-}[W^0(p,q)]\cap\Sigma_\mathfrak{S}=\phi$. Consider the surface $W^{+}(p,q)\equiv \partial\mathcal{I}^{+}(q)\cap\Sigma_\mathfrak{S}$. By definition $W^{+}(p,q)$ lies to the future of $W^0(p,q)$ and thus by the Second Law of horizons on $\partial\mathcal{I}^{+}(q)$, it has lesser area than that of $W^0(p,q)$. Thus using the minimality of $M(S)$ on $\Sigma_\mathfrak{S}$, one can write ${\rm Area}[W^0(p,q)]>{\rm Area}[W^{+}(p,q)]>{\rm Area}[M(S)]$.\end{proof}
    
 %   (ii) $\Sigma_\mathfrak{S}$ lies to the past of $W^0(p,q)$, that is $\mathcal{I}^{+}[W^0(p,q)]\cap\Sigma_\mathfrak{S}=\phi$. Now instead consider the surface $W^{-}(p,q)\equiv \partial\mathcal{I}^{-}(p)\cap\Sigma_\mathfrak{S}$. Using Second Law of horizons on $\partial\mathcal{I}^{-}(p)$ and following similar argumenets as that of case (i), we can show that ${\rm Area}[W^0(p,q)]>{\rm Area}[W^{-}(p,q)]>{\rm Area}[M(S)]$. Thus in both the cases, ${\rm Area}[W(p,q)]>{\rm Area}[M(S)]$.

%(iii) If $W^0(p,q)$ cuts the $\Sigma_\mathfrak{S}$ and is not strictly to the future or past of it, but contains pieces that are to the past as well as pieces that are to the future, then we can apply the above argument to each piece separately, sum the result, and still get the same inequality. If there is a piece of the causal surface that lies on $\Sigma_\mathfrak{S}$, then that can go into the sum as it is. 
    
    Since the relative HRT surface $\chi_c(S)$ has been proved to be equivalent to $M(S)$, the above result holds for $\chi_c(S)$ as well. 
    
    Note that for stating and proving this property, we needed $S$ to be SAS; an analogous statement does not exist for naive sub-regions. This is because the definition of a causal surface anchored to screen relies on it being cut out by an ACD.

\item {\bf Definition 3.3:} The {\em classical relative entanglement wedge} $EW_c(\cal{A})$ corresponding to a sub-region $\cal{A}$ is defined as the bulk domain of dependence of a partial Cauchy surface bounded by the relative maximin surface $M(S)$/ relative HRT surface $\chi_c(S)$ and the sub-region $\cal{A}$.

    \item For naive sub-regions $\mathcal{A}$ and $\mathcal{B}$ on the holographic screen such that $\mathcal{A}\supset \mathcal{B}$,
     
     (a) $EW_c(\mathcal{A})\supset EW_c(\mathcal{B})$ with $\chi_c(S_{\mathcal{A}})$ spacelike separated from $\chi_c(S_{\mathcal{B}})$ where $S_X\equiv\partial X$. This is a classical version of entanglement wedge nesting.
     
     (b) there exists a  $\Sigma_\mathfrak{S}\subset\mathcal{C}_\mathfrak{S}(S_{\mathcal{A}})\cap \mathcal{C}_\mathfrak{S}(S_{\mathcal{B}})$ on which both $\chi_c(S_{\mathcal{A}})$ and $\chi_c(S_{\mathcal{B}})$ are minimal.   
      
      The proofs of these results follow the same way as Theorem 17 of \cite{Wall}.
    
    %\item {\em Let $\mathfrak{C}(p_i,q_i)$ with $i=A,\ B,\ C$ such that $\mathfrak{C}(p_i,q_i)\cap\mathfrak{C}(p_j,q_j)\subset\mathcal{M}/{\rm Int}(\mathcal{M})\ \forall\ i,j=A,\ B,\ C\ {\rm and}\ i\neq j$.\footnote{Such constraints on the asymptotic causal diamonds provides us with the corresponding allowed subregions $\mathcal{A}_i$ with $i=A,\ B,\ C$ which are disjoint but may share special anchoring surfaces.} Then the relative version of the strong subadditivity property holds, namely}
    \item {\bf Strong Sub-Additivity:} For disjoint naive sub-regions $\mathcal{A}$, $\mathcal{B}$ and $\mathcal{C}$ on the holographic screen (which may share a boundary), the relative version of the strong subadditivity property holds, namely
    \begin{equation}        {\rm Area}[\chi_c(S_{\cal{A}\cal{B}})]+{\rm Area}[\chi_c(S_{\cal{B}\cal{C}})]\geq {\rm Area}[\chi_c(S_{\cal{A}\cal{B}\cal{C}})]+{\rm Area}[\chi_c(S_{\cal{B}})]     \end{equation}
    \begin{proof} 
    
    (i) Using the previous property, we know that there exists a spacelike slice $\Sigma_\mathfrak{S}\subset\cal{C}_\mathfrak{S}(S_{\cal{A}\cal{B}\cal{C}})\cap \cal{C}_\mathfrak{S}(S_{\cal{B}})$ on which both $\chi_c(S_{\cal{A}\cal{B}\cal{C}})$ and $\chi_c(S_{\cal{B}})$ are minimal surfaces and $\chi_c(S_{\cal{A}\cal{B}\cal{C}})$ is nowhere inside of $\chi_c(S_{\cal{B}})$.%everywhere outside
    
    (ii) One can shoot null congruences from $\chi_c(S_{\cal{A}\cal{B}})$ and $\chi_c(S_{\cal{B}\cal{C}})$ which intersect $\Sigma_\mathfrak{S}$ at $\Tilde{\chi}_c(S_{\cal{A}\cal{B}},\Sigma_\mathfrak{S})$ and $\Tilde{\chi}_c(S_{\cal{B}\cal{C}},\Sigma_\mathfrak{S})$ respectively. Using the Raychaudhuri equation in conjunction with the NCC and the generic condition along with the fact that $\chi_c(S_{\cal{A}\cal{B}})$ and $\chi_c(S_{\cal{B}\cal{C}})$ are extremal surfaces (and hence have $\theta=0$), one can argue that the null congruences will have $\theta<0$. Hence $\Tilde{\chi}_c(S_{\cal{A}\cal{B}},\Sigma_\mathfrak{S})$ and $\Tilde{\chi}_c(S_{\cal{B}\cal{C}},\Sigma_\mathfrak{S})$ have lesser area than and are homologous\footnote{This is because $\Tilde{\chi}_c(S_{\cal{A}\cal{B}},\Sigma_\mathfrak{S})$ and $\Tilde{\chi}_c(S_{\cal{B}\cal{C}},\Sigma_\mathfrak{S})$ are connected by $\chi_c(S_{\cal{A}\cal{B}})$ and $\chi_c(S_{\cal{B}\cal{C}})$ respectively via null congruences shot from the former.} to $\chi_c(S_{\cal{A}\cal{B}})$ and $\chi_c(S_{\cal{B}\cal{C}})$ respectively.
    
    %(iii) From the argument for strong subdditivity on the spacelike slice $\Sigma_\mathfrak{S}$, we have\\ ${\rm Area}[\Tilde{\chi}_c(S_{\cal{A}\cal{B}},\Sigma_\mathfrak{S})]+{\rm Area}[\Tilde{\chi}_c(S_{\cal{B}\cal{C}},\Sigma_\mathfrak{S})]\geq {\rm Area}[\chi_c(S_{\cal{A}\cal{B}\cal{C}})]+{\rm Area}[\chi_c(S_{\cal{B}})]$ and thus relative version of strong subadditivity is proved. 
(iii) On the Cauchy slice $\Sigma_\mathfrak{S}$, one can split $\Tilde{\chi}_c(S_{\cal{A}\cal{B}},\Sigma_\mathfrak{S})\equiv f+e$ and $\Tilde{\chi}_c(S_{\cal{B}\cal{C}},\Sigma_\mathfrak{S})\equiv d+g$ as shown in the Figure \ref{SSA}. 
\begin{figure}[h]\centering
\hspace{-5mm}
\includegraphics[width=80mm,trim = {0 0cm 0 5cm}, clip]{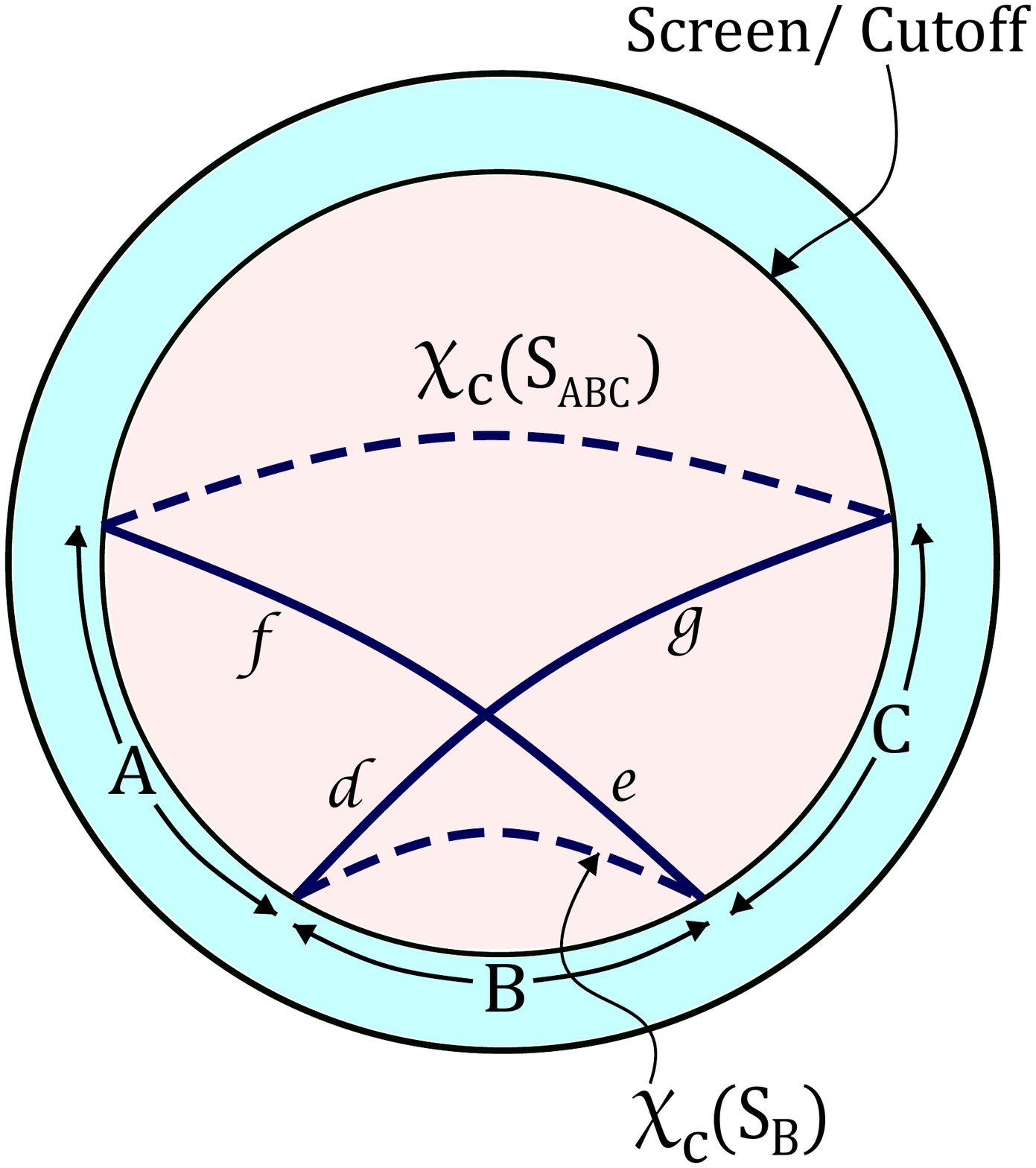} \\
\vspace{-2cm}
\caption{Strong sub-additivity, on one spatial slice.}
\label{SSA}
\end{figure}
Since $\chi_c(S_{\cal{A}\cal{B}\cal{C}})$ and $\chi_c(S_{\cal{B}})$ are minimal area surfaces on $\Sigma_\mathfrak{S}$, we have
   \begin{subequations}
    \begin{equation}
    \label{SSA1}
        {\rm Area}[f]+{\rm Area}[g]\geq {\rm Area}[\chi_c(S_{\cal{A}\cal{B}\cal{C}})]
    \end{equation}
    \begin{equation}
       \label{SSA2}
        {\rm Area}[d]+{\rm Area}[e]\geq {\rm Area}[\chi_c(S_{\cal{B}})]
    \end{equation}
    \end{subequations}
    This is possible because the surfaces $f+g$ and $d+e$ are anchored to $S_{\cal{A}\cal{B}\cal{C}}$ and $S_{\cal{B}}$ respectively. This is a repetition of the argument in \cite{Headrick}. Adding \eqref{SSA1} with \eqref{SSA2} and rearranging the areas on the LHS, one can argue that ${\rm Area}[\Tilde{\chi}_c(S_{\cal{A}\cal{B}},\Sigma_\mathfrak{S})]+{\rm Area}[\Tilde{\chi}_c(S_{\cal{B}\cal{C}},\Sigma_\mathfrak{S})]\geq {\rm Area}[\chi_c(S_{\cal{A}\cal{B}\cal{C}})]+{\rm Area}[\chi_c(S_{\cal{B}})]$ and thus relative version of strong subadditivity is proved. \end{proof}
    
    \end{itemize}

%\subsubsection{Discussion}

Our proof of strong sub-additivity was for naive sub-regions. We expect that a similar statement can also be proved for SAS sub-regions, modulo one caveat: to usefully state strong subadditivity, we need the SAS's of all three ACDs relevant for the proof above to lie on the same $t={\rm fixed}$ slice of some asymptotically flat coordinate system. We have not been able to prove this statement in full generality (see footnote \ref{allACD}), even though we strongly suspect it is true. But it is easy enough to convince oneself of its validity in one special case.  This is when the spacetime has no time dependence. There is a timelike Killing vector in the bulk in this case and by choosing it to match up with the Minkowski time at the boundary, it can be chosen as the asymptotically flat time coordinate. Because time translation is an isometry, it should be clear that waists of symmetric causal diamonds lie on $t= {\rm const}$ slices\footnote{This is most instructively illustrated by considering a 1+1 geometry that is spatial translation {\em non}-invariant, but time translation invariant.}. Note that since we aim to investigate black holes at various epochs of Hawking radiation as in \cite{Penington}, and in each of them the geometry is effectively static as far as the determination of entanglement wedges are considered, this is in principle enough for the purposes of deriving the Page curve. Nonetheless, in this paper have presented the naive sub-region version of the proof because (as we emphasized in the last section) SAS sub-regions and naive sub-regions approximate each other better and better as the cut-off becomes large. 

A second point implicit in our proof is that we have used a representative of the extremal surface on other slices,  eg. $\Tilde{\chi}_c(S_{\cal{A}\cal{B}},\Sigma_\mathfrak{S})$. Implicit in this is the assumption that such representatives exist. This can be proved by first observing that the slice  is anchored to pass through the screen sub-region $\mathcal{A}$ with $t= {\rm fixed}$ on the screen. Any Cauchy slice that is anchored to $\mathcal{A}$ and on the interior of the screen must cut the inward directed codimension-1 null surface shot out from the HRT surface. Note that the HRT surface together with screen sub-region defines the classical relative entanglement wedge.

If we were to phrase the SSA proof in terms of SAS sub-regions, this argument can be phrased as a variation of the statement 6 (e) in \cite{Wall}. The key observation in the SAS case is that $\Sigma_\mathfrak{S}$  must by construction intersect the relative casual wedge\footnote{This is the domain of dependence of the region between $W^0(p,q)$ and the screen sub-region $\mathcal{A}$ itself.} by an argument analogous to the one in the previous paragraph. But we also know that  classical relative entanglement wedge defined above must contain the relative casual wedge, because the relative HRT surface lies outside the causal surface. This means that from the fact that $\Sigma_\mathfrak{S}$ cuts the causal wedge, it must follow that it also cuts the null congruence shot from the relative HRT surface. This provides us with the required representative in the SAS version of the proof. Note that this is a version of argument 6 (e) in \cite{Wall}, but with a slight twist because here the screen has no intrinsic causal structure and therefore we have to work instead with the (relative) bulk causal wedge. Since we are restricting our Cauchy slices to pass through the entire $t= {\rm fixed}$ slice at the boundary, and not merely through the anchoring surface, these arguments are all more automatic that they are in AdS. In our case we can fully dispense with the relative causal wedge (the analogue of the boundary domain of dependence in AdS) and instead work directly with the relative classical entanglement wedge to make the argument. In \cite{Wall} the fact that only the boundary of the sub-region was fixed, and not the whole sub-region, made argument 6 (e) slightly less trivial.

The key point in all of these arguments is that we are crucially using asymptotic flatness and the fact that the screen is large, so that we can employ slightly fancier versions of the observations regarding Minkowski space in \cite{CK}. The presence of an asymptotically Minkowski structure gives us more tools than the purely causal structure based arguments used by others in similar contexts.

\section{A Phase Transition at the Page Time in Flat Space}

The key observation of \cite{Penington, Almheiri} is that in AdS/CFT, there is a phase transition in the entanglement wedge  after the Page time. This observation  leads to the realization that the Page curve turns around. Our goal is to make a similar statement in the context of flat space quantum gravity. In order to do this, we will first find it useful to clarify the relationship between fine-grained and coarse-grained quantities. The former can be viewed as statements in the full quantum gravity Hilbert space, while the latter are statements in bulk effective field theory. The relationship between the two is not too transparent even in the AdS/CFT literature, and the relevant quantities have some interesting distinctions in flat space. We will phrase the discussion in the AdS/CFT language, and make appropriate translations and comments as we proceed. 
%The point of this exercise is to state things in a form that lends itself to translation to the flat space context,  after appropriate identifications of the quantities are made. We will also emphasize some key points which make the translation possible. A key ingredient in the argument is the connection between the fine-grained and the coarse grained Hilbert spaces via the generalized entropy, and one of our goals in the previous sections was to argue that this is indeed a well-defined quantity even in flat space. 

%At the end of the sub-section, we will also explain gow to translate this discussion into the corresponding flat space discussion. 

%expect analogous statement to apply more generally, with appropriate ACD-based generalizations of the notion of a boundary sub-region.

\subsection{Fine-Grained and Coarse-Grained Page Curves}

\begin{figure}[h]\centering
\hspace{-5mm}
\includegraphics[height=80mm,width=100mm,trim = {0 0cm 0 10cm}, clip]{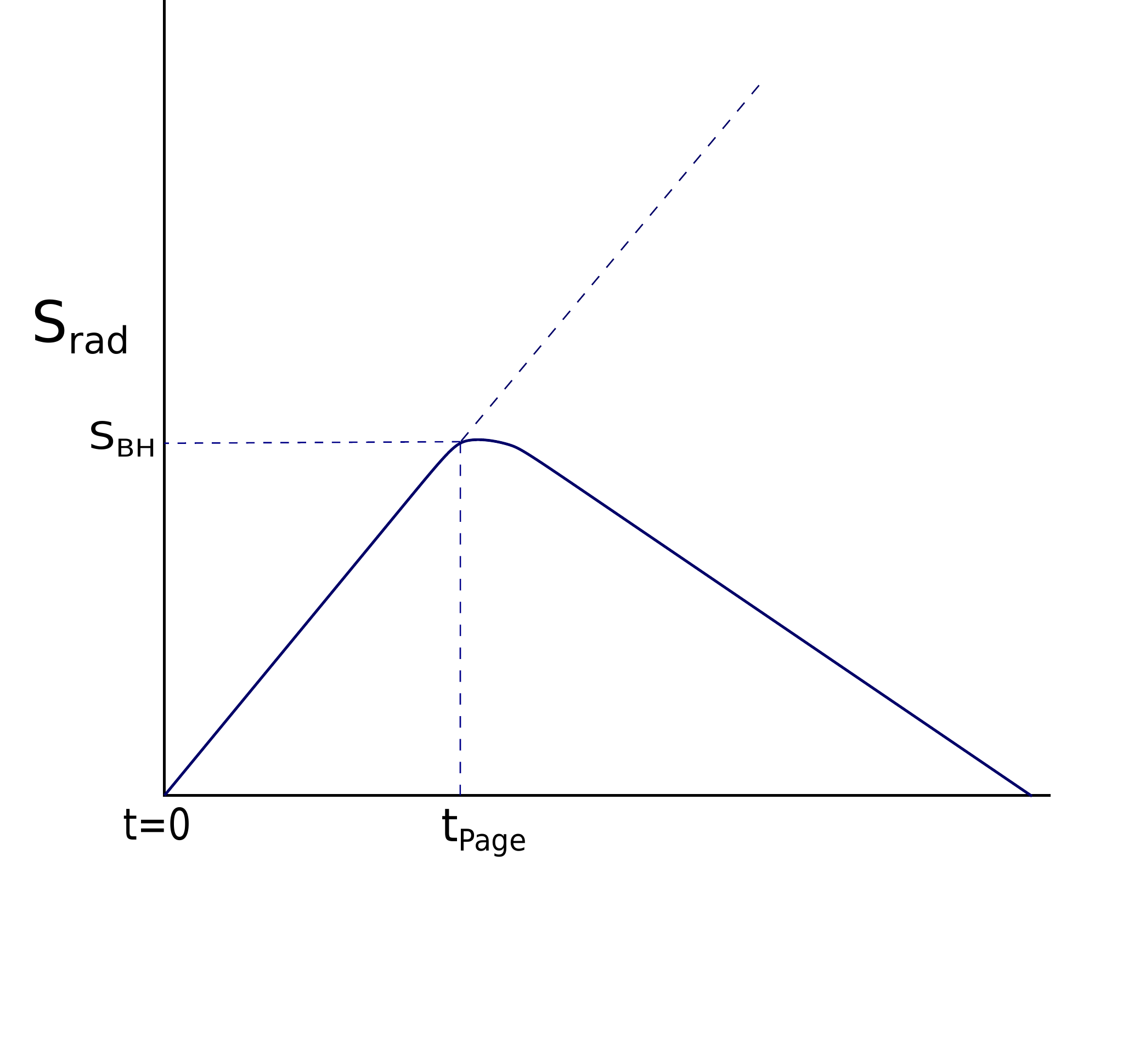} \\
\vspace{-1.5cm}
\caption{The mandatory Page curve.}
\label{PAGE}
\end{figure}

In \cite{Penington}, the system\footnote{In \cite{Penington}, the factor ${\cal H}_{sink}$ was called ${\cal H}_{rad}$, but it should not be confused with a radiation bath or a heat reservoir with which the CFT is in thermal equilibrium, so we prefer to call it a sink. It is a huge Hilbert space (aka available phase space) to which the Hawking radiation is eternally lost.} under study is 
\bea
{\cal H}_{total}={\cal H}_{CFT} \otimes {\cal H}_{sink}. \label{CFT-sink}
\eea
The initial state one considers is a black hole that has formed by collapse, after the formation stage, in the CFT Hilbert space ${\cal H}_{CFT}$. But the CFT is coupled to the sink Hilbert space ${\cal H}_{sink}$ as well. The black hole Hawking radiates, and ${\cal H}_{sink}$ is to be understood as the sink to which the radiation leaks off to from the CFT Hilbert space.  The state of the system at any given time $| \Psi\rangle$ has support on both tensor factors, and there are two natural reduced density matrices one can define using them:
\bea
\rho_{CFT}={\rm Tr}_{sink}| \Psi\rangle \langle \Psi |, \ \  \rho_{sink}={\rm Tr}_{CFT}| \Psi\rangle \langle \Psi |
\eea
The entanglement entropy (von Neumann entropy) of either of these reduced density matrices is the same because the total state is pure. It is the time evolution of this entanglement entropy that we are interested in knowing  as the black hole evaporates. We will call this the fine-grained entropy of the CFT/radiation:
\bea
S_{CFT}=- {\rm Tr}_{CFT} \left( \rho_{CFT} \ln \rho_{CFT}\right) = S_{rad}=- {\rm Tr}_{sink} \left(\rho_{sink} \ln \rho_{sink} \right) \label{S}
\eea
Page \cite{Page1, Page2} has argued very generally, that the plot of this quantity should look schematically like figure \ref{PAGE}, the so-called Page curve. 
It should start around zero, increase, turn around, decrease and get back to around zero. 

The discussion above is completely general and is phrased in terms of the fundamental degrees of freedom, ie., the degrees of freedom of the CFT and the sink Hilbert space. But Hawking evaporation is naturally phrased {\em not} directly in terms of these fine-grained degrees of freedom, but in terms of the bulk effective field theory around the background of a black hole (where the black hole is to be thought of as a high energy state in the CFT). Information paradox can be viewed as the statement that when one computes the entanglement entropy we described above using effective field theory around a black hole, instead of doing the turnaround that Page curves are supposed to do, $S_{rad}^{naiveEFT}$ seems to relentlessly increase\footnote{We include the superscript $naiveEFT$ to emphasize that here we are including only the semi-classical bulk contribution to $S_{rad}$.}. This version of the information paradox is what the arguments of \cite{Penington, Almheiri} resolve. 

It is useful to define a more general notion of entanglement entropy than the one we introduced above, that exploits the natural locality structure of the CFT Hilbert space. We consider sub-regions of the CFT as tensor sub-factors of the CFT Hilbert space, and we compute the entanglement entropies of these sub-regions by defining their reduced density matrices\footnote{\label{thermal-subregion}This discussion is usually phrased for just the CFT Hilbert space without the ${\cal H}_{sink}$ factor, and we will adopt that language in this paragraph. Reduced density matrices for sub-regions in the CFT with respect to the full ${\cal H}_{total}$ can be defined, but as far as we know they have not been systematically discussed in the literature. In a flat space context, we will develop this idea further in a later sub-section. In this sub-section, we will discuss the special case when the ``sub"-region is the entire spatial slice of the CFT, momentarily.}. A key ``janus" relation that connects the fine-grained and coarse-grained descriptions is the observation that the fine-grained CFT sub-region entropy is given by the generalized entropy, see \eqref{tempSgen}, of the minimal bulk QES that is anchored to the boundaries of that sub-region:
\bea
S_{CFT}[\partial \chi]= S_{gen} [\chi] \label{micmac}
\eea
This statement is believed to hold to all orders in bulk perturbation theory. Note that the right hand side is defined in bulk EFT and the left hand side is a CFT quantity. %The key relation that connects the fine-grained and coarse-grained descriptions is the observation that the generalized entropy we defined for boundary-anchored bulk surfaces can be viewed as entanglement entropies of sub-regions in the CFT.

In \cite{Penington, Almheiri} the time dependence of this object during the Hawking evaporation of a black hole formed by collapse was studied, with two further caveats:
\begin{itemize}
\item the CFT ``sub"region one considers is the entire spatial slice of the CFT,  
\item the state is taken to live in the full ${\cal H}_{total}$ and not just ${\cal H}_{CFT}$. 
\end{itemize}
This means that now we are computing the $S_{CFT} = S_{rad}$ defined in \eqref{S} as a function of time, but using bulk effective field theory via adaptations of \eqref{micmac}. Note that adaptations are indeed necessary because \eqref{micmac} is defined within ${\cal H}_{CFT}$, but our black holes state lives in the coupled system ${\cal H}_{CFT} \otimes {\cal H}_{sink}$. To do this, the idea of {\em entanglement wedge complementarity} was argued to hold. For our purposes, this simply means that a QES in the bulk of the black hole geometry determines the entanglement wedges of both the degrees of the freedom in the CFT, as well as that of the radiation that is already in the sink: the region between the QES and the boundary of AdS is the entanglement wedge of the CFT and its complement is the wedge of the sink. 

Note that the key point where flat space quantum gravity differs from AdS is that in the former, $i^0$ does not have a useful notion of sub-region associated to them. This is associated to the fact that the CFT is a local theory, whereas we do not expect (naive) locality in the hologram of flat space. So at this stage, it may seem that we do not have a way to proceed, because the janus formula \eqref{micmac} is defined only for CFT sub-regions. Remarkably, there are a few observations that emerge in flat space, to save the day. The first, is that even though sub-regions do not exist, the existence of the ACD and the shadow of the ACD on the conformal boundary emerge as proxies for sub-regions. Second, at finite cut-off, ACDs help us define approximate sub-regions and therefore connect with our AdS intuition. In fact, we will see later that these sub-regions are naturally interpreted as sub-regions of a system coupled to a sink. Third, to make the \cite{Penington} arguments, we only need the entanglement wedge of the {\em full} system, not specific proper sub-regions. This makes life simple because even though there are numerous conceptual parallels between CFT sub-regions and shadows of ACDs, at a technical level they are very different objects, so we need not have been so lucky. Fourth, the entanglement phase transition we discuss below will be crucially related to the area of the horizon, and not directly to asymptotic quantities. This means that the argument can go through to flat space. 

Let us first describe the general picture of how we expect $S_{rad}$ to evolve, and then we will describe the various quantities more concretely in the context of flat space black hole evaporation. What we expect based on very general principles is that the fine-grained $S_{CFT}=S_{rad}$ should follow the Page curve. We will focus on $S_{rad}$ for concreteness. We wish to compute this quantity in bulk effective field theory -- we expect it to also produce the Page curve, because except at the very final stages of evaporation all bulk curvatures are small and we expect EFT to hold. The question is: how should one compute this entropy in EFT? The naive Hawking calculation would determine $S_{rad}$ by simply computing the entanglement entropy between the modes that have exited the geometry up to that time (outgoing modes that have gone into the sink) and the modes that have gone into the horizon (ingoing modes that fell into the black hole)\footnote{Note that this is a complete set of modes for the bulk effective field theory.}. 
\begin{figure}[h]\centering
\hspace{-5mm}
\includegraphics[angle=0,width=75mm,trim = {0 0cm 0 3cm}, clip]{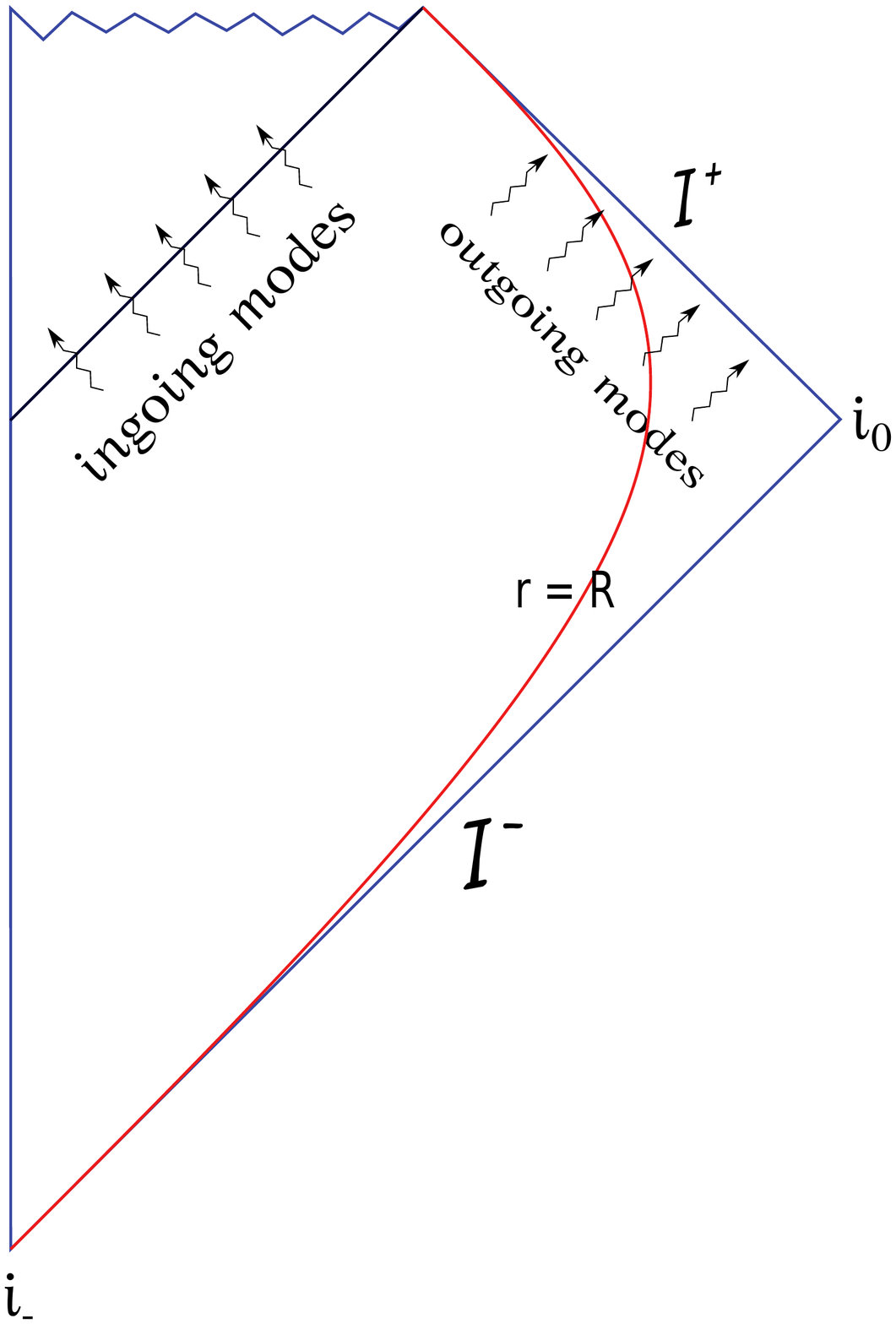} \\
\vspace{-1cm}
\caption{The outgoing and ingoing modes are entangled, but the former exit any finite screen due to Hawking evaporation.}
\label{modes}
\end{figure}
Doing this would amount to not paying attention to the entanglement wedge structure of the holographic theory, simply assuming that the entire bulk is in the entanglement wedge of the CFT (ie., the minimal QES is empty), and as a corollary keeping only the bulk piece in the generalized entropy
\bea
S_{gen}[\chi] = \frac{A[\chi]}{4 G} + S_{bulk}[\chi] \label{tempSgen}
\eea
when using  \eqref{micmac} to compute $S_{CFT}=S_{rad}$. Note that when the black hole interior is in the entanglement wedge of the CFT, the entanglement between the interior modes and the modes already in the sink is the dominant contribution to $S_{gen}$ and therefore the putative  $S_{rad}=S_{bulk}$. As the black hole keeps evaporating and more modes reach the sink, this can only increase: we have the information paradox. 

But what happens if we use both terms in \eqref{tempSgen} to first determine the minimal QES to identify the correct entanglement wedges? It is clear that in the early stages of Hawking evaporation, the entanglement between the black hole interior and the sink is small, because most of the radiation is yet to happen. This makes it plausible that $S_{gen}[\chi]$ is going to be minimized by minimizing the area term, and on any Cauchy slice, this is going to be just the empty surface. This means that in the early stages of Hawking radiation, the argument we described in the previous paragraph is correct and we see that $S_{rad}$ keeps rising. But something interesting happens around the Page time, where $S_{rad} \sim S_{BH}$. Note that beyond this time (by our own previous argument), $S_{bulk}$ corresponding to the empty QES becomes bigger than the $A[\chi]/4 G$ of a surface of (approximately) horizon radius. Note also that the moment the minimal QES is near/outside the horizon, the interior modes are in the entanglement wedge of the sink. The modes that exited the geometry as outgoing modes (and therefore purify the interior modes), are already in the sink. This means that $S_{bulk}[\chi]$ of this candidate QES is negligible compared to that of the empty QES, whose $S_{bulk}[\chi]$ is of the order of Bekenstein-Hawking entropy because we are now beyond the Page time. This means that it is favorable for the minimal QES to switch to this new near-horizon QES. This is the promised entanglement wedge phase transition. As the black hole keeps radiating, the size of the horizon shrinks and along with it, the $S_{rad}=S_{gen}[\chi] \sim A[\chi]/4G$ of this new minimal QES\footnote{Let us emphasize again, that our notation distinguishes between $S_{rad}$ and $S_{rad}^{naiveEFT}$. The latter is what Hawking would have calculated, and it keeps increasing even after the Page time. Sometimes this quantity is called $S_{rad}$ in the literature. One reason we have tried to be precise about fine-grained and coarse-grained quantities is to avoid this type of ambiguity.}. We are now in the falling part of the post-Page-time Page curve, and have resolved one version of the information paradox.

Before the Page time, the minimal QES is the empty surface, which is interpreted to mean that the entanglement wedge of the CFT degrees of freedom (as captured by the asymptotic spatial boundary) contains the entire interior of the black hole. But after the phase transition at the Page time, the new entanglement wedge is defined by a finite radius surface close to the horizon and therefore the degrees of freedom in the interior are no longer reconstructible from the boundary/CFT. By entanglement wedge complementarity, the interior degrees of freedom are in the Hilbert space of the Hawking radiation that is now in ${\cal H}_{sink}$. 

\subsection{Two Formulations for Flat Space Hawking Evaporation}\label{formulation}

To build the connection between this picture and flat space, the first thing we will need is to identify a tensor factorization of the form  
\bea
{\cal H}_{BH} \otimes {\cal H}_{sink}. \label{FQG-sink}
\eea
%^\mathfrak{S}
The first factor is to be viewed as a suitably holographic Hilbert space in which the black hole lives, and the second is the sink to which Hawking radiation can leak off to. If such a factorization cannot be identified, it is meaningless to talk about the conventional Page curve. The key question then is, is there such a natural tensor factorization? %If such a tensor factorization cannot be meaningfully identified, there will be no sense in which a conventional Page curve will exist. 

% The reason why we have associated a superscript $\mathfrak{S}$ to it is to emphasize that the Hilbert space we are using is to be associated to the holographic screen that we introduced, in some manner. But how exactly should this be done? How should we think about this ${\cal H}_{FQG}^\mathfrak{S}$? This is what we will try to answer in this section. 

%We will see that the flat space reconstruction wedge is naturally defined in the full flat space quantum gravity, ie., it  extends all the way to the conformal boundary. To be more specific, it is constructed without any reliance on the holographic screen and therefore it has a Platonic character to it. The holographic screen on the other hand, is man-made. Despite this, the latter was useful to us from a few different points of view, especially to make connections with entropies of sub-regions and thereby to make the parallels with AdS manifest. See also  discussions in \cite{CK, Budha} for other uses of the holographic screen. 

The fact that the screen and the conformal boundary are distinct in flat space means that now we need to re-think the picture of Hawking evaporation. We will establish the existence of entanglement wedge phase transitions and the Page curve in what follows, but before doing so, will now pause to discuss how to formulate the Page curve problem in flat space. We will see that there are (at least) two ways to do this. In this and the next section, we will describe and relate the two points of view. The first formulation is what we feel should qualify as the ``conventional" formulation of the flat space Page curve. But we will also present another formulation that we have not seen elsewhere, which we think is of interest. % Despite its possible novelty to the reader, we feel that the second interpretation may provide insights on holography and quantum gravity in general spacetimes. See \cite{CK, Budha} for related discussions.

%\footnote{The bulk gravity in the setting of \cite{Raghu} is coupled to a CFT. This CFT extends from the dynamical AdS bulk to non-dynamical Minkowski piece.} 

Let us first go back to the results of \cite{Penington, Almheiri} which couple the CFT to a large radiation sink. This allowed a certain formulation of the information paradox to be resolved in AdS. The fact that one is  letting the Hawking radiation leak out of the AdS/CFT system is very suggestive of flat space where we may expect the radiation to ``leak off to infinity". It was further noted in a 1+1 dimensional asymptotically AdS setting \cite{Raghu} that one can model the sink by simply attaching a 1+1-dimensional non-gravitating Minkowski space to the boundary of AdS. This suggestion, and indeed the original idea of \cite{Penington, Almheiri} which couples the CFT to a sink, already capture some of the intuition of flat space black hole evaporation.  It seems immediate that our flat space holographic screen should be analogous to the AdS boundary in this language, but it comes with some subtleties. Let us discuss these subtleties in the form of the two different formulations.

\begin{itemize}
\item {\bf Formulation 1:} The most direct flat space adaptation of the ``AdS gravity + Minkowski sink" picture presented in \cite{Raghu} is to simply view the holographic screen as providing a (possibly approximate) tensor factorization \cite{CK}. This is the tensor factorization between the inside-the-screen-region with dynamical gravity where the evaporating black hole lives, and an outside-region with a nearly non-dynamical metric that extends all the way to the conformal boundary. The radiation that goes through the screen into the outside-region is to be viewed as having gone into the sink. This is analogous to the set up in \cite{Raghu} where the AdS boundary was the analogue of the screen.  See also \cite{Thor, Iizuka, Iizuka2, Strominger} for variations in explicitly flat space settings. In \cite{Raghu}, the boundary of AdS is a place where gravity decouples, so the Minkowski space they attach to the boundary of AdS does not contain dynamical gravity.  In our flat space screen \cite{CK} however, the metric also gets a transparent boundary condition. Gravity is weak but still dynamical there.  At the end of section 4.5 we will come back to the connection with \cite{Raghu}.

%When the screen is large however, as we argued earlier, the metric can be treated as approximately non-dynamical for various purposes.

The advantage of this point of view is that it immediately connects with our mental picture that radiation ``leaves the system" in asymptotically flat space. The disadvantage is that  since there is no real decoupling of gravity on the screen,  one needs to be careful in interpreting the sink Hilbert space as being distinct from the approximate gravity Hilbert space - after all, gravity is holographic. It should also be noted that even though the region beyond the screen is approximately flat, at the end of Hawking evaporation it will contain {\em all} of the matter and information that was there in the original black hole, and therefore it is far from empty. 

We believe despite these worries that the picture is reasonable. The reason is that even though all the radiation from the black hole is present in the asymptotic region after Hawking evaporation, it is also infinite in ``space" and therefore the backreaction is arbitrarily small\footnote{In AdS, note that because of the finite propagation time to the boundary, there will be reflection back from the boundary in finite time. }. This means that the approximate tensor factorization between the interior and the exterior of the screen may be reasonable enough for the purposes of the Page curve argument. Let us emphasize that the screen region and beyond is at arbitrarily low curvature during the entirety of the Hawking evaporation process. So we expect that it can be replaced by the Hilbert space of a weakly coupled (essentially free!) bulk EFT\footnote{For the purposes of the black hole evaporation problem, we can treat gravitons as free fields in this region.}. This is the sink Hilbert space.

%The extremal surface calculations we have done in the previous section (which will be interpreted in the next subsection), should be taken as evidence for this. 

%The approximation we are making when doing this is that we are assuming that flat space quantum gravity allows an approximate tensor factorization into a theory definable on the holographic screen at a finite but large cut-off, and the approximately Minkowski region outside the screen. The latter gets interpreted as the sink. 

In fact, in a later sub-section, we will argue that ACDs provide a natural way to interpret the total space of this factorized Hilbert space as the full Hilbert space of flat space quantum gravity defined with respect to the conformal boundary. In other words,
\bea
{\cal H}_{QG} \approx {\cal H}_{QG}^\mathfrak{S} \otimes {\cal H}_{out} \label{1vs2}
\eea
where the first factor on the right corresponds to the interior of the screen, and the second the exterior which is interpreted as the sink. The right hand side is what defines \eqref{FQG-sink} in Formulation 1. In fact, considerable evidence that the first factor should be viewed as holographic was already presented in \cite{CK} using ACDs. The fact that the observations we make in this paper (eg., the screen sub-regions satisfy strong sub-additivity) are possible, strengthen this case. These are all hints that this approximate tensor factorization is a reasonable one in flat space for the systems under consideration. We will see later that there exists a natural way to connect the two sides of \eqref{1vs2} by using ACD-inspired ideas and viewing the screen as a regulator.  We can make the approximation better by increasing the size of screen. Throughout the paper, we will present further arguments and circumstantial evidence to strengthen the claim that this tensor factorization is natural in flat space. 

%It also raises the possibility that there exists a limit where one can send the screen size to infinity (together with appropriate other scalings of various quantities) under which the correlation functions one defines on the screen have well-defined non-trivial limits.

Note that the screen size is a largely arbitrary dimensionful parameter in the problem. It has similarities to the AdS scale. Note also that small and large black holes exist for flat space with box boundary conditions \cite{Bala}. For large black holes, the Hawking radiation rate in the zone region and near-screen region may be harder to cleanly disentangle, but for small black holes we expect no such issues.

\item {\bf Formulation 2:} While the above picture is quite reasonable, we feel that it may also be useful for some purposes (eg., for investigations of formal aspects of holography and quantum gravity) to treat the full flat space quantum gravity Hilbert space as a single entity. After all, the full quantum gravity is usually defined in terms of the asymptotic region and the conformal boundary at infinity, and not by excluding those regions in some approximation and treating them as part of a sink. This asymptotic approach is useful, eg., when we treat the observables in flat space as S-matrix elements. The price we pay for this definition is that in this language, our mental picture that ``radiation leaves the system" does not apply, because the system is defined to include the radiation. Note that this does {\em not} mean that the picture presented in Formulation 1 is incorrect. All it is saying is that we are no longer looking at any  notion of a ``sub-system" that contains the black hole (now we are working with the ${\cal H}_{QG}$ on the left hand side of \eqref{1vs2}), and therefore tautologically we will not be seeing the Page curve. Let us also note in passing that the time it takes for signals to reach the conformal boundary of flat space is infinite\footnote{The light ray traverses curves with zero proper time, so what we mean by this statement is this -- if we were to take an inertial observer in the bulk of Minkowski space, and measure the time interval (as measured by his clock) it takes for him to send out a signal to a reflector and receive it back, this interval will diverge as the reflector is moved to the asymptotic boundary. This is not the case in AdS.}.

%Hawking radiation originated in the quantum gravity Hilbert space, and since there is no other tensor factor, it never leaves ``The System" (aka Penrose diagram). 
%In other words now we are working with the ${\cal H}_{QG}$ on the left hand side of \eqref{1vs2}. 
%because (a) the volume of the sink and (b) the propagation time to the conformal boundary are infinite for any finite value of the cut-off, we do not believe this is the correct way to think about this limit. 

%The reader may worry that in the limit where the cut-off goes to infinity, the left hand side of \eqref{1vs2} should be viewed as a limit of the right hand side where the second tensor factor becomes ``smaller and smaller". If that is the case, it may seem reasonable to treat the left hand side as a leaky limit of just the first factor on the right (suggesting potential non-unitarity). But

So if we are going to treat the entire flat space quantum gravity as ``The System", can we formulate information paradox here\footnote{We will assume here that this full system has a unitary description. The radiation is always present in the full system. We will also see in the next subsection a natural way to relate/regulate entanglement entropies in the two pictures, which strongly suggests that the region outside the cut-off purifies the region inside. Note also that quantum gravity in flat space defined with respect to the conformal boundary is widely believed to be unitary, for many other reasons.} in a way that parallels the work of \cite{Penington, Almheiri}? This would require us to come up with a way to couple flat space quantum gravity to an external sink in a natural way. How are we to do this? A usual philosophy might be to couple the sink to the null boundary. At the formal level this can certainly work, and everything we say below (including the Page curve) will go through there as well. But we will phrase the problem in a slightly different way.

To draw some inspiration, let us first go back and look at the set up of \cite{Penington} from a holographic perspective. The idea there is that the Hawking radiation that gets to the conformal boundary of AdS\footnote{Note that this happens in AdS in finite time, at least for massless particles.}, goes over into a sink Hilbert space. To make this happen, we have to couple the CFT to another system: we have to modify the CFT path integral to include the sink, as well as terms that couple CFT operators and sink operators. Note that from the AdS/CFT perspective, the sink operators that couple to CFT operators will look like sources for the CFT, and therefore from the bulk perspective they will be boundary values of various bulk fields. In other words, in order to couple the Hawking radiation to a sink, what we have to do is to do the ``bulk plus sink" path integral with boundary values of bulk fields constructed from appropriate sink operators. Since the sink dynamics are not of much interest other than the fact that its Hilbert space size should be suitably large, the precise details will not matter. What we emphasize however is the fact that the boundary values of the bulk fields provide us a way to couple AdS quantum gravity to a sink, and this will allow us to extract Hawking radiation from the bulk. This is the set up to formulate the version of the information paradox considered in \cite{Penington}. Note that all we needed for this was just the most basic element of the AdS/CFT correspondence, namely that the boundary values of bulk fields are sources for the dual field theory.

In order to have a similar formulation, the first thing we may look for is a holographic correspondence that could work in flat space. Such a prescription that works to all orders in bulk perturbation theory was given in \cite{Budha}. It was suggested that instead of viewing bulk fields as being specified by boundary values, they should be viewed as being specified by sources on codimension-1 screens (eg., our holographic screen) together with prescribed fall-off conditions of bulk fields at infinity\footnote{Note that the screen encodes a hologram of the entire spacetime in this language \cite{Budha} and not just the region inside the screen. Let us repeat what we already mentioned once in the Introduction - the role of the screen in Formulation 1 and 2 are quite different. The difference is loosely parallel to the distinction between normalizable and non-normalizable modes in AdS/CFT.}. The bulk partition function would be a functional of the codimension-1 sources in this picture. Multiple things were shown to work out very naturally when one does this, which were difficult to realize in previous efforts. First, it was shown that this approach reduces to the standard boundary value prescription in AdS\footnote{Note that without this, we would be sacrificing the successes of the standard AdS/CFT prescription. The match happens because boundary limits of bulk codimension-1 sources turn out to be boundary values of bulk solutions near the boundary of AdS.}. Second, it was shown that the Euclidean prescription naturally analytically continues to the causal Feynman propagator in Lorentzian signature. Third, a natural proxy for the AdS normalizable mode emerges in Lorentzian signature, in the form of a homogeneous mode. Fourth, the extrapolate and differential dictionaries match. Fifth \cite{Ajay}, a well-defined variational principle exists for gravity {\em without} adding boundary terms once one specifies the asymptotic fall-offs and codimension-1 sources, and there exist natural ways to get a finite on-shell action \cite{Ajay}. 

It turns out that this prescription is in fact also naturally compatible with  the picture we presented above for extracting Hawking radiation into a sink. The difference from the AdS discussion of \cite{Penington} is that now the partition function (of flat space quantum gravity) is to be thought of as a functional of these codimension-1 sources instead of boundary values. These sources can be viewed as built from sink operators as before\footnote{Note that the sink need not live {\em on} the screen, the source on the screen is just a knob that one can use to couple the quantum gravity to the sink.}. This gives us an operational way to couple flat space quantum gravity to a Hawking radiation sink. Note also that since the screens of \cite{Budha} are naturally at finite radius, Hawking radiation reaches the screen at finite time. This is crucial for the viability of the proposal because extracting Hawking radiation from the conformal boundary instead of the screen would be conceptually challenging\footnote{Even though to be fair, since we are extracting radiation into a sink {\em from} the quantum gravity Hilbert space in Formulation 2, extracting from the screen or the boundary, are both likely only of formal interest. }, to say the least. This is another place where the convergence of the screen and the boundary in AdS, makes things different in flat space.

This formulation also leads to an understanding of the Page curve, just like Formulation 1. The systems under consideration are somewhat different in both cases and the information paradox is formulated differently. The first formulation is more natural if we want to take the point of view that Hawking radiation escapes off to infinity from a black hole that is localized somewhere in the bulk of flat space. This is possibly the picture that most closely captures the intuition that  black hole evaporation is in many ways like a burning lump of coal. The second formulation would be useful if we want to view Hawking radiation as being extracted out from the flat space quantum gravity Hilbert space (as defined via asymptotic flatness near conformal boundary), into a sink which is a whole another system. It is possible to view this as a Hawking radiation absorbing boundary condition at the screen.  We suspect that this set up maybe useful for understanding formal questions about flat space holography. Let us also emphasize again, that Formulation 2 will go through with minor changes, if we decided to extract radiation not at the screen, but at the null boundary. We will discuss this as well, when we discuss the Page curves.

The Hilbert space of Formulation 2 is closer in spirit to the one in \cite{Penington}:
\bea
{\cal H}_{total} = {\cal H}_{QG} \otimes {\cal H}_{sink}
\eea
where the first factor now stands for the full theory. We may choose to write the first factor as ${\cal H}_{QG}^\mathfrak{S}$ to emphasize that it is through a coupling on the screen that we couple it to the second (sink) factor -- but we will not do so to avoid confusion with the right hand side of \eqref{1vs2}.  Note that much like in \cite{Penington} and unlike Formulation 1, the sink Hilbert space is an entirely different system now.

We will argue below that both these setups lead to Page curves. This is unsurprising because in both cases radiation can escape to a sink, it is only the precise definition of the sink that differs.

\end{itemize}

\subsection{Quantum Extremal Surfaces and Entanglement Wedges}\label{rec}

To discuss the Page curve, we need to consider Quantum Extremal Surfaces in flat space, which define entanglement wedges. In AdS, these are surfaces that are anchored to the boundary which extremize the  generalized entropy as we noted before. They capture entanglement entropies in the CFT of sub-regions to which the QES is anchored to. In asymptotically flat space, our claim is that ACDs provide a way to encode analogues of boundary sub-regions\footnote{Or if one wants to deal with data based exclusively at the conformal boundary, one can work with the {\em shadow} of an ACD \cite{future}. This is the intersection of the ACD with the conformal boundary. We will not emphasize this point of view here because the holographic data labeled by either object is the same.}. We will now present evidence that there is indeed a QES that one can associate to an ACD. The claim is that for every ACD, one can associate a minimal QES that asymptotically approaches its causal surface\footnote{There is also the homology constraint that the QES should satisfy with respect to the intersection of the ACD and $i^0$. In the following we will often suppress the qualifier ``minimal" when discussing the minimal QES. Note also, since we are working with single ACDs, causal surfaces and causal waists are the same.}. The domain of dependence of the region bounded by the minimal area QES and $i^0$ on the side of the ACD, will be called the {\em entanglement  wedge} of the ACD. We will discuss this object in more detail elsewhere \cite{future}, here we will settle for presenting some of its properties to get some intuition. 

\begin{itemize}
\item Clearly, the QES of an ACD in Minkowski (and therefore empty) space is just its causal surface. As discussed in \cite{CK} these are just hyperplanes that stretch all the way across in ``diametrically" opposite directions on the spatial slice to the celestial sphere. In 3+1 dimensions these correspond to planes on a spatial 3-dimensional space, and in 2+1 dimensions they correspond to straight lines on a spatial 2-plane. These facts should be compared to the observation that causal surfaces and extremal surfaces coincide in empty AdS for spherical subregions.
\item When there is matter in an asymptotically flat geometry, we expect that asymptotically the (quantum) extremal surfaces will still reduce to those of Minkowski space. These are the causal surfaces of Minkowski ACDs which are hyperplanes. But deep in the bulk matter will cause extremal surfaces to deviate, just like they do in AdS. Because gravity is attractive, we expect them to ``bend outward" as we will prove in this section. That gravity is attractive is realized via the Null Curvature Condition (NCC) classically, and via the Generalized Second Law (GSL) in all order perturbation theory.    
\item The QES splits the spatial slice into two disjoint pieces. This property is crucial in our ability to define a bulk entanglement entropy associated to the surface. Again this is true in AdS.
\item A related fact is that one can associate a complementary ACD to any ACD. This is the analogue of a complementary sub-region for a sub-region on the boundary of AdS. Both the ACD and its complement share the same QES. See figure \ref{com1}, \ref{com2}. 
\begin{figure}[h]\centering
\hspace{-5mm}
\includegraphics[angle=0,width=100mm]{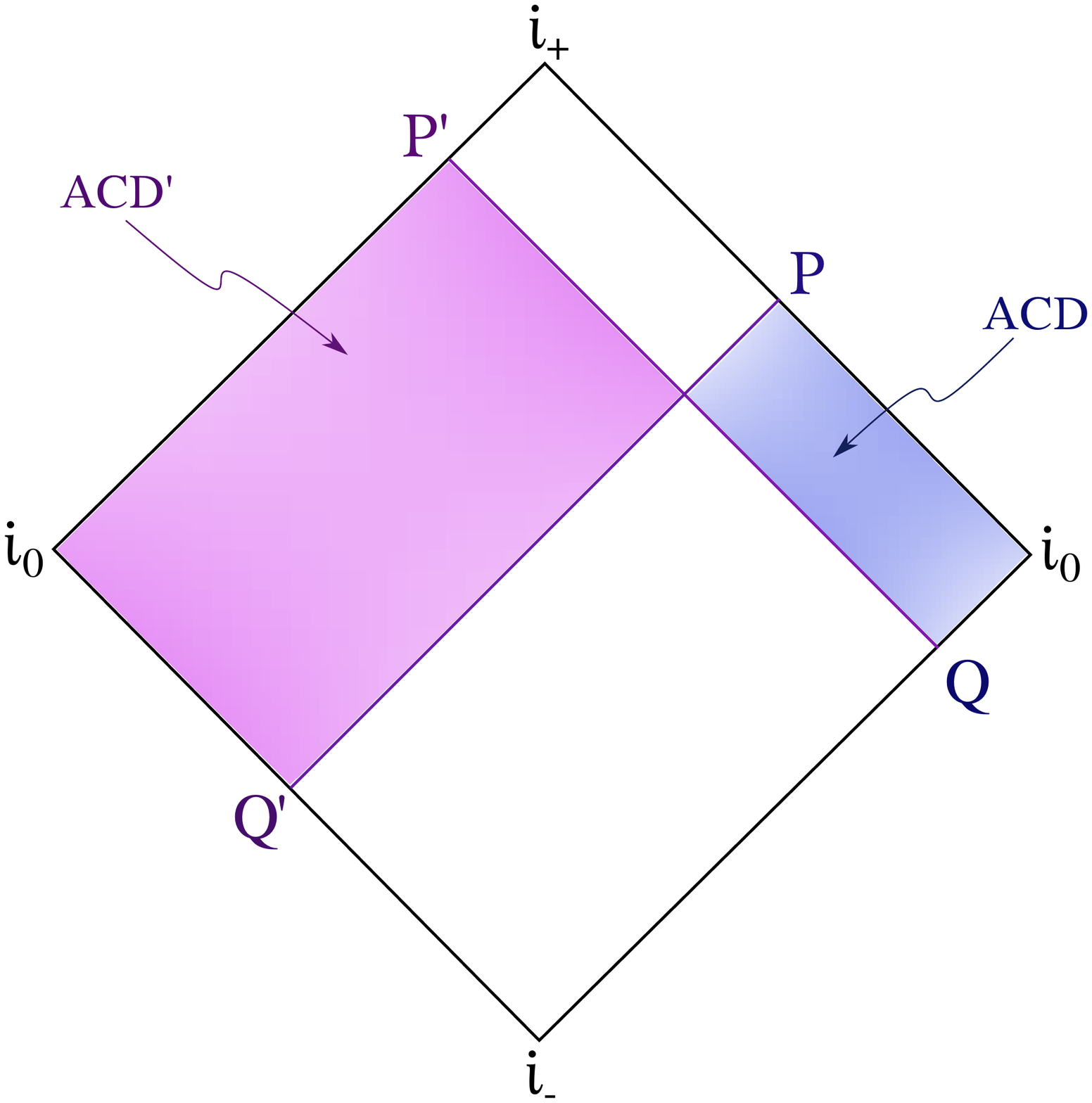} \\
\vspace*{-3cm}
\caption{Complementary ACDs, schematic side view in conformal coordinates. In general, they need not meet anywhere in the bulk, if there is matter.}
\label{com1}
\end{figure}
\begin{figure}[h]\centering
\hspace{-5mm}
\includegraphics[angle=0,width=75mm]{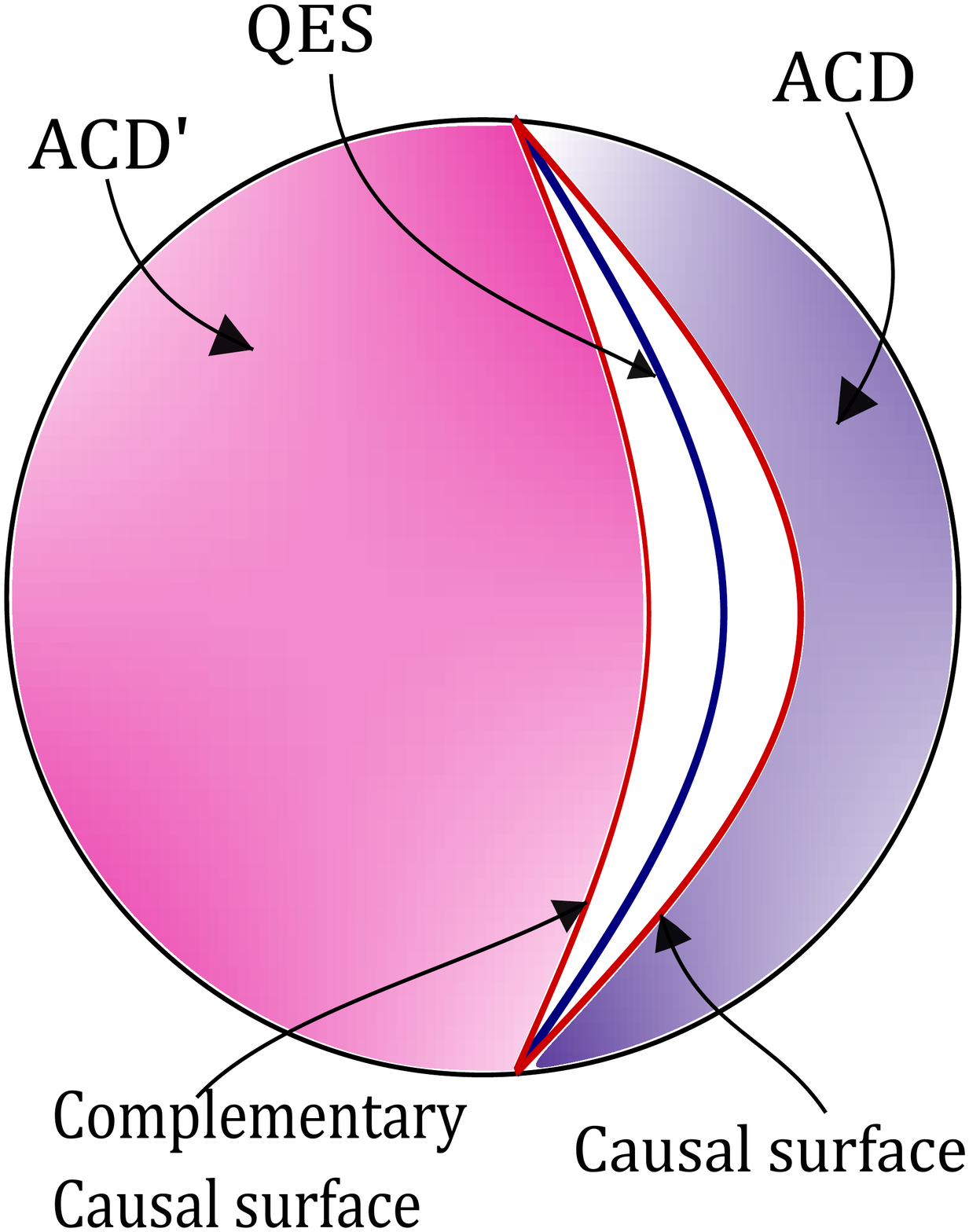} \\
\vspace*{-1cm}
\caption{Complementary ACDs, top view in conformal coordinates. Asymptotically all three curves coincide.}
\label{com2}
\end{figure}
When the spacetime is Minkowski, the QES also coincides with the causal surfaces of the two ACDs, which are generically otherwise not coincident. Again, these statements are true in AdS \cite{CHI}. The vertices of the complementary ACDs are anti-podally related. The fact that the word ``antipodal" appears in other contexts in flat space holography (eg., \cite{StromingerReview}) has not escaped us \cite{future}.  The fact that Rindler wedges come in pairs in Minkowski space, is a trivial example of ACD complementarity.

%But in AdS, more is true. General subregions formed by arbitrary unions (and not just spherical subregions) have complementary subregions in AdS because they are just complements of sets. This might seem like a tall order for ACDs, but remarkably it is not. The complement of a union is the intersection of the complements, and as can be checked, this is in fact true for ACDs as well. 

\item When there is matter, the QES lies deeper into the bulk than the causal surface and is spacelike separated from it. This is a suggestion that the bulk reconstructible region is larger than the ACD \cite{EngelWall} in flat space. These are properties shared by the QES in AdS as well, and this explains the name entanglement wedge (which we have adapted from AdS).

We will prove  this below as an illustration of the methods involved. This is a variation of some of the classical results proved in an earlier section, but the tool for the proof is the Generalized Second Law, and not Null Curvature Condition (cf., Introduction and \cite{EngelWall}).  
\\
% of an allowed subregion $\mathcal{A}$
{\bf Theorem 4.1:}
The Quantum Extremal Surface $\chi_q$ associated with an ACD $\mathfrak{C}(p,q)$ cannot intersect the ACD. and is spacelike separated from it's causal waist $W(p,q)$. (We will assume genericity in this proof, see eg. \cite{Wall, EngelWall}. The proof for the non-generic case follows straightforwardly as in \cite{EngelWall}.).
\begin{proof}
One can shoot null congruences from $\chi_q$ in four normal directions -- future outward, past outward (away from the ACD), future inward and past inward (towards the ACD). The spacetime region bounded by the null congruences in the future outward (inward) and past outward (inward) directions will be referred to as the exterior ${\rm Ext}(\chi_q)$ (interior ${\rm Int}(\chi_q)$). Note that our choice of exterior and interior is opposite to that of \cite{EngelWall}. We assume that $W(p,q) \cap {\rm Ext}(\chi_q) \neq \phi$ and then prove the theorem by contradiction, like in \cite{EngelWall}. We call $p$, and $q$, the vertices of the ACD.

We start by continuously shrinking the ACD $\mathfrak{C}(p,q)$ by moving the points $(p,q)$ to $(p_0,q_0)$ such that (a) the causal waist $W(p_0,q_0)$ of the new ACD $\mathfrak{C}(p_0,q_0)$ lies in ${\rm Int}(\chi_q)$ and (b) either $\partial\mathcal{I}^{-}(p_0)$ or $\partial\mathcal{I}^{+}(q_0)$ touches $\chi_q$ at points $\{y\}$. Without loss of generality, we can choose $\partial\mathcal{I}^{-}(p_0)$ to be the surface which touches $\chi_q$. At any one of the points $y_0\in\{y\}$, we can functionally differentiate $S_{gen}$ for $N(\chi_q)$ and $\partial\mathcal{I}^{-}(p_0)$ with respect to their shared normal direction $n^a$ at $y_0$. Note that $N(\chi_q)$ is the null congruence shot from $\chi_q$ in the future and past inward directions. Using Theorem 2.1 of \cite{EngelWall}, we can write
\begin{equation}
   \frac{ \delta S_{gen}(N(\chi_q)) }{\delta n^a} k^a  \geqslant \frac{\delta S_{gen}(\partial\mathcal{I}^{-}(p_0)) }{\delta n^a} k^a
\end{equation}
where $k^a$ is the future inward pointing null vector normal to $N(\chi_q)$ and $\partial\mathcal{I}^{-}(p_0)$ at the coincident point $y_0$.
From the definition of quantum extremal surfaces, we know that $\frac{\delta S_{gen}(N(\chi_q)) }{\delta n^a} k^a = 0$ and so we obtain,
\begin{equation}
   0 \geqslant \frac{\delta S_{gen}(\partial\mathcal{I}^{-}(p_0)) }{\delta n^a} k^a
\end{equation}
The Generalized Second Law (GSL) states that the generalized entropy of $\partial\mathcal{I}^{-}(p_0)$ (which is a causal horizon) is nondecreasing in time. In the generic case, we will have a strict inequality as follows.
\begin{equation}
\label{GSL}
   0 < \frac{\delta S_{gen}(\partial\mathcal{I}^{-}(p_0)) }{\delta n^a} k^a
\end{equation}
and hence we have reached a contradiction. This implies that the initial assumption was incorrect and hence $W(p,q) \cap {\rm Ext}(\chi_q) = \phi$. Moreover due to the strict inequality of \eqref{GSL}, $\chi_q$ does not lie on $\partial\mathcal{I}^{-}(p_0)$ or $\partial\mathcal{I}^{+}(q_0)$ or both (in which case, it coincides with $W(p_0,q_0)$) and thus $\chi_q$ never intersects the ACD $\mathfrak{C}(p,q)$.
\end{proof}

\item Given the previous bullet point, one could worry that the following (superficially) problematic situation could arise, when there is a cut-off/holographic screen. Let us work with 2+1 dimensions for simplicity. Consider the QES corresponding to an ACD, and consider a (different) ACD\footnote{Assuming it exists.} passing through the points where the QES cuts the screen, figure \ref{CONFIGflat}. Outside the cut-off, the extremal surface looks like it is {\em inside} another causal surface, which might seem worrisome.

\begin{figure}[h]\centering
\hspace{-5mm}
\includegraphics[angle=0,width=70mm]{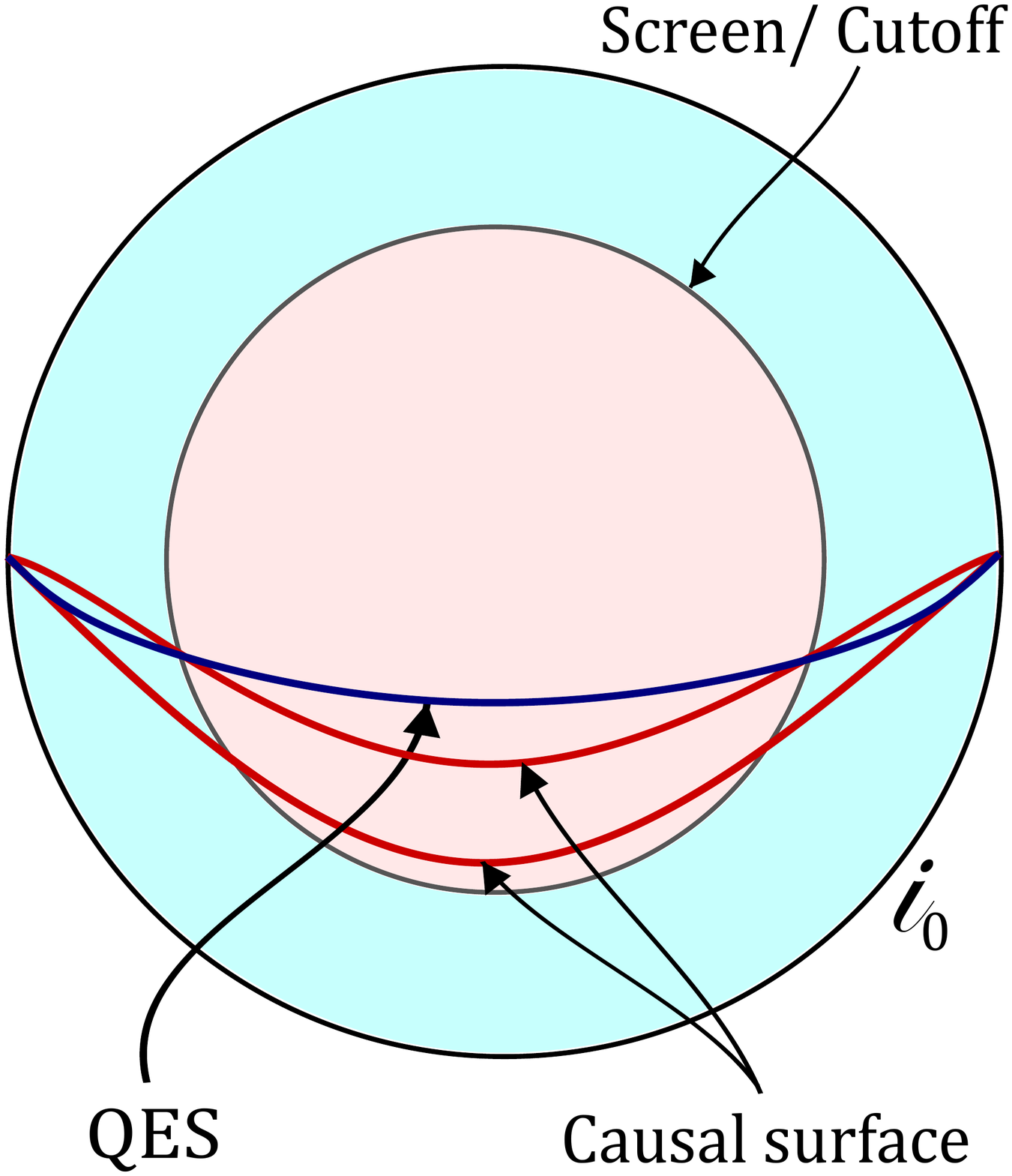} \\
\vspace*{-1cm}
\caption{To compare the relative positions of causal surfaces and extremal surfaces meaningfully, we need to make sure that they are anchored at the spatial boundary. Note that the phenomenon we see here has a precise analogue in AdS as well, next figure.}
\label{CONFIGflat}
\end{figure}
In fact, there is no contradiction: we will leave it as an exercise for the reader to figure out why that is the case. But to emphasize that this is not restricted to flat space, let us draw  what the analogous picture looks like in AdS. See figure \ref{CONFIGAdS}. 
\begin{figure}[h]\centering
\hspace{-5mm}
\includegraphics[angle=0,width=100mm]{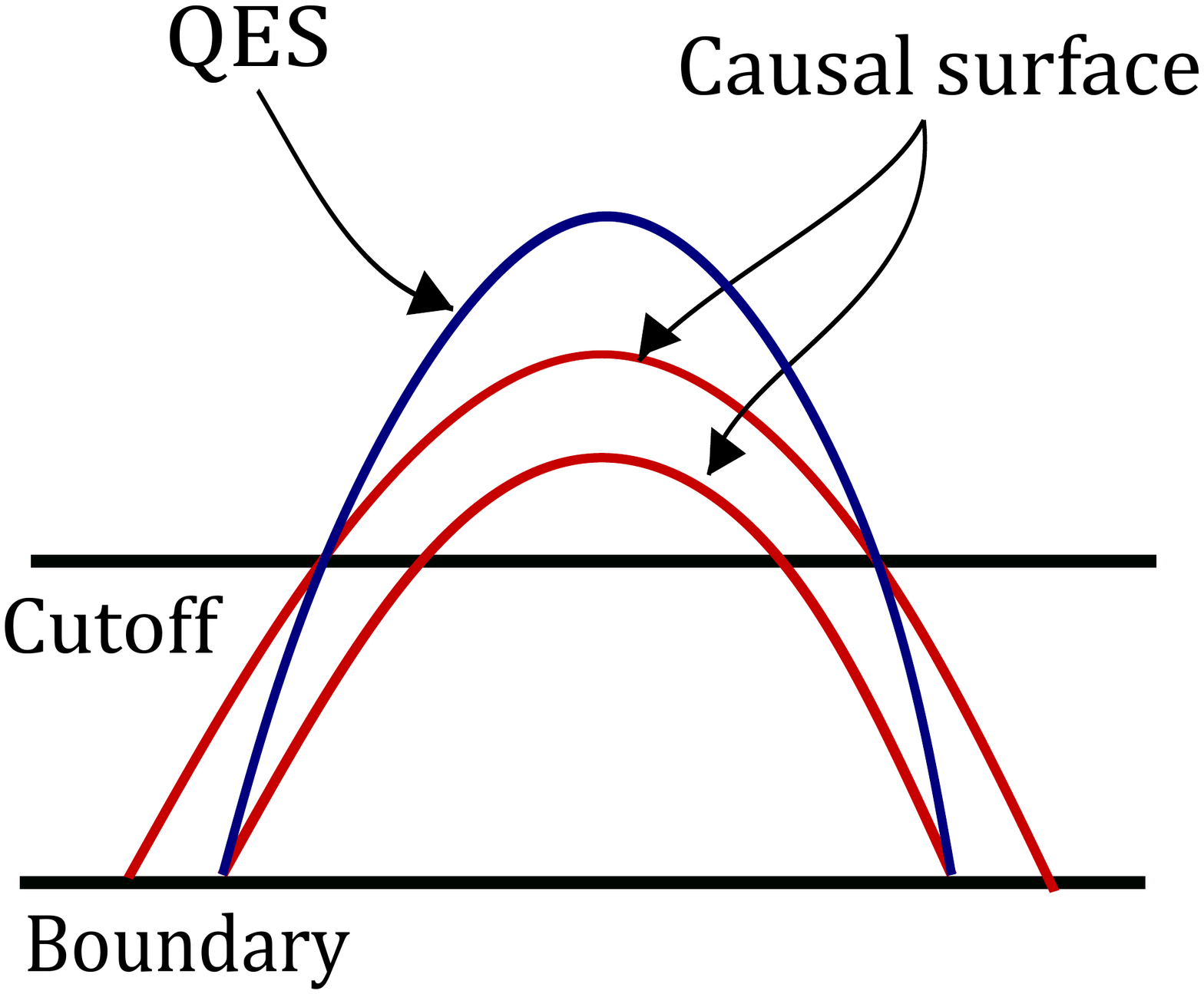} \\
\vspace*{-3cm}
\caption{Even in AdS (parts of) extremal surfaces can be inside causal surfaces, if their anchoring surfaces are at finite cut-off. }
\label{CONFIGAdS}
\end{figure}
%Note also that the temptation to conflate the causal surfaces of non-trivial spacetimes with those of Minkowski space must be resisted, when interpreting these figures.

Hints/answer: The cut-off is crucial to create this seeming problem.   When one considers the full entanglement wedge, causal surfaces are always inside their respective entanglement wedges.  Points that are shown to intersect in the figures \ref{CONFIGflat} and \ref{CONFIGAdS}, need not always. The spacetime is only {\em asymptotically} flat. Thinking in higher dimensions is instructive. %There is no contradiction if one only considers anchoring at the conformal boundary.  
\item Note that the statements we have made above can also be phrased at a classical level by defining a classical minimal area extremal surface, which will be the analogue of an HRT surface. This is also in perfect analogy with AdS, and indeed was the historical prequel to the QES and entanglement wedge in that context. 
\item Just like in AdS, the QES in flat space suffers from IR divergences. We will see that a natural prescription for regulating these divergences presents itself, and it lets us tie together Formulations 1 and 2 in a very intuitive way. 
\end{itemize}

The parallels with AdS are striking, so it might give the impression that what has happened here is a trivial adaptation of the usual AdS lore. So let us emphasize the one place where flat space holography is deeply different from that in AdS. This has to do with the fact that there is no obvious analogue of a sub-region at $i^0$. This is physical: a fixed angular difference corresponds to an infinite physical distance on the celestial sphere, but more importantly, extremal surfaces of ACDs asymptotically reach diametrically opposite points. Note that this is distinct from AdS, where extremal surfaces can be anchored to arbitrarily small spherical boundary sub-regions. So unlike in AdS, sub-regions on $i^0$ are not very useful here because the latter are most useful when one can associate a local theory to them. One manifestation of this is the fact that $i^0$ is a point after compactification. A second manifestation (which we will not delve into here) is that HRT/QES surfaces corresponding to unions of ACDs that are not proper subsets of each other will contain the entire spacetime in their entanglement wedges. This is a clear indication in our view that the hologram of flat space is a non-local theory. A related fact is that it is only when we introduce a cut-off (like in the previous section) that we see strong sub-additivity and related features one expects from sub-regions. In the concluding section, we will give a brief overview of the emergent picture, but we will leave a more detailed discussion for later \cite{future}.

%Assigning areas  to these extremal surfaces makes most sense when there is a holographic screen. 

\subsection{Entropy of a Subregion, with Sink}

In discussing flat space Page curve, we noted two formulations in a previous subsection. In Formulation 1 the interior of the holographic screen/cut-off is treated as an approximate definition of the relevant black hole system (and the region outside is a sink to which Hawking radiation leaks off). In Formulation 2, we treat the entire Penrose diagram as the definition of the quantum gravity, and extract radiation out from the system by coupling it to an external sink.  The entanglement wedge as we have defined above is most natural in the second formulation. To discuss the first formulation, we will also need an approximate definition of entanglement wedges for the latter. We expect this definition to be consistent with the approximate tensor factorization \eqref{1vs2}:
\bea
{\cal H}_{QG} \approx {\cal H}_{QG}^{\mathfrak{S}} \otimes {\cal H}_{out}. \label{factor}
\eea
Here the left hand side stands for the full quantum gravity Hilbert space as defined by the asymptotic boundary, the first term on the right hand side is the approximate description of quantum gravity within the cut-off/screen,  and the third term is the exterior -- to be interpreted  as the sink. 

It is a non-trivial fact about the holographic/gravitational nature of the first sub-factor above, that sub-regions defined on it satisfy strong sub-additivity via their extremal surfaces. Indeed we have seen in an earlier section that the area of a suitable extremal surface anchored to a sub-region of a (large) holographic screen had this property. This should have a natural connection with sub-region entropy \cite{RT}, but what precise quantity is it computing? Note that the situation here is different from that in AdS, because now there is no decoupling at the screen. As suggested by Formulation 1, we should view the exterior region as a sink, an extra tensor factor. In the discussion about the fine-grained and coarse-grained Page curves earlier, we noted that one can in principle define entanglement entropies of sub-regions when the system is coupled to a sink (see footnote \ref{thermal-subregion}). In flat space, with the above tensor factorization, we see that this is precisely the situation we have. It is eminently plausible therefore that the area of the extremal surface anchored to a sub-region has something to do with the entanglement entropy of the sub-region where the reduced density matrix is computed by tracing over both the rest of the screen, as well as the sink. In other words, the object we are trying to calculate is
\bea
S(R_0) \equiv {\rm Tr}_{sink, \tilde R_0}\ |\Psi\rangle \langle \Psi| 
\eea
where $R_0$ stands for a sub-region on the screen, $\tilde R_0$ its complement on the screen. The state $|\Psi\rangle$ is a state in the full Hilbert space ${\cal H}_{QG}$ in \eqref{factor}. We will be interested in evaluating this quantity when the state $|\Psi \rangle$  has a semi-classical bulk EFT description. A natural proposal for this entanglement entropy in bulk EFT is 
\bea
S(R_0) = {\rm min}\left\{ {\rm ext}\left(\frac{A[\chi]}{4G}+S_{bulk}[R_0 \cup \chi] \right)\right\}  \label{screenF}
\eea 
where $\chi$ is a bulk surface on the interior of the cut-off, anchored to $\partial R_0$ and homologous to $ R_0$ on the screen. Essentially, the instruction is that we should minimize the sum of the bulk entanglement entropy of quantum fields inside the region surrounded by $R_0$ and the bulk surface $\chi$ (across $R_0 \cup \chi$), plus the area of $\chi$ divided by $4G$. Let us emphasize also that while the sink is explicitly present in the definition of the left hand side, on the right, it simply shows up as a part of the spacetime in the bulk EFT description. Note that if the spacetime ended at the screen, as in the reflecting boundary situation at the AdS boundary, there is no entanglement across $R_0$ and we end up with the usual generalized entropy formula. One can formally think of \eqref{screenF} as being equal to 
\bea
{\rm min}\left\{ {\rm ext}\left(S_{gen}[\chi]+S_{bulk}[R_0] \right)\right\}. 
\eea
Let us emphasize however that bulk entanglement entropy is defined for closed bulk regions, so this is just a mnemonic. But it does help us see that the new contribution is\footnote{Note: This is a bulk calculation, so the area is of the sub-region, not the boundary of the sub-region. } 
\bea
\sim S_{bulk}[R_0] \sim \frac{{\rm Area}(R_0)}{\epsilon_{UV}^{d-1}}.  \label{AreaScreen}
\eea
As familiar from field theory calculations of entanglement entropy \cite{CasiniHuerta} this term comes with a UV cut-off dependence. Since the bulk EFT cut-off is hierarchically smaller than the Planck scale, this ensures that the term is sub-leading.  This also gives us a natural understanding why our classical extremal surfaces anchored to the screen satisfy strong sub-additivity: they are to be viewed as the leading classical approximations to \eqref{screenF}. Let us note also that the bulk area law \eqref{AreaScreen} above by itself  satisfies strong sub-additivity\footnote{This is trivial, but unlike in the conventional case, one of the ``sub-regions" can also be the sink. So we present the quick check in an Appendix.}, so it is natural that the full quantum object also does.

Note that \eqref{screenF} is perfectly well-defined and finite as it stands\footnote{With a finite hierarchically smaller scale than Planck scale, as the UV cut-off for EFT.}, and it makes sense in Formulation 1. The  definition also gives us a plausible candidate for an entanglement wedge anchored to the cut-off. This will simply be the domain of dependence of the region bounded by the sub-region $R_0$ and its associated (quantum) extremal surface $\chi$ obtained from \eqref{screenF}. When we discuss the information paradox,  we will be interested in the entanglement entropy of the entire screen slice. In that context, at the beginning of Hawking evaporation, instead of zero we will have the area of the screen (cf., \eqref{AreaScreen}) as a hierarchically small piece in the entanglement entropy. This can be viewed as a  shift in the datum of the Page curve, and this is again entirely physical: the inside-the-screen region is not isolated. Finally, let us also point out that in the context of Hawking evaporation, we expect $S_{bulk}[R_0]$ to start out hierarchically small like we mentioned above, but to steadily increase  until the Page time when it equals the Bekenstein-Hawking entropy. This is because the semi-classical state is changing: it is steadily building up entanglement entropy due to the fact that outgoing modes are exiting the screen while the ingoing modes (that fall into the horizon) are not, and the two are entangled. After the Page time, due to the phase transition, the sink and the region inside the new non-trivial QES are both in the entanglement wedge of the sink - so this term drops down again to hierarchically small values. 

The fact that this type of a prescription is possible should be viewed as evidence that there is an approximately holographic description of the interior on the screen. We do not know explicitly what the dual theory on the screen is, but it is ``sufficiently local" that the screen sub-regions satisfy strong sub-additivity. We say that this holographic description is approximate for a few different, but related, reasons. Firstly, there is the fact that we are working within a cut-off of radius $R$, and therefore \eqref{large} suggests that this is not a complete description of all the states possible in an asymptotically flat quantum gravity theory (eg., very large black holes of the order of the size of the cut-off are not part of the description). A related fact is that gravity is weakly dynamical at the screen. A second point is that the metric is only asymptotically flat. As we noted earlier, this means that sub-regions cut out by ACDs are only approximately the same as the (naive) sub-regions on the $t=0$ slice of the screen. Another related fact is that the holographic system is now attached to a sink.

%The entanglement wedges defined with respect to screen sub-regions are complementary\footnote{This statement has some exceptions: an obvious example, is after the Page time of an evaporating black hole. What we mean is that the bulk entanglement entropy contribution across the screen is subleading in $G$, so it does not affect entanglement wedge complementarity of interior sub-regions for many interesting states, even though the system is not in a pure state.}, but the SAS of the complement of an ACD is only approximately complementary to the SAS of the ACD. 

\subsubsection{Entropies for ACDs: Renormalization}

How do we connect the above screen dependent picture to the entanglement wedge of the full quantum gravity defined with respect to the asymptotic boundary? In particular, in order to formulate the entanglement wedge phase transition argument from the conformal boundary (formulation 2), we need to associate entanglement entropies to (shadows of) ACDs. These are the objects that substitute boundary sub-regions at the conformal boundary.

One option is to try to view the screen dependent approach as a regulator for the full quantum gravity. We can consider the SAS of an ACD at some finite cut-off, and evaluate \eqref{screenF} for its approximate sub-region. This will be the regulated version of the entropy of the ACD. If we take the cut-off blindly to infinity, it is immediately clear that the screen dependent contributions will have power law divergences in the screen radius, and therefore some form of regulation is needed. What is a natural prescription for renormalization? This question might seem at first glance hopeless because analogous calculations in AdS usually have a flavor of holographic renormalization, and in flat space these ideas are undeveloped. But all is not lost -- long before the advent of holographic renormalization, Gibbons and Hawking \cite{GH} managed to do a similar calculation by doing a background subtraction to get a sensible result. They were trying to compute the action, and we are trying to compute entanglement entropies, but the philosophy is similar. We will compute the same entropy \eqref{screenF}, but computed in Minkowski space $S_0[R_0]$ with the same screen sub-region\footnote{Note that the screen lives in an asymptotically flat chart, so the coordinates are meaningful also in Minkowski space. Note however that for an asymptotically flat spacetime, neither the approximate sub-region nor the sub-region induced on Minkowski space this way need be exactly spherical at finite cut-off.}, and subtract that from our definition of $S[R_0]$. This gives a quantity that we  now expect to be independent of the cut-off:
\bea
S^{ren}[R_0]=S[R_0] - S_0[R_0]. \label{sub}
\eea
To give an example, the HRT surface in the Minkowski 3+1 case for a screen sub-region bounded by a circle is a planar region in the bulk bounded by the same circle \cite{CK}. So the subtracted quantity $S_0[R_0]$ in this case is simply the sum of the area of the plane within the circle and the area of the spherical screen region bounded by the circle. The resulting quantity now has a chance of being finite\footnote{Without a more detailed analysis, we cannot be certain that the subleading divergences are also removed by this prescription. But note that parallel quantities are perfectly well-defined in AdS, see eg., \cite{CHI, Boo}.} as we take the screen size to infinity. 

Let us observe a few things about this prescription. Firstly, note that one of the things it does is to include the bulk entanglement entropy from the sink. This should be clear for the $S_{bulk}[R_0]$ piece: subtracting it out means that now we are including the fields on the sink side as well, thereby reducing the entanglement.  This is a natural thing to do, because we should include the part of the entanglement wedge that is outside the cut-off, when we are trying to work with constructs that are natural from the perspective of the conformal boundary. Similarly, for the contribution from the extremal surface, subtracting the Minkowski background is tantamount to keeping track of the entanglement entropy that goes over and beyond that of the ground state. It automatically means that for Minkowski space, the entanglement entropies of ACDs one defines this way are identically zero. Since there is no useful spacelike sub-region on $i^0$, it is not entirely clear to us whether this is a feature of the ground state of flat space holography, or an artefact of the renormalization prescription\footnote{But note that background subtraction in AdS also has the same feature that the regulated ground state does not have entanglement, see, eg. \cite{CHI,Boo}. It may be possible to treat the contribution from the extremal surface using a different regulator so that one gets a non-vanishing result for the entropy.  %But entanglement entropies are defined for tensor factors/sub-regions, so subtracting the ``bulk entanglement entropy across $R_0$ in Minkowski space" is not a very operationally meaningful statement. 
We suspect that a detailed study of empty AdS with a cut-off could be instructive here for guessing the most natural prescription. In particular, there may exist specific subclasses of asymptotically flat geometries which are better suited for background subtraction than the most general ones (see section 3.3 of \cite{Sorce} for a related discussion in AdS). We will leave  a detailed investigation into alternate regulators for future work.}. Note in particular that for excited states, we have non-zero entanglement entropies, and therefore we can still formulate the entanglement wedge phase transition argument. In any event, the punchline of these discussions is that there exists plausible ways of defining entanglement wedges and entanglement entropies to both sub-regions on the screen as well as to ACDs anchored to the asymptotic boundary. 

%As we discussed in the previous section, the area of a codimension-2 surface anchored to a SAS or to a $t=0$ screen subregion is a well-defined idea in a classical spacetime for a sufficiently large cut-off. Therefore we phrased the discussion there largely in terms of surfaces that were defined inside the cut-off surface. But to see the phase transition at the Page time, we need to go beyond classical extremal surfaces, and the way to do this is to work with the generalized entropy instead of the area of a surface. This object involves a bulk entanglement piece, and is therefore most naturally defined \cite{EngelWall} across codimension two surfaces that separate a bulk spatial (Cauchy) slice into two disconnected pieces (sometimes somewhat misleadingly called the interior and exterior). In particular, if the surface ends abruptly at a bulk point, defining a bulk antanglement entropy {\em across} it becomes ambiguous. That is, while the generalized entropy $S_{gen}[\chi]$  across a surface $\chi$ that separates the Cauchy slice into dissconnected regions can be meaningful (once IR divergences -if any- are appropriately regulated), it is unclear how one might go about defining this quantity for a surface $\chi$ that ends on a bulk point.  This means that we need a way to modify the discussion we had in the previous section when were dealing with classical entropy (aka area) of screen-anchored surfaces. Moreover, there should be a natural way for this modification to connect with the previous classical discussion. 

Let us make a few comments about the connection between extremality on the screen vs that on the boundary. %One can extend all surfaces $\chi$  anchored on a SAS (and to be varied over), along the waist of the causal diamond beyond the cut-off. See figure \ref{SCS}. This means that the extremal surfaces that we define will now only be required to be stationary under variations of the surface inside the cut-off. 
It should be clear that for large cut-offs this will approach the exact quantum extremal surface defined directly with respect to the ACD, ie., we expect
\bea
\lim_{R \rightarrow \infty} \frac{S^{ren}[R_0]-S_{ACD}[\chi_0]} {S_{ACD}[\chi_0]} = 0,
\eea
where $R$ is the cut-off size and $\chi_{0}$ stands for the true QES.

%\begin{figure}[H]\centering	\hspace{-5mm}	\includegraphics[angle=0,width=80mm]{drawing11.pdf} 	\vspace*{-1cm}	\caption{The dashed red curve bounded by $\partial A$ and its blue continuation to $i^0$ denote the location of the causal surface. $\partial A$ stands for the SAS. The curvy blue line is the surface to be varied. We only consider variations within the cut-off, when defining extremality for surfaces anchored on the cut-off. For large cut-offs, outside the cut-off the extremal surface approximately coincides with the causal surface.}\label{SCS} \end{figure}

%The above discussion gives us natural ways to define extremal surfaces that split a Cauchy slice into two pieces, while being anchored in an appropriate sense to the SAS and to the conformal boundary. Together with the fact that SAS can approximate subregions on the screen, this gives us a way to relate true entanglement wedges to screen-based entanglement wedges. defined in these two ways. To complete the discussion however, we will also need to address the fact that the surfaces that we are considering above are infinite, and therefore $S_{gen}$ suffers from IR divergences. 

The discussions above make it plausible that one can associate entropies to ACDs by suitable regularization and renormalization of quantities at the screen. However fully watertight discussion of some of these issues is unavailable even in AdS, so we will not aim to resolve them here \cite{future}. In \eqref{sub} we simply subtracted the Minkowski contributions, but it would be interesting to see if  a subtraction where answers for Minkowski ACDs are finite and non-zero, is possible. It will also be nice to prove that the subtraction/renormalization guarantees the positivity of the entropy. For the Page curve purposes of this paper however, we will not worry about these issues too much, and simply assume that such a prescription can be found. The key point that a QES without boundary anchors has finite entropy even without renormalization is the fact we will use, for Formulation 2. Let us emphasize also that for Formulation 1, the quantities are manifestly finite, so there is no subtleties of renormalization at all. 

Finally, to whet the reader's appetite, let us say a word about how to interpret the entanglement entropy of an ACD. The complement of an ACD is a perfectly well-defined quantity. This means that the entropy of an ACD is exactly analogous to the entropy of a spherical boundary sub-region in AdS/CFT: one should  think of the entropy as a result of tracing over the complement. We will observe here that we are associating an entanglement entropy to a covariant object - note that typically one associates entanglement entropies to tensor factors of a Hilbert space. Of course, if the tensor factor is a sub-region in a quantum field theory, we can trivially also associate entropy to the domain of dependence of that sub-region. Here on the other hand, we only have direct access to the covariant object \cite{future}. Let us also repeat one fact we briefly alluded to before, because it may be a bit surprising -- the entanglement wedges of a union of ACDs where one is not a proper subset of the other, is the entire spacetime. This is a demonstration of non-locality. What is surprising about flat space is not that it is non-local, but that it still allows a remarkable object called the ACD that has many features of a spherical boundary sub-region. A related fact is that when we described regulation/renormalization in this section, we only dealt with single ACDs and their entropies. This is because the HRT surfaces of unions of (non-nested, non-complementary) SAS sub-regions go off to $i^0$ as the cut-off is taken to infinity. This is because of the above fact -- entanglement wedges of non-trivial unions of ACDs contain the entire spacetime. 

%Let us also note a couple of words about unions of ACDs: it is possible to argue that the entanglement wedge associated to the union of two ACDs is the entire  

%\footnote{A detailed study of how to regulate boundary anchored extremal surfaces seems clearly necessary in flat space.}. We can consider the SAS of an ACD\footnote{We will only discuss this calculation for a single ACD here, this is the analogue of a spherical sub-region in AdS. Just as in AdS for general sub-regions, we expect that modulo technical complications, this can be generalized to unions of ACDs/SAS's. } 

\subsection{The Flat Space Entanglement Wedge Phase Transition}

The entirety of the spatial slice of the AdS boundary where the CFT lives, played a crucial role in the discussions of \cite{Penington, Almheiri}. This is a proxy for the ``the system", or the Hilbert space of AdS quantum gravity. Whether the black hole interior is in its entanglement wedge after the Page time was the key question. In order to formulate information paradox in flat space, we need the equivalent object in flat space. 

We wish to consider the black hole at $t=0$ in some asymptotically flat chart and also at a later epoch $t=t_e$ of Hawking radiation extraction (which is after the Page time). In principle, we could consider the spatial slice on the screen cut out by the constant-$t$ slices, and view them as the analogue of the AdS boundary. This is natural for Formulation 1, and is indeed what we will do in that context. But for Formulation 2, being more intrinsically tied to the conformal boundary, this is not entirely satisfactory.  So how should we go about this? One may think that the entanglement wedge of the full system is easiest defined via ACDs whose vertices are at future and past timelike infinity. This is certainly true in Minkowski space. It is conceptually reasonable more generally as well, because we expect even black hole spacetimes (after evaporation is complete) to have the Penrose diagram of Minkowski space. But it leads to awkwardness because we are interested in using more conventional black hole Penrose diagrams to locate and understand entanglement wedges (as in \cite{Penington}). The trouble is that in such Penrose diagrams, symmetric\footnote{Remember that symmetric ACDs are the natural building blocks of the notion of a spherical sub-region at a specific time slice \cite{CK}.} ACDs anchored at future/past timelike infinity need not cover the ``entire spacetime". 
  %The reason is that on-screen sub-regions are an approximate idea: entanglement wedges in Formulation 2 are more directly determined in terms of the ACD data on the conformal boundary.
\begin{figure}[h]\centering
\hspace{-5mm}
\includegraphics[angle=0,width=100mm]{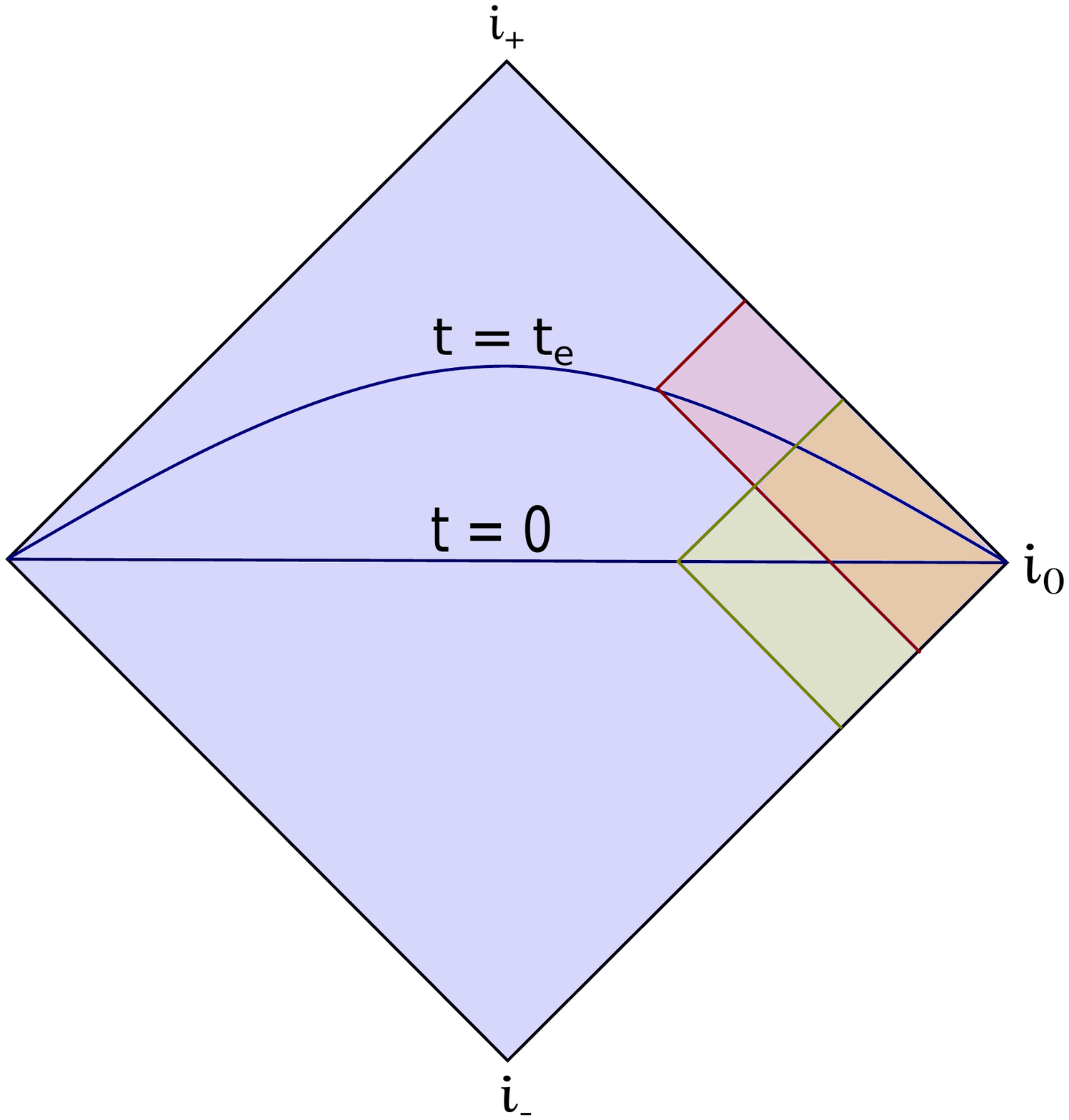} \\
\vspace*{-3cm}
\caption{Symmetric ACDs can be defined at different times. Both ACDs in the picture here can be symmetric, but the associated times are different.}
\label{drawing-1}
\end{figure}
A simple way past this problem is to remember that vertices of an ACD define not one, but two ACDs. The obvious one, and its complement. This means that one can use the complement of a trivial (ie., vanishing) symmetric ACD to capture the analogue of the entirety of the theory Hilbert space. By doing this, we have bypassed the reliance on the future and past timelike infinity. Note further that precisely because we are using symmetric ACDs, this object knows about the asymptotic time, it is exactly analogous to the AdS boundary at various times. This is because symmetry of the ACD is defined with respect to the time slice in question. A symmetric ACD at $t=t_e$ will look skewed in a Penrose diagram where $t=0$ is the diagonal Cauchy slice, see Figure \ref{drawing-1}.

With all this scaffolding in place, it is easy to see that before the Page time, the minimal QES is trivial. This is intuitively obvious because to exclude the interior modes and thereby get rid of the $S_{bulk}$ coming from their entanglement with the sink modes, the QES will have to be around horizon size. But before the Page time, $A[\sim{\rm horizon}]/4G$ is greater than $S_{bulk}$, so it is clear that this trade-off is not worth it for the minimal QES. A more precise version of this argument can also be made using classical maximin reasoning parallel to the AdS case.

\begin{figure}[h]\centering
\hspace{-5mm}
\includegraphics[angle=0,width=100mm]{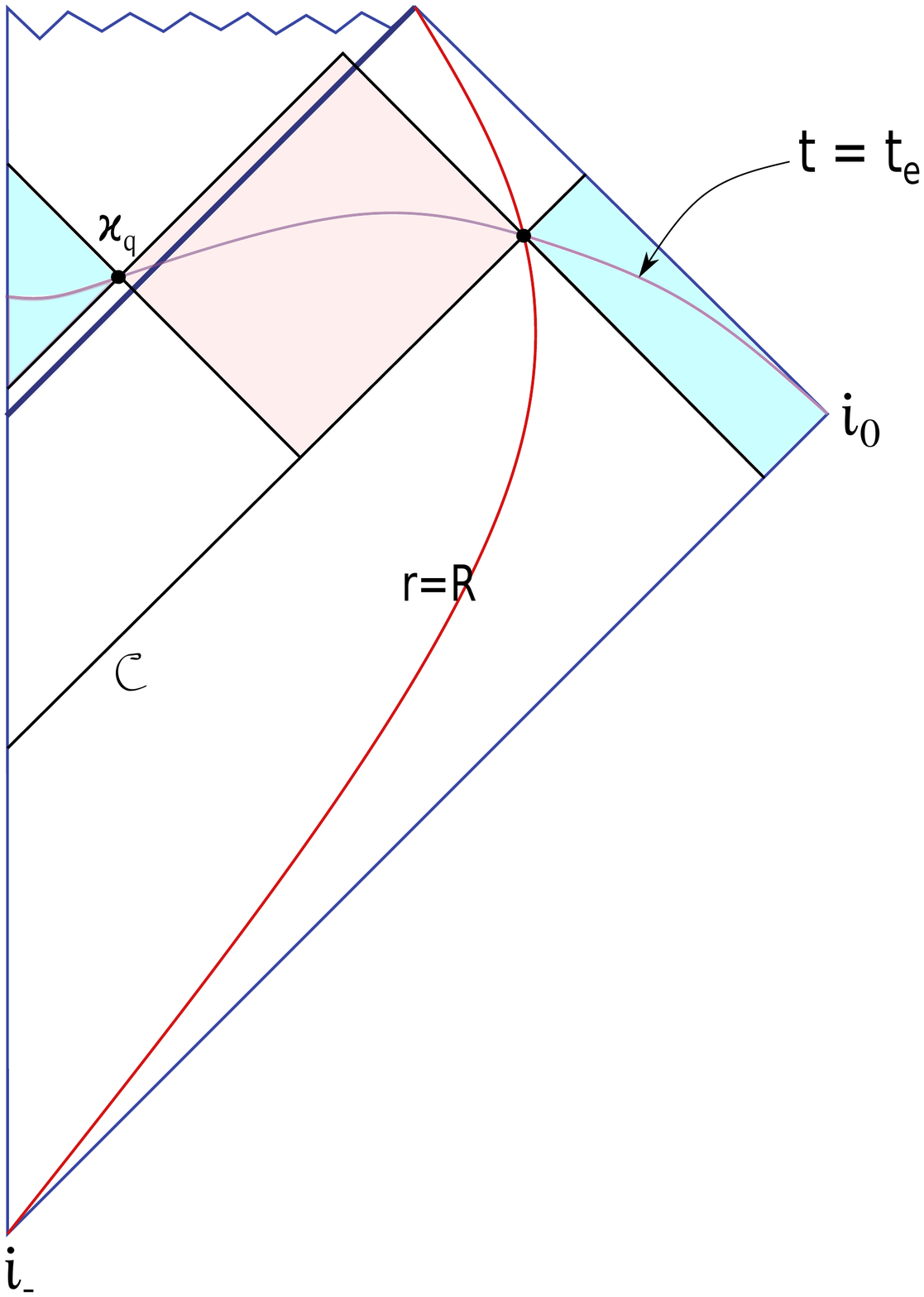} \\
\vspace*{-1cm}
\caption{The (relative) entanglement wedges in Formulation 1. The screen wedges are pink, the radiation wedges are blue, demonstrating the island.}
\label{drawing-2}
\end{figure}

\begin{figure}[h]\centering
\hspace{-5mm}
\includegraphics[angle=0,width=100mm]{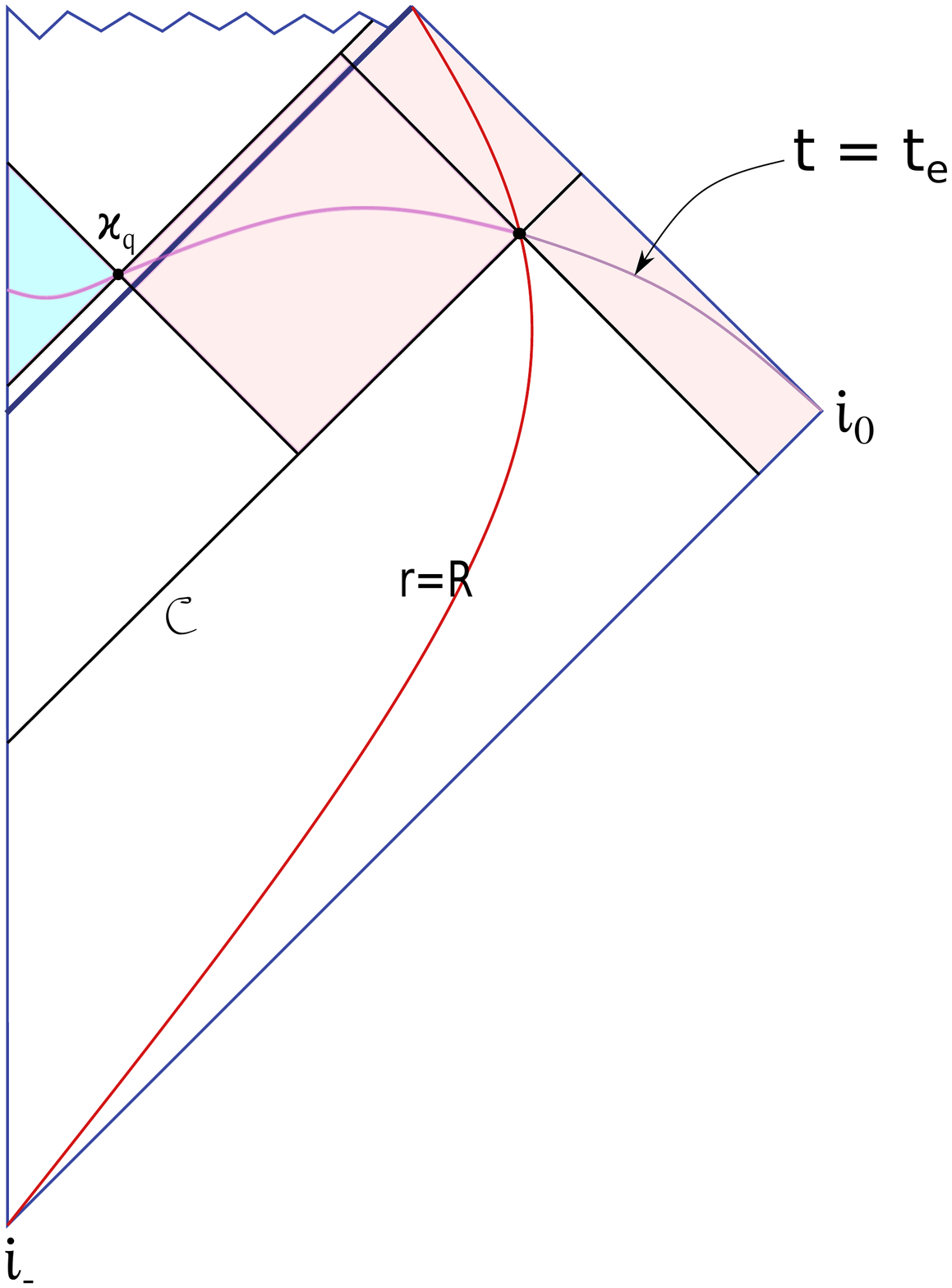} \\
\vspace*{-1cm}
\caption{The asymptotic entanglement wedge of the conformal boundary is pink (because of radiation extraction at $t=t_e$ on the screen, only part of it is accessible), the radiation wedge is the island. The rest of the radiation is in an external sink, and not shown.}
\label{drawing-3}
\end{figure} 
Let us discuss the post-Page time case in slightly more detail. To do this, it is important that we consider the two formulations we discussed in the previous subsection, separately. The first formulation is parallel to the discussions in \cite{Raghu0, Raghu, Almheiri, Thor, Iizuka, Iizuka2, Strominger}. The calculation of \cite{Penington} identifying the location of the minimal QES after the Page time goes through exactly unchanged in both formulations, because these arguments are only based on near-horizon physics and the epoch of radiation extraction\footnote{Note that since the minimal QES is near the horizon, we can use effectively two dimensional formulas, exploiting the fact that near-horizon geometry has a Rindler structure \cite{Penington}.}. We can in fact set up the calculation exactly parallel to how it is done in \cite{Penington} by first considering the classical maximin surface on the past of the lightcone of the holographic screen\footnote{Note that in both formulations, it is at the screen that the radiation goes into the sink Hilbert space, even though the details of the latter are different in the two formulations.} at $t=t_e$, so we will not repeat the details. This identical calculation has now been done enough times in the literature with minor variations. 

%This automatically explains the results of \cite{Thor, Iizuka, Iizuka2}. 

The one point perhaps worth noting here is that the epoch of radiation extraction is the time recorded on the screen in both formulations. In Formulation 1, since the sink simply amounts to free-streaming of the radiation beyond the screen in a nearly flat space, the entanglement wedge of the boundary is a bit different structurally from the one in \cite{Penington} and analogous to the one in \cite{Raghu0, Raghu}. Note that we defined the screen entanglement wedge of the within-screen-system to be the domain of dependence of the region between the minimal QES and the $t=t_e$ slice of the screen. This construction of the sub-region entanglement wedges for the screen was motivated earlier. 

In Formulation 2, radiation going into the sink results in Hawking radiation being extracted out of the Penrose diagram (or equivalently, quantum gravity Hilbert space). This means that one has to remove the past light cone (of the present epoch on the screen) from the entanglement wedge of ${\cal H}_{QG}$. This is because of the absorbing boundary conditions at the sink -- past evolution is not deterministic because the radiation is lost into the sink. A similar point was noted in AdS as well \cite{Penington}. We have indicated the various entanglement wedges that arise in the two Formulations, in the figures. Note that the entanglement wedges of the two systems are quite different, but they both have the ``island" behind the horizon. %Formulation 2 is intrinsic to the conformal boundary while Formulation 1 is based on the cut-off surface. It will be very interesting to see the precise connection between flat space quantum gravity in these two formulations. 

How should one think of the two Formulations in relation to each other? Let us start with formulation 1. The picture here should be compared to that in \cite{Almheiri, Raghu} where the boundary of AdS was coupled to a Minkowski space, and transparent boundary conditions for fields was assumed. Transparent boundary conditions mean that there are no boundary conditions: the field is allowed to propagate however it wants, and the boundary conditions are imposed elsewhere in the geometry as necessary. In a path integral language, this means that unlike what one typically does in AdS (which is to fix the boundary values of fields) one integrates over them in the path integral. Note that since the boundary value of bulk fields are interpreted as sources, this is precisely according to our discussion about sources in section \ref{formulation}: we are integrating over sources in order to couple the system to the sink. One key difference between \cite{Almheiri,Raghu} and our Formulation 1 is that the former keep the metric fixed at the boundary, and only let the other fields fluctuate: their coupling to the sink is strictly non-gravitational. This is because they are in AdS. What we have argued in this paper in various ways is that when describing the evaporation of a black hole in flat space, we can treat gravitons also in a similar manner at the screen. Essentially, we are treating the screen as a transparent boundary, including for gravitons. 

In formulation 2 on the other hand, the picture is different. Now we are viewing the quantum gravity system of the Penrose diagram as being sourced by a codimension one source on the screen. When we extract Hawking radiation into an external sink, we are to understand that we integrate over this source (automatically, because we integrate over the sink fields). So the role of the screen is quite different in the two formulations. \begin{figure}[h]\centering
\hspace{-5mm}
\includegraphics[angle=0,width=100mm]{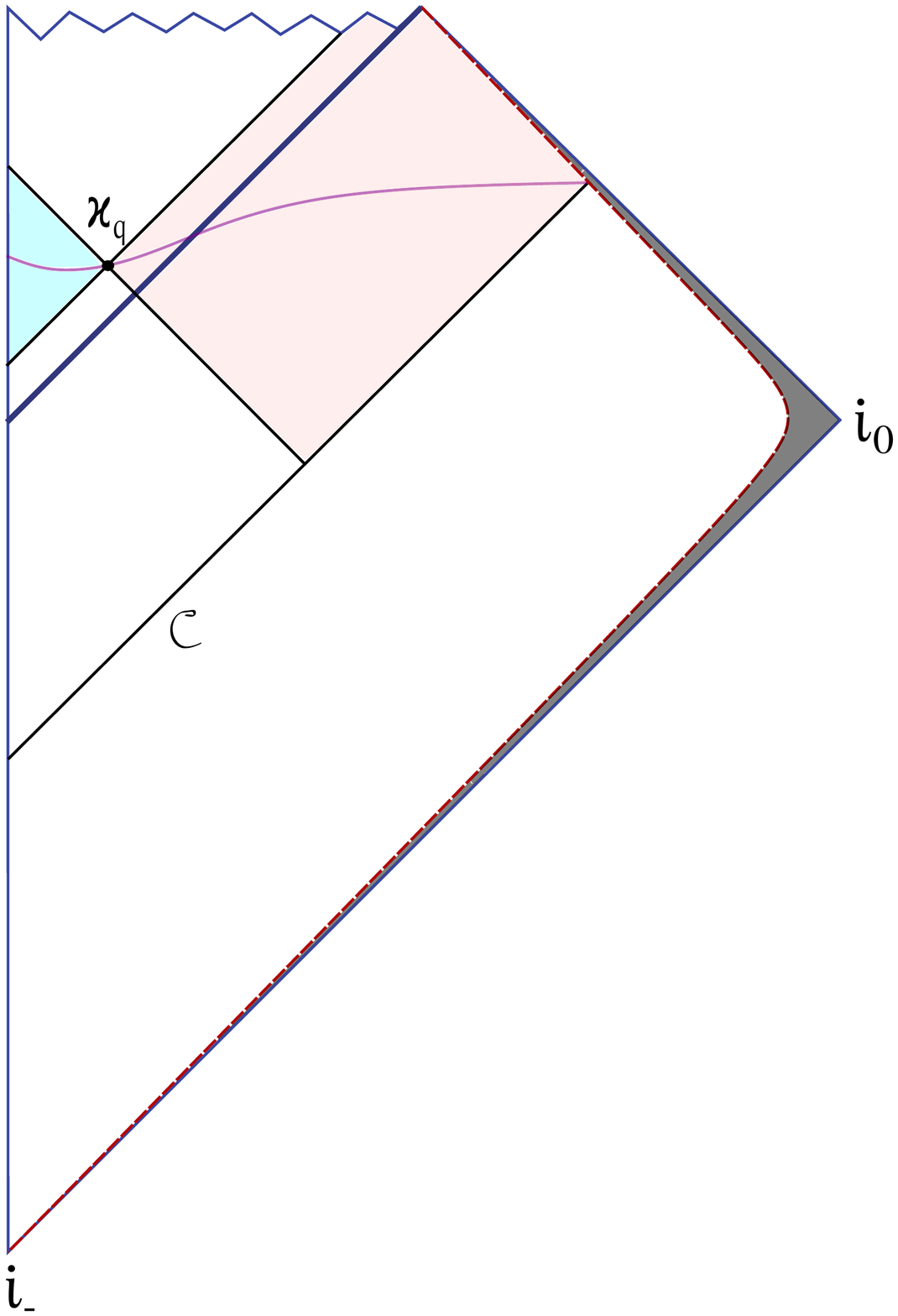} \\
\vspace*{-1cm}
\caption{Formulation of information paradox with observer at null boundary as a limit of Formulation 2 -- assuming such a null infinity limit exists (which is something we have not addressed).}
\label{nullex}
\end{figure} 
Let us also take a moment to note that the two formulations we have considered here naturally correspond to the two modes that are present in the Lorentzian version of holography discussed in \cite{Budha}: the source mode and the homogeneous mode. In Formulation 1 here, the source is set to zero, and in the second, it is turned on. Note that in the AdS case discussed in \cite{Penington, Almheiri} there was no natural formulation of the information paradox corresponding to setting the source to zero. This is because in AdS, the Page curve cannot be obtained in this set up, without coupling the system to an external sink. 

Finally, let us note that one picture for formulating the information paradox is to view future null infinity as the location of the observer, in analogy with the boundary of AdS. Heuristically, we can view this as a limit of Formulation 2 where our timelike screen has been taken to ``infinity", assuming that such a limit exists. See figure \ref{nullex}. The point is that at null boundary we will have to couple suitable boundary values of fields to external systems to take out the radiation, therefore the interpretation will be identical to Formulation 2. One may worry that since signals take infinite time to reach the boundary, observers at null infinity are ill-defined. While the precise physical interpretation of an observer at null infinity is somewhat unclear, formally this may be reasonable because we are working with fields and not particles. A good analogy to keep in mind is that in AdS, massive particles do not reach the boundary, but sources and expectation values of massive fields are well-defined at the boundary. We will conclude this section by noting two recent papers which may be relevant for the viability of null infinity-based approaches. In \cite{Laddha}, it is argued that all the information at future null infinity is actually contained near $i^0$. This can be viewed as a limit of the statement that the entire information about a spatial slice in holography is contained at the boundary of that slice -- future null infinity of the Minkowski causal diamond is a limiting Cauchy slice. A null infinity based approach has also very recently been used in \cite{Strominger} to argue for the Page curve in 1+1 dimensions.

\section{Concluding Comments}

The work of \cite{Penington, Almheiri} was set in AdS, but it was already clear that the information paradox in flat space may have a resolution along similar lines. The technical reason behind this is that the calculation of the island after the Page time \cite{Penington} depends largely only on near horizon data\footnote{That the radiation was leaving the system all through the black hole evolution is also implicitly used, but the details of this did not matter.}. The results of our paper in that regard are therefore unsurprising. See related recent work in \cite{Raghu, Thor, Iizuka, Iizuka2, Strominger} which draw similar conclusions. What we have tried to focus on instead is to identify the structures in flat space that allow us to make these expectations precise, and help us go beyond a mere analogy. To do this, we must necessarily work in higher dimensions. Perhaps surprisingly, we were able to identify compelling versions of (quantum) extremal surfaces and entanglement wedges using the asymptotic boundary of flat space. Suitable versions of these objects also exist when we introduce a holographic screen, and through the machinery of ACDs we were able to connect these two pictures. An immediate consequence of our work is that these results are {\em not} merely tied to toy models in 1+1 dimensions.

In other words, what we have done in this paper is to point out the existence of bulk entanglement objects in flat space which have well-known counterparts in AdS. While emphasizing the parallels, we have also emphasized some crucial ways in which the story is different in flat space. Among them the fact that flat space has a natural tensor factorization in many contexts \cite{CK} played an important role.  We strongly suspect that these observations have ramifications for flat space quantum gravity that go far beyond providing a connection with the AdS information paradox. We have not dwelled on this aspect in this paper, not because we think it is insignificant, but because there is too much to say \cite{future}. In the following, keeping with the title of this paper, we will discuss points related to the information paradox, but before doing so let us give a very brief summary of flat space holography as we have encountered here. 
\begin{itemize}
\item Asymptotic Causal Diamonds (or equivalently their shadows) are a potentially useful covariant characterization of the ``true" tensor factors of flat space quantum gravity. This is much like CFT sub-regions are in AdS/CFT.
\item ACDs are in numerous ways analogous to spherical boundary sub-regions in AdS/CFT, as was explored in \cite{CK} and this paper.
\item One important feature that distinguishes them from CFT sub-regions is the fact that the entanglement wedge of a union of ACDs (where one is not a subset of the other), is the entire spacetime. This is a manifestation of the non-locality of the hologram. 
\item Despite this non-locality, the entanglement entropy of an ACD is a well-defined quantity. This is because an ACD and its complement naturally cover the entire Hilbert space. This can be viewed as a type of Rindler split \cite{QFTBH} -- not just of Minkowski space, but of general asymptotically flat space, cf. figure \ref{com1}.
\item On top of the observations listed in the abstract of \cite{CK}, in this paper, we showed that one can associate a natural notion of entanglement wedge to an ACD. This is what we used to argue for the existence of the Page curve by coupling the flat space quantum gravity Hilbert space to a sink. (Formulation 2.)
\item Another useful idea that we introduced is that of the holographic screen \cite{CK}. It captures the fact that for certain classes of states, we expect that the dynamics in the asymptotic region can be described by a bulk non-gravitational EFT (which is viewed as including free gravitons) on a fixed background. 
\item An isolated Hawking-radiating Schwarzschild black hole is a system that we expect can be described in this way. Far from the black hole, during its entire evaporation process, the curvatures are low and we expect EFT to work. More detailed arguments are presented throughout the paper.
\item One can define sub-regions on the screen using ACDs. This can be done exactly, in the Minkowski case \cite{CK} and approximately, in the asymptotically flat case. Appropriately defined QES-like surfaces associated to these sub-regions, cf. \eqref{screenF}, lead to strong subadditivity. Since the gravitational dynamics in this class of states is contained within the screen, this can be viewed as a consequence of the general arguments that went into the proof in \cite{Headrick}.
\item    Many related properties including quantum error correction were noted in \cite{CK} for sub-regions on the screen, by exploiting the connection to ACDs. We believe these observations further strengthen the plausibility/utility of the screen-based tensor factorization of {\em bulk} dynamics.   
\item The sub-region entropy \eqref{screenF} and the associated new type of QES, lets us define screen-based entanglement wedges. This entanglement wedge undergoes a phase transition, as Hawking radiation leaks off into the sink Hilbert space outside the screen. This explains the Page curve. (Formulation 1.)
\end{itemize}

\subsection{When is the Information Paradox Resolved?}

A corollary of our work is that it has allowed us to fairly confidently identify the relevant tensor factorizations that may play a role in the flat space information problem. A key step in formulating the information paradox as a Page curve problem is to do this identification in a compelling way. We need to find one tensor factor to contain the black hole and the other to act as a sink. In the formulation of \cite{Penington, Almheiri} in AdS, the sink was provided by hand. 

In the flat space case, the widely held expectation that the evaporation of the black hole should result in the Page curve \cite{Page1, Page2} tacitly assumes the existence of such a tensor factorization. It should be emphasized that this assumption is motivated by local theories (eg., a burning lump of coal), and that it is heuristic in the context of quantum gravity. But in this paper, we noted how this can make sense: this is our Formulation 1, where we observed that it may be meaningful to have an approximate tensor factorization between regions within the screen and those outside. Some preliminary evidence was presented for this view in \cite{CK}, and in this paper we strengthened it further. We also extended the evidence in \cite{CK} that the regions within the cut-off had an approximate {\em local} holographic dual. We could, eg., define notions of extremal surfaces anchored to screen sub-regions which satisfy strong sub-additivity\footnote{In fact its is possible to demonstrate \cite{Vyshnav} that many other (holographic and non-holographic) entanglement entropy inequalities hold here as well.}. We gave a new definition of sub-region quantum entropy \eqref{screenF} using these ideas, which automatically incorporated the fact that the exterior of the screen has a natural interpretation as a sink. With the split into an interior and a sink, we could now formulate and demonstrate the emergence of the Page curve.

%For these claims to be sensible, a tensor factorization of the system must implicitly exist.

These observations are in direct tension with the claims of the recent paper \cite{Laddha}. The authors of \cite{Laddha} argue that the holographic nature of gravity completely invalidates the usual structure of the Page curve in both AdS and flat space, and that instead of rising and falling, the Page curve stays flat throughout\footnote{Even though we disagree with them, it should be emphasized that \cite{Laddha} is one of the few papers that actually addresses some of the key issues, and we found it thought-provoking.}. It has also been suggested that since the formation and evaporation of a small AdS black hole is dual to the formation and decay of a deconfined plasma ball, one should look to the latter process in the CFT for insight. However, we are unsatisfied with the tensor factorization discussed on p. 24 of \cite{Laddha}. This splits the CFT into two regions; one with the plasma ball and the other, its complement. Treating the local tensor factor structure of the CFT as the one relevant for the black hole Page curve, we find not-so-compelling\footnote{Note also that this is different from the tensor factorization of \cite{Penington, Almheiri} which couples the entire CFT to a sink. Let us also emphasize that a Hilbert space can have multiple (possibly approximate) factorizations.}. We feel that the {\em apparent} locality in the holographic direction must play an important role. Radiation can ``leave the black hole region" in flat space\footnote{Or small black hole in AdS.}, and where it does, curvatures can be arbitrarily small throughout the evaporation. A context where this fact does not have a satisfying understanding, cannot be the right setting for the interesting Page curve. In other words, to meaningfully understand the Page curve, we need to identify a reasonable tensor factorization in a bulk EFT compatible language, or give strong evidence that such a formulation cannot exist. The latter point of view will have the burden of finding an alternate explanation for the mundane bulk intuition that radiation can ``leave the black hole region". The question here is not really about the true fine-grained degrees of freedom, which we all agree live at infinity. 

In this paper (see also \cite{CK}), we noted various reasons to think that our Formulation 1 eqn. \eqref{1vs2}, offers  a suggestive bulk tensor factorization. We agree with \cite{Laddha} that if we view the entire flat space quantum gravity as a single tensor factor, we will not find the tent-shaped Page curve. This should be compared to our Formulation 2, where the entire theory is treated as a single tensor factor. It is clear that this is analogous to the case of AdS without a sink, and just like done by \cite{Penington, Almheiri}, we had to couple an external sink to this system in order to extract the radiation and reproduce the Page curve\footnote{It has further been suggested to us that the bulk Page curve in AdS should only be viewed as the bulk dual of a boundary plasma ball Page curve, without direct relevance to the black hole information paradox. Quite apart from the previous discussions, we feel that a bulk calculation of a Page curve begs for a bulk understanding.  What we have done in this paper demonstrates this even more starkly -- we found that there are very natural {\em bulk} objects in flat space that act as proxies for AdS entanglement wedges and the like, and that they too exhibit a phase transition at the Page time. But since now we are in flat space, there cannot possibly be a dual plasma ball interpretation.}.

%Note that all one has to do to {\em not} see the Page curve is  ignore the tensor factorization.

Let us also note a related point in AdS. We can consider AdS with a cut-off, like we did for flat space. This raises the possibility: can we treat the outside-the-cut-off region of AdS as a sink? Note that even in AdS the backreaction in the asymptotic region is small, see eg. \cite{Bousso}. But the crucial point that makes AdS different is that the propagation time from the cut-off to the boundary is finite. This means that one has to worry about boundary conditions to complete the problem, and if it is reflective (as in AdS/CFT without a sink) then we cannot truly treat the AdS asymptotic region as a sink: the radiation will go back into the system after reflection. The way out is that we have to add an explicit leaky boundary condition at the asymptotic boundary, which would be tantamount to adding a sink to the AdS/CFT system like in \cite{Penington, Almheiri}. This issue is naturally bypassed in flat space because the boundary is at ``infinity", and therefore it is automatically reasonable to think of the asymptotic region as a sink. 

 %In fact even with the right tensor factorization, we feel that the bulk Page curve found by \cite{Penington, Almheiri} requires a bulk understanding. In our view, the story does not merely end with as the bulk dual of the (trivial) boundary Page curve for a plasma ball coupled to a sink. 

Another caveat in discussing information paradox is that perhaps fully resolving it means going beyond the Page curve. While we believe the arguments of this paper give strong evidence (beyond that already presented in \cite{CK}) for a natural tensor factorization in flat space quantum gravity, this is not quite the same as answering the question: who is the observer who has access to this factorization? There is a case to be made that more sophisticated versions of the information paradox that go beyond the Page curve, should be phrased in terms of an observer. Perhaps a suitable infaller, perhaps one who can collect the Hawking radiation, perhaps something else. It also brings up questions about whether this observer is to be thought of as living within the Hilbert space of the quantum gravity, or is he/she supposed to be an external system to which the quantum gravity is coupled to. All of these questions are interesting and at least some of them are possibly profound, but some basic questions we believe are indeed tied to the Page curve -- one question being, are there any natural tensor factorizations in the system, that contain the black hole in one of the factors? Once a compelling answer is found, one can discuss unitarity by looking at the (lack of) turnaround of the Page curve. The results of \cite{Penington, Almheiri} and our work in this paper should be viewed in this context. 

Even though we will not address the question of actual measurements of Hawking radiation in detail, we will now present a set of comments, which may provide a heuristic picture of how to think about this. Along the way, we hope to strengthen the possibility  that information does indeed ``come out" of a flat space black hole. 
\begin{itemize}
\item The first observation is that small black holes in large-$N$ theories may be understandable in terms of deconfinement of a sub-matrix of size $M$ in the $SU(N)$ gauge theory. There is some evidence for this, see \cite{Berenstein1, Masanori1, Santos, Yaffe, Berenstein2, Masanori2} for some discussions. 
The off-diagonal modes that couple the $M \times M$ sector and the rest of the $N \times N$ are claimed to be heavy, which makes it plausible that they can thermalize among themselves and have the interpretation as small black holes. Further evidence (and some caveats) can be found in the references.
\item The second observation is that in AdS/CFT the AdS scale $R_{AdS}$ is related to the gauge group rank $N$ via Maldacena's  formula
\bea
\frac{R_{AdS}^4}{l_P^4} = N
\eea 
where $l_P$ is the Planck length. Note that this extremely basic formula  already contains a hint that looking at sub-matrices with smaller than $N$ size, might have something to do with looking at (black holes) smaller than the AdS scale\footnote{This picture may (speculatively) also provide a partial understanding of the role of boundary gauge invariance in the holographic duality. A connection with holographic RG also must be involved. An interesting fact about holographic RG that is relevant for our suggestion here is that on sub-AdS scales, it is a UV/UV correspondence \cite{Sam}.}. Together with the previous bullet point, this suggests that one could think of (semi-classical) black holes in AdS as lying in a range 
\bea
1 \ll \frac{R^4}{l_P^4} \lesssim N \label{BHrange}
\eea
where $R$ can be thought of loosely as the size of (a box containing) the black hole.
\item The third observation is that together, these two points suggest that Hawking radiation from a small black hole into the ambient asymptotically AdS spacetime, could be viewed as the leakage of heat from the deconfined $M \times M$ sub-matrix to the ``remaining" confined degrees of freedom. For taking this point of view, it may be natural to treat the zone region of the black hole also as part of the deconfined degrees of freedom. Note that this is natural, because the outside-the-horizon region of a large AdS black hole should not be viewed as having anything to do with confined degrees of freedom. 

\item The fourth observation is that in our eq. \eqref{large}, while the length scale $R$ shares many features of the AdS length scale, there is one sense in which it is different. This is in the fact that it is not a parameter of the underlying theory, and can be tuned. This is unlike $R_{AdS}$ in AdS/CFT, where it is simply the 't Hooft coupling (upto an $\alpha'$ factor). This means that in flat space we can crank up $R$ arbitrarily large if we want, and look at bigger and bigger length scales. This means that we can view this as a generalization of \eqref{BHrange} to flat space
\bea
1 \ll \frac{R^4}{l_P^4} < \infty. 
\eea
\item The above fact, ties in with the observation that one can think of flat space quantum gravity as the limit of AdS/CFT where we take $R_{AdS}$ in string units to infinity, while holding $g_s$ fixed, which amounts to holding $l_P$ above fixed while sending $N$ to infinity. 
\end{itemize}

These observations suggest that black holes which live inside our flat space screen of size $R$, Hawking radiate to a bigger region and perish, for the same reason that small black holes Hawking radiate and wither away into the rest of AdS -- they lose energy to the confined degrees of freedom. In flat space, in the AdS/CFT language the gauge group rank of the theory is formally infinite, and therefore all black holes, no matter what the size of the screen it takes to contain them, evaporate away. In this picture, flat space black holes are the partially deconfined phases of an $SU(\infty)$ theory. Note that it is satisfying for this partial deconfinement picture that the UV/IR relationship is reversed on sub-AdS scales (and flat space). This loosely ties in with our expectation that deconfined phase is tied to high energies/temperatures etc., as well as the expectation that smaller black holes have higher temperature. Indeed, the origins of the UV/UV arguments in \cite{Sam} can be traced to the specific heats of black holes.

Note that the deconfinement of a subset of color-space degrees of freedom  also makes it comprehensible why the (approximate) tensor factorization we presented may be meaningful, even though we have not relied on a factorization in the physical space of the CFT/hologram. With the new tensor factorization, the problem is now again much like that of a burning lump of coal, and information can come ``out" of the black hole due to Hawking radiation. If this picture is true, we have a setting to understand our Formulation 1 and the Page curve. We hasten to add however that there seem to be some questions regarding the partial deconfinement picture of small black holes that are not yet fully resolved \cite{Santos, Yaffe}. Also, for what we are suggesting to work, the picture will have to be much more generally true than ${\mathcal N}=4$ SYM on $\IR \times S^3$ and black holes on AdS$_5 \times S^5$. Our goal here is merely to present a scenario where our factorization may emerge.

After that speculative digression on sub-matrix deconfinement, let us make one final point before we conclude. Let us note how the firewall question is indirectly addressed by this web of recent ideas. The entanglement between the interior and exterior modes that is crucial in these discussions, is a result of the assumption that the horizon is smooth (which goes into the derivation of Hawking radiation). Note also that the interior being in the entanglement wedge of the radiation after the Page time is a necessary condition for one resolution of the firewall paradox -- the one where we declare that the early radiation and the interior are the ``same". This is explicitly realized when the entanglement wedge phase transition happens. So it seems to us that these results should be viewed as providing some evidence for the smoothness of the horizon\footnote{The firewall we are discussing here is the usual one -- the one in the context of probe quantum fields at a black hole horizon. There is another kind of firewall that has recently been noted \cite{Himanshu}, which arises due to backreacting classical fields that cause gravitational collapse. It has been argued that such scalar fields traverse a greater than ${\cal O}(1) $ field range during their evolution, which is believed by some, to lead to a breakdown of EFT. This could be viewed as a type of firewall argument.}.
 
In the above discussion, we have tried to emphasize our prejudices and potential pitfalls. If one grants these prejudices, our arguments demonstrate that entanglement wedge phase transitions \cite{Penington, Almheiri} are a key to the information paradox, not just in AdS but also in the ``real world".

%Finally, let us conclude the discussion of \cite{Laddha} on a point where we share their concern. The arguments we have presented, as well as the arguments in \cite{Penington, Almheiri}, rely on coupling the Hilbert space of a quantum gravity to an external sink Hilbert space and then assuming that the Hawking radiation can be be extracted out into the sink from the quantum gravity Hilbert space. Is this really how one should formulate the bulk version of the information paradox, in the sense that a single observer can access it? We are not yet sure about this, but many illustrious authors seem to think it is. What we are reasonably conident about, is that this is one formulation of the information paradox, one certainly having to do with Hawking radiation: whether

\section{Acknowledgments}

We thank Ben Craps, Oleg Evnin, Samir Mathur, Onkar Parrikar, Charles Rabideau, Suvrat Raju, Ashoke Sen and Amitabh Virmani for discussions. We especially thank Madhur Mehta for early collaboration on this project, and Vyshnav Mohan for related collaborations and discussions.

\appendix

\section{Induced Domain of Dependence $\mathfrak{D}_{\mathfrak{S}}(p,q)$ on the Screen}
In this appendix, we would like to plot the simplest case of a region corresponding to the induced domain of dependence $\mathfrak{D}_{\mathfrak{S}}(p,q)=\mathfrak{C}(p,q)\cap\mathfrak{S}$. For the sake of simplicity, we work with the cylindrical polar coordinate system $(t,r,\phi)$ in (2+1)-dimensional empty Minkowski spacetime. The bulk metric corresponding to this system is
\begin{equation}
    ds^2=-dt^2+dr^2+r^2d\phi^2
\end{equation}
In this coordinate system, consider two points $(t_0, R_c,\phi_1)$ and $(t_0, R_c,\phi_2)$ on the screen $\mathfrak{S}$ at $R=R_c$, at the same time $t=t_0$. The equation of the planes passing through these two points and elevated at angles of $\pm\pi/4$ can be written as
\begin{equation}
\label{A.2}
\begin{aligned}
   &\big[-(\sin{\phi_2}-\sin{\phi_1})(r\cos{\phi}-R_c\cos{\phi_1})+(\cos{\phi_2}-\cos{\phi_1})(r\sin{\phi}-R_c\sin{\phi_1})\big]\\ &\mp\big[(\cos{\phi_2}-\cos{\phi_1})\sin{\phi_0}-(\sin{\phi_2}-\sin{\phi_1})\cos{\phi_0}\big](t-t_0)=0
   \end{aligned}
\end{equation}
Here $\phi_0=\frac{\phi_1+\phi_2}{2}$ can be understood as the azimuthal angle corresponding to both the points $\ p \in \mathfrak{I}^+, \ q \in  \mathfrak{I}^-$. We wish to identify the curves corresponding to the intersection of these planes with the screen $\mathfrak{S}$ and thus we substitute $r=R_c$ in \eqref{A.2}. We can also set $t_0=0$ without loss of generality. 
\begin{figure}
        \begin{subfigure}[h]{0.45\textwidth}
                \centering
                \includegraphics[width=.85\linewidth]{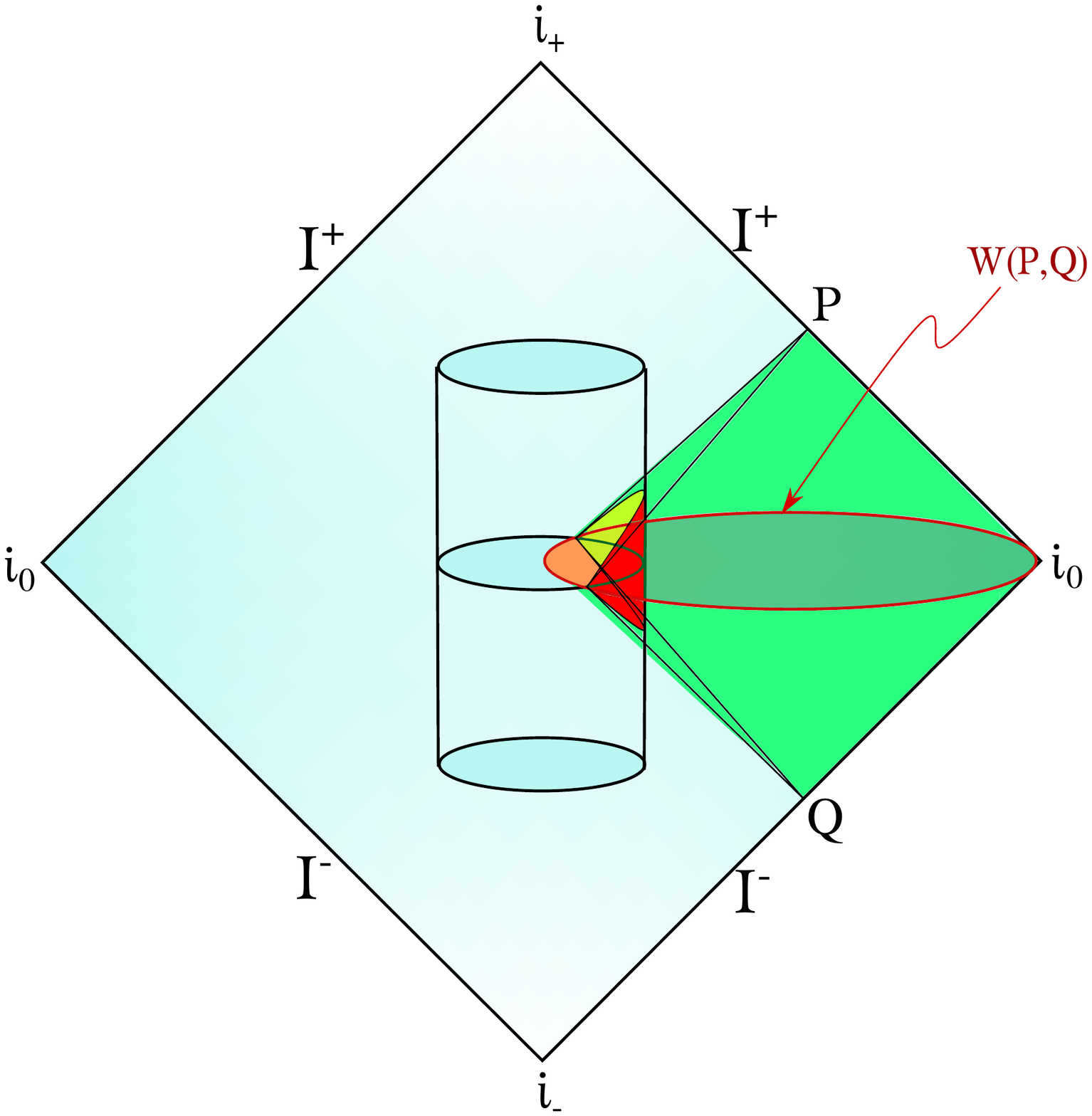}
                \vspace{-2cm}
                \caption*{}
                \label{schematic_Dind}
        \end{subfigure}%
        \begin{subfigure}[h]{0.45\textwidth}
                \centering
                \includegraphics[width=.85\linewidth]{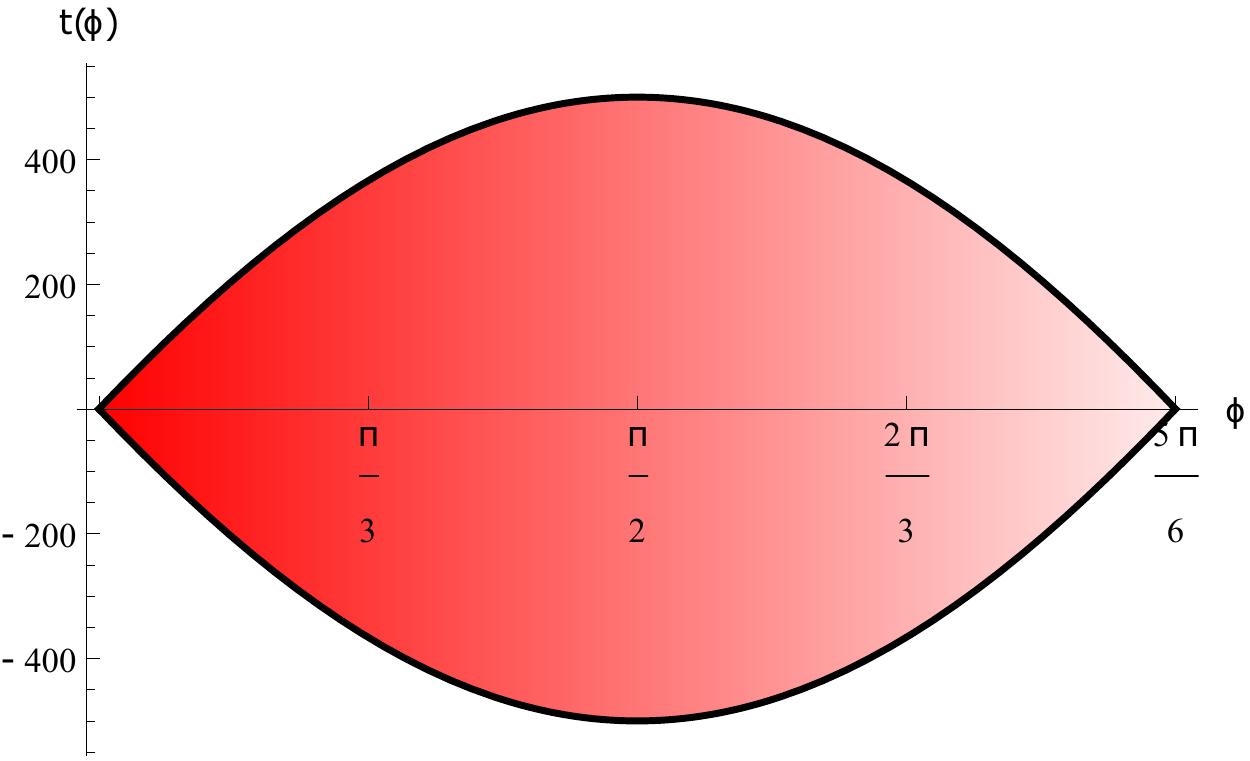}
                \vspace{-1.5cm}
                \caption*{}
                \label{plot_Dind}
        \end{subfigure}%
        \caption{Left panel: A schematic plot of the region $\mathfrak{D}_{\mathfrak{S}}(p,q)$ (shaded in red) enclosed by the two curves obtained by intersecting the two planes inclined at angles $\pm\pi/4$ with the holographic screen $\mathfrak{S}$. Right panel: Numerically constructed region $\mathfrak{D}_{\mathfrak{S}}(p,q)$ (shaded in red) enclosed between the two curves (thick black) obtained by substituting $\phi_1=\pi/6,\ \phi_2=5\pi/6,\ \phi_0=\pi/2\ {\rm and} \ R_c=1000$  in \eqref{A.3}. }\label{Dind}
\end{figure}

On solving we obtain,
\begin{equation}
\label{A.3}
    t(\phi)=\pm R_c\frac{(\sin{\phi}-\sin{\phi_1})-\alpha(\cos{\phi}-\cos{\phi_1})}{\sin{\phi_0}-\alpha\cos{\phi_0}}, \ {\rm where}\  \alpha=\frac{\sin{\phi_2}-\sin{\phi_1}}{\cos{\phi_2}-\cos{\phi_1}}
\end{equation}
The region enclosed by these two curves is the desired induced domain of dependence $\mathfrak{D}_{\mathfrak{S}}(p,q)$ on the screen $\mathfrak{S}$ (See Figure \ref{Dind}).

It is straightforward to check that the ``slope" of this curve $\frac{1}{R_c}\frac{dt}{d\phi}$ on the screen at any point is less than $\pi/4$. This is an indication that the effective propagation on the screen is superluminal, as expected by the fact that we are working at finite cut-off and the ``screen theory" is therefore non-local \cite{Budha}.

\section{Gravitational Boundedness vs Asymptotic Flatness}

The holographic screen we considered in this paper is to be viewed as a useful way to ``isolate" a black hole. Our primary interest is of course only in  black holes, but we suspect there may be a generalization to include arbitrary gravitationally bound systems. We will make some preliminary comments in this direction in this Appendix. 

Is it possible to formalize the notion of a system, that can interact with the asymptotic region via gravitational waves and Hawking radiation, but is otherwise gravitationally bound? The results of this paper suggest (but do not require) that this may be a well-defined limit of quantum gravity when describing the dynamics of certain classes of states. Note that this idea is slightly stronger than the usual notion of asymptotic flatness. This is because conventional asymptotic flatness allows, eg., a pair of black holes moving apart from each other at or above their mutual escape velocity \cite{Amitabh}\footnote{The discussion of \cite{Amitabh} is phrased in the rest frame of one of the black holes.}. This means that no box, no matter how large, will be able to contain the system if one waits long enough. The ultimate reason for this is that asymptotic flatness is defined for spatial and null infinities, and things escape from large boxes close to timelike infinity. 

Condition (c) in our definition (2.2) is almost certainly broken for the two black hole example above, but this can be easily remedied. Consider non-black massive objects that are gravitationally unbound, and they will also be able to escape the box. This can happen at arbitrarily small energies, so a cut-off on energy scale like our \eqref{large} also will not change the conclusion. For single black holes, by choosing the frame, we can make sure that they stay within the box\footnote{Note however that changing frames for massive spacetimes requires a boost charge that is infinite \cite{Traschen}. So they are best viewed as super-selection sectors.}. But if the system is not gravitationally bound, there is no such frame. Note that from the perspective of the goals of this paper, these issues are tangential: as long as we can treat the black hole whose evaporation we are interested in as isolated, we are in business. 

Finally let us observe the situation captured by AdS. In AdS, massive particles are bound, but null radiation can reach the boundary. In other words, it is precisely parallel to the situation we outlined in this Appendix.

\section{Strong Subadditivity for the Bulk Area Term in \eqref{screenF}}

For our proposal \eqref{screenF} to have any chance of being correct, a necessary condition is that it must satisfy strong subadditivity (SSA). If we ignore subtleties due to corner terms etc., it is possible to show that the leading contributions in \eqref{screenF} do satisfy SSA. Here we will demonstrate it for the leading term \eqref{AreaScreen} in the bulk area contribution across the screen sub-region:
\begin{equation}
    S_{bulk}(R_0) \approx \dfrac{{\rm Area}(R_0)}{\epsilon^{d-1}},
\end{equation}
where $\epsilon$ is the UV cut-off. Note that when all three tensor factors involved are screen sub-regions, this leads to nothing new. But when one of the tensor factors is the sink Hilbert space, it leads to a new check. When considering the sink, this term is proportional to the area of the whole screen. Let it be denoted by $S$. For a union of sink with a cutoff sub-region, this term will be the area of the complement of that sub-region on the screen. If we take strong sub-additivity in the form $S(AB) + S(BC) \geq S(ABC) + S(B)$ with  distinct tensor factors $A$, $B$ and $C$, the case when $A,B,C$ are all sub-regions of the screen is obviously saturated. When $A,C$ are sub-regions on the screen and $B$ is the sink, we get 
\bea
\dfrac{{\rm Area}(A^c) + {\rm Area}(C^c)}{\epsilon^{d-1}} \geq \dfrac{{\rm Area}((A+C)^c) + {\rm Area}(S)}{\epsilon^{d-1}}.
\eea
Here too inequality is saturated\footnote{We have introduced the notatation $X^c$ for the complement of a region $X$ on the screen.}. It is also trivial to check that when $A$ is the sink and $B,C$ are sub-regions on the screen, the inequality is again satisfied (but not saturated for finite area $C$).

\end{document}